%% file: template.tex
\pdfoutput=1

\documentclass[a4paper, 12pt]{article}

\input{Packages.tex}

\title{Математические модели распространения COVID-19}
\author{
\begin{tabular}[t]{c@{\extracolsep{8em}}c} 
О. И. Криворотько & С. И. Кабанихин \\
НГУ & ИМ СО РАН \\ 
Новосибирск, Россия & Новосибирск, Россия \\
krivorotko.olya@mail.ru & ksi52@mail.ru
\end{tabular}
}

\date{}

\makeatletter         
\def\@maketitle{
	\begin{figure}[ht]
	   \minipage{0.76\textwidth}
			\label{EscudoUNAM}
	   \endminipage
	   \minipage{0.32\textwidth}
			\label{EscudoFC}
		\endminipage
	\end{figure}
	
	\vspace{5cm}
	
\begin{center}
{\Huge \bfseries \sffamily \@title }\\[4ex] 
{\Large  \@author}

\vspace{12cm}

Новосибирск, 2022
\end{center}}
\makeatother

\begin{document}

\maketitle

\abstract{В работе приведена классификация и анализ математических моделей распространения \nom{COVID-19}{коронавирусная инфекция, вызванная штаммом вируса SARS-CoV-2} в различных группах населения: семья, школа, офисы (3-100 человек), регионы (100-5000 человек), города, области (0.5-15 миллионов человек), страны, континенты и Земной шар. Рассмотрены основные группы моделей (основанные на анализе временных рядов, дифференциальные, имитационные, а также их комбинации). В основе первой группы лежит анализ временных рядов на основе методов фильтрации, регрессионных и сетевых моделей (Раздел 2). Вторая группа основана на уравнениях (Раздел 3): системы обыкновенных дифференциальных уравнений, стохастические дифференциальные уравнения, уравнения в частных производных. Третья группа (Раздел 4) -- имитационные модели, включая клеточные автоматы и агентно-ориентированные модели. Четвертая группа моделей (Раздел 5) основана на комбинации нелинейных марковских цепей, оптимального управлении, объединенных в рамках теории игр среднего поля. В силу новизны и сложности заболевания COVID-19 параметры большинства моделей, как правило, неизвестны, и это приводит к необходимости рассматривать и решать обратные задачи. В работе приведен анализ основных алгоритмов решения обратных задач эпидемиологии: стохастическая оптимизация, природоподобные алгоритмы (генетический, дифференциальной эволюции, роя частиц и т.п.), методы усвоения, анализа больших данных и машинного обучения.}

\vspace{20mm}
\begin{center}
    {\bf Рецензенты:}
    \vspace{2mm}
    
    доктор физико-математических наук\\ \textit{Г. А. Бочаров} (ИВМ РАН им. Г.И. Марчука)
    
    \vspace{2mm}
    
    доктор физико-математических наук\\ \textit{А. В. Булинский} (МГУ им. М.В. Ломоносова)
    
    \vspace{2mm}
    
    член-корреспондент РАН\\ \textit{А. А. Шананин} (МФТИ)
\end{center}

\newpage
\tableofcontents

\newpage
\renewcommand{\nomname}{Список сокращений} 
\printnomenclature[5em]
\addcontentsline{toc}{section}{Список сокращений}

\newpage
\input{text/Introduction}

\newpage
\input{text/1time_series}

\newpage
\input{text/2equation_based_model}

\newpage
\input{text/3CA_ABM}

\newpage
\input{text/4MFG_model}

\newpage
\input{text/5algorithms}

\newpage
\input{text/6conclusion}

\newpage
\input{text/7references}

\input{text/Appendix}

\input{text/Authors}

\end{document}

%% file: Packages.tex
\usepackage{lscape}		

\usepackage{cmap}						
\usepackage[T2A]{fontenc}				
\usepackage[utf8]{inputenc}				
\usepackage[english, russian]{babel}	

\usepackage{amssymb,amsfonts,amsmath,cite,enumerate,float,indentfirst} 
\numberwithin{equation}{section}
\numberwithin{figure}{section}
\numberwithin{table}{section}
\usepackage{bm} 

\usepackage{enumitem}
\usepackage{esvect}

\usepackage{indentfirst} 

\usepackage{comment}
\usepackage{algorithm} 
\usepackage[noend]{algpseudocode}

\usepackage[usenames]{color}
\usepackage{color}
\usepackage{colortbl}
\usepackage[dvipsnames]{xcolor}

\usepackage{longtable}					
\usepackage{multirow,makecell,array}	
\usepackage{array,tabularx,tabulary,booktabs} 
\usepackage{hhline}

\usepackage[singlelinecheck=off,center]{caption}	
\usepackage{soul}									
\usepackage{fixltx2e}                               

\usepackage{cite} 

\usepackage[plainpages=false,pdfpagelabels=false]{hyperref}
\definecolor{linkcolor}{rgb}{0.9,0,0}
\definecolor{citecolor}{rgb}{0,0.6,0}
\definecolor{urlcolor}{rgb}{0,0,1}
\hypersetup{
    colorlinks, linkcolor={linkcolor},
    citecolor={citecolor}, urlcolor={urlcolor}
}

\usepackage{graphicx}		
\graphicspath{{images/}}	
\usepackage{rotating}
\usepackage{wrapfig}


\usepackage{nomencl}
\makenomenclature    

\newcommand*{\nom}[2]{#1\nomenclature{#1}{#2}}

\usepackage{fancyhdr}
\fancypagestyle{plain}{%
   \fancyhead{} 
}

%% file: text/Introduction.tex
\section{Введение}
В работе изложены математические основы программного комплекса COVID-19, созданного в СО РАН совместно с коллегами из РФЯЦ-ВНИИТФ им. академ. Е.И.~Забабахина (Снежинск), ФИЦ КНТ (Красноярск), МФТИ и МГУ (Москва), а также коллегами из Болгарии, Великобритании, Казахстана, Китая, США. В основе комплекса программ лежит комбинация трех основных типов моделей: SIR, АОМ и ИСП.

В работе рассматриваются математические модели, которые используются для анализа и прогнозирования развития пандемии COVID-19, включая анализ временных рядов, уравнения математической физики и имитационные модели, различные их комбинации и обобщения. В силу новизны и сложности заболевания COVID-19 параметры большинства математических моделей, как правило, неизвестны, и это приводит к необходимости рассматривать и решать обратные задачи. Основные проблемы моделирования распространения COVID-19:
\begin{enumerate}
    \item[1.] Данные для решения обратной задачи являются неполными и зашумленными, а также представляют собой большие данные (ежедневные сводки о заболевших, заразившихся, вакцинированных и т.д.).
    \item[2.] Параметры меняются со временем: контагиозность $\alpha (t)$, вероятность появления тяжелых случаев $1-\beta (t)$, смертность $\mu (t)$, процент бессимптомных случаев $\alpha_E(t)$ и т.д.
    \item[3.] Процесс распространения COVID-19 существенно изменяется при введении или отмене ограничительных мер (ношение масок, социальная дистанция, перевод на удаленный рабочий режим, закрытие школ, предприятий, районов и городов), появление каждого нового штамма (альфа, бета, гамма, дельта, \ldots).
\end{enumerate}
С учетом 1.-3. решение обратных задач требует новых подходов и методов, включая байесовский, стохастическую оптимизацию, природоподобные алгоритмы (генетический, имитации отжига, глубокие нейронные сети) и другие методы машинного обучения и искусственного интеллекта.
В работе будет приведен краткий обзор и анализ работ по указанных новым подходам.

Статья организована следующим образом. В разделе~\ref{sec_time_series} изложены методы анализа временных рядов, которые составляют основу статистических данных в эпидемиологии, а именно: статистические методы (подраздел~\ref{sec_regression}), методы на основе машинного обучения (подраздел~\ref{sec_ML_models}) и на базе фильтрации (подраздел~\ref{sec_filtr_models}). В разделе~\ref{sec_equation_models} приведен обзор математических моделей моделирования вспышки COVID-19, в основе которых лежат системы обыкновенных дифференциальных уравнений (подраздел~\ref{sec_ODE_models}) и уравнений в частных производных (подраздел~\ref{sec_conv-diff}). Ключевой характеристикой описания распространения эпидемии является индекс репродукции вируса $\mathcal{R}_0$, показывающий количество возможных инфицированных в результате контакта одного инфицированного с восприимчивой популяцией. Алгоритм вывода $\mathcal{R}_0$ в общем случае, а также для SEIR-HCD модели, представлен в разделе~\ref{sec_reproductionNo}. В разделе~\ref{sec_CA-ABM} приведено описание имитационных моделей, а именно: в Разделе~\ref{sec_cellular_automata} описаны математические модели, основанные на теории клеточных автоматов, в Разделе~\ref{sec_ABM} -- агентно-ориентированные модели (АОМ)\nomenclature{АОМ}{агентно-ориентированная модель}. Подробное построение агентной модели для описания распространения COVID-19 приведено в подразделе~\ref{sec_ABM_COVID}. Раздел~\ref{sec_MFG_model} посвящен моделям игр среднего поля (ИСП)\nomenclature{ИСП}{игра среднего поля (mean-field game)} с управляющим уравнением первого порядка. В разделе~\ref{sec_KFP} описано уравнение Колмогорова-Фоккера-Планка (КФП)\nomenclature{КФП}{уравнение Колмогорова-Фоккера-Планка}, определяющее распределение агентов в моделях эпидемиологии, а в разделе~\ref{sec_GJB} -- уравнение Гамильтона-Якоби-Беллмана (ГЯБ)\nomenclature{ГЯБ}{уравнение Гамильтона-Якоби-Беллмана}, отвечающее за оптимальное управление. В разделе~\ref{sec_algorithms} приведены численные алгоритмы решения прямых (подраздел~\ref{sec_DirectPr}) и обратных (подраздел~\ref{sec_InversePr}) задач эпидемиологии. Результаты численных расчетов проанализированы в разделе~\ref{sec_num_results}. Также в подразделе~\ref{sec_num_conclusion} сформулированы выводы и в~\ref{sec_future_work} -- направления дальнейших исследований.

В приложении~\ref{sec_app_data} изложено описание используемых эпидемиологических данных распространения COVID-19 в Новосибирской области, используемые в численных расчетах. В приложении~\ref{sec_app_software} приведено описание комплекса программ и характеристики численных расчетов.

В качестве иллюстрации в работе приведены результаты расчетов по трем моделям: \nom{SEIR-HCD}{камерная модель, основанная на системе из 7 обыкновенных дифференциальных уравнений, где $E$~-- бессимптомные носители инфекции (Exposed), $H$~-- госпитализированные (Hospitalized), $C$~-- критические случаи (Critical), нуждающиеся в подключении аппарата ИВЛ, $D$~-- умершие (Death)} -- камерная модель, основанная на системе из 7 обыкновенных дифференциальных уравнений (Раздел~\ref{sec_R0-SEIR-HCD}), агентная (Раздел~\ref{sec_ABM_COVID}) и модель игры среднего поля (Раздел~\ref{sec_num_res_MFG}). 

%% file: text/1time_series.tex
\section{Модели, основанные на анализе временных рядов}\label{sec_time_series}
В разделе будут рассмотрены регрессионные и сетевые модели, методы фильтрации и их взаимосвязи~\cite{Kondratiev_2013}, на основе анализа наиболее достоверных статистических данных -- количество ежедневных \nom{ПЦР}{полимеразная цепная реакция -- экспериментальный метод молекулярной биологии, способ значительного увеличения малых концентраций определённых фрагментов нуклеиновой кислоты (ДНК) в биологическом материале}-тестов $T(t)$ и индекс самоизоляции $a(t)$. Время $t$ во всех моделях измеряется в днях.

\subsection{Введение}
Особенностью функций $T(t)$ и $a(t)$ является повторяемость подъемов и спадов по времени с изменяющейся амплитудой. Например, $T(t)$ возрастает по вторникам и ослабевает к понедельнику, а $a(t)$ ослабевает в период выходных, праздников и отпусков.

\begin{sidewaysfigure}
    \centering
    \includegraphics[width=\textwidth]{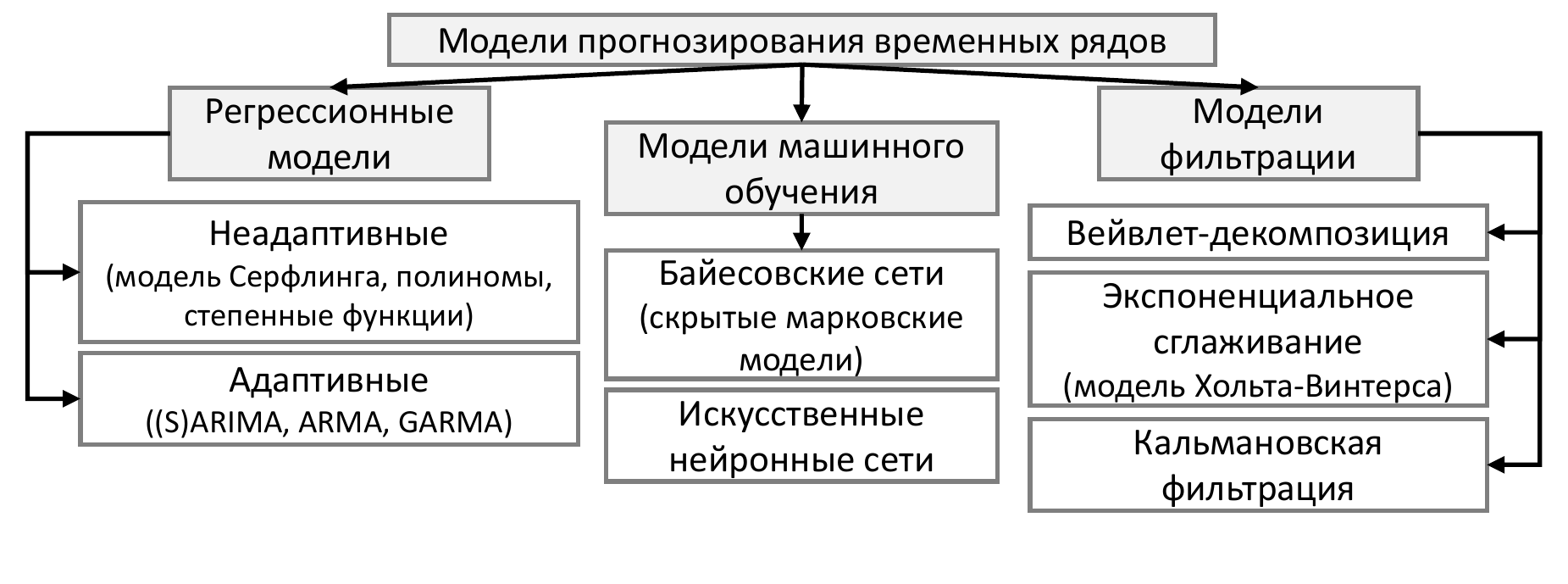}
    \caption{Диаграмма моделей прогнозирования временных рядов.}
    \label{fig:diag_time-series}
\end{sidewaysfigure}

Методы математической статистики и машинного обучения помогают обрабатывать, анализировать эпидемиологические данные и проводить краткосрочное прогнозирование поведения $T(t)$ и $a(t)$ при отсутствии резких изменений ситуации (введение ограничительных мер, мутации вируса)~\cite{Zaharov_2021}. В разделе будут приведены регрессионные, сетевые модели, а также методы фильтрации и их взаимосвязи, диаграмма которых приведена на рис.~\ref{fig:diag_time-series}.

Отметим, что в данном разделе мы рассматриваем модели, которые используют только значения наблюдаемых показателей $T(t)$ и $a(t)$.

\subsection{Регрессионные модели}\label{sec_regression}
Регрессионные модели подразделяются на неадаптивные модели, для оценки параметров которых используются все имеющиеся данные, и адаптивные, значения параметров которых рассчитываются на основе скользящего окна наблюдений~\cite{Burkom_2007}.

Вид регрессионной зависимости для неадаптивной модели выбирается исходя из свойств анализируемого временного ряда. В отдельных случаях можно применять полиномиальные или степенные функции, но чаще модель должна учитывать сезонный характер заболеваемости (например, модель Серфлинга~\cite{Serfling_1963}).
Неадаптивные регрессионные модели учитывают всю предысторию заболеваемости, но игнорируют локальные колебания эпидемических показателей, поэтому появление нового штамма вируса и введение ограничительных мер в регионе снижают значимость данных предыдущего периода (устаревшие данные).

Адаптивные регрессионные модели используют ограниченный отрезок временного ряда и поэтому более чувствительны к изменению ситуации. Важную роль при использовании адаптивных моделей играет ширина скользящего окна (в пределах нескольких месяцев) --
число последних наблюдений, на основе которых оценивают параметры модели.

В регрессионных моделях предполагается, что невязка модели (ошибка предсказания) -- независимая случайная величина, имеющая нормальный закон распределения с нулевым математическим ожиданием и постоянной дисперсией~\cite{Kondratiev_2013}. Одной из проблем прогнозирования является наличие существенной автокорреляции невязок. Регрессионная модель может быть дополнена, а прогноз -- уточнен, например, с помощью авторегрессионных моделей.

\subsubsection{Авторегрессионная модель прогнозирования временного ряда}
Для построения прогнозов сезонных рядов, например, количества проведенных ПЦР-тестов в регионе, использовалась авторегрессионная модель \textsc{SARIMA} являющейся модификацией модели \textsc{ARIMA} (AutoRegressive Integrated Moving Average)~\cite{Boks_1974}, которая описывает одномерные временные ряды с сезонной компонентой~\cite{SARIMA_2017}. \textsc{ARIMA} является расширением моделей тип \textsc{ARMA} для нестационарных временных рядов, которые можно сделать стационарными взятием разностей некоторого порядка от исходного временного ряда (так называемые интегрированные или разностно-стационарные временные ряды).

Модель \textsc{ARIMA}$(p,d,q)$ для нестационарного временного ряда $T(n)$ имеет вид:
\begin{equation*}
    \triangle^d T(n) = c + \sum_{i=1}^p a_i \triangle^d T(n-i) + \sum_{j=1}^q b_j \epsilon(n-j) + \epsilon(n).
\end{equation*}
Здесь $\epsilon(n)$ -- стационарный временной ряд белого шума, $c, a_i, b_j$ -- параметры модели, $\triangle^d$ -- оператор разности временного ряда порядка $d$, гарантирующий стационарность ряда (последовательное взятие $d$ раз разностей первого порядка — сначала от временного ряда, затем от полученных разностей первого порядка, затем от второго порядка и т.д.). Обычно, при построении модели \textsc{ARIMA} порядок разностей ограничивается числом $d=2$.

Алгоритм прогнозирования временного ряда следующий:
\begin{enumerate} 
  \item Применяем преобразование Бокса-Кокса \cite{Box_1964} для уменьшения дисперсии.
  \item Вычисляем сезонную разность (сдвиг на 7 дней) первого порядка.
  \item Вычисляем вторую разность (сдвиг на 1 день) ряда, полученного в пункте 2.
  \item Проверяем стационарность ряда из п. 3 критерием Дики-Фуллера.
  \item Передаем соответствующие проделанным действиям параметры в модель \textsc{ARIMA}$(p,d,q)$ и подбираем остальные на основе минимизации информационного критерия Акаике. В качестве данных передается ряд из п. 1.
  \item Полученная модель с настроенными гиперпараметрами используется для дальнейшего прогнозирования. Полученный результат подвергается обратному преобразованию Бокса-Кокса.
\end{enumerate}

\begin{figure}[h!]
    \centering
    \includegraphics[width=\textwidth]{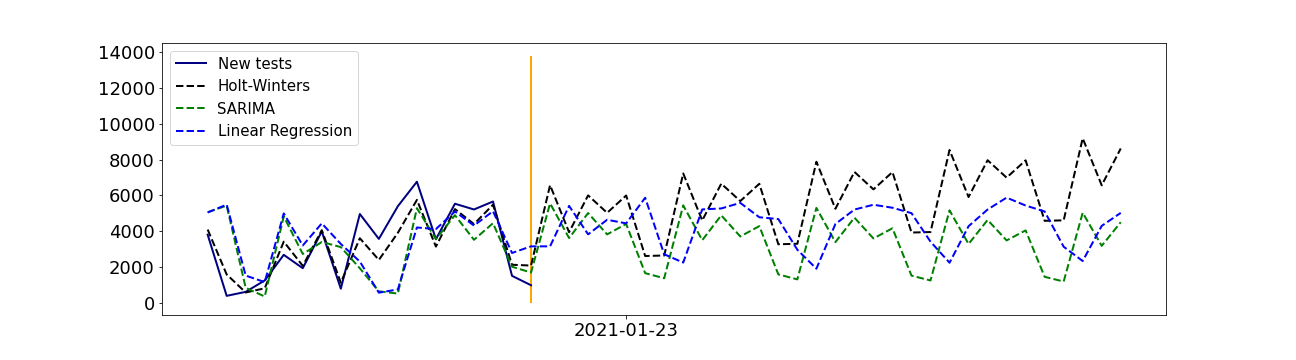}
    \caption{Результаты предсказания временного ряда ежедневно проводимых ПЦР-тестов $T(t)$ в Новосибирской области (сплошная синяя линия) на месяц вперед с 18.01.2021 (вертикальная оранжевая линия) моделями Хольта-Винтерса (черная пунктирная линия), линейной регрессии (синяя пунктирная линия) и \textsc{SARIMA} (зеленая пунктирная линия).}
    \label{fig:forecasting}
\end{figure}

Был построен прогноз временного ряда количества ежедневных ПЦР-тестов в Новосибирской области с 18.01.2021 по 17.02.2021 на основе исторических наблюдений с 05.05.2020 по 17.01.2021 с помощью моделей \textsc{SARIMA}, Хольта-Винтерса \cite{HoltWinters_1987} (подробнее см.~Раздел~\ref{sec_filtr_models}) и машинного обучения линейной регрессии (см.~Рис.~\ref{fig:forecasting}). \textsc{SARIMA} показала на предсказаниях разных промежутков в среднем не лучшие, но наиболее устойчивые результаты (усредненная абсолютная ошибка для модели \textsc{SARIMA} равна 1454.73, для модели Хольта-Винтерса -- 1646.49, для модели линейной регрессии -- 1586.34).

\subsection{Модели на основе машинного обучения}\label{sec_ML_models}
Машинное обучение является мощным инструментом для поиска взаимосвязи между входными и выходными данными в случаях, когда аналитическое исследование затруднительно. Применение эвристических подходов для раннего обнаружения эпидемиологических рисков в некоторых случаях позволяет улучшить качество прогнозирования.

\begin{figure}[!ht]
    \begin{minipage}[h]{0.45\linewidth}
    \center{\vspace{11mm}\includegraphics[width=\textwidth]{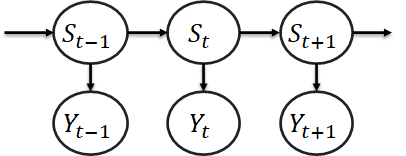}\\[26pt] а)} 
    \end{minipage}
    \hfill
    \begin{minipage}[h]{0.45\linewidth}
    \center{\includegraphics[width=\textwidth]{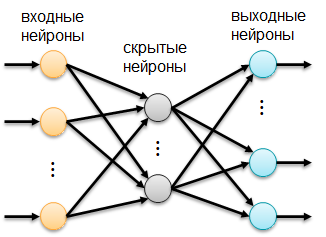}\\ б)} 
    \end{minipage}
    \caption{Схема зависимостей в скрытой марковской модели (а) и ИНС с одним скрытым слоем (б).}
    \label{fig:ANN_Markov-chain_scheme}
\end{figure}

Одним из представителей моделей машинного обучения являются динамические байесовские сети -- ориентированный граф, вершины которого соответствуют переменным модели, а ребра -- вероятностным зависимостям между ними, которые заданы определенными законами распределения~\cite{Sebastiani_2006}. После обучения как на большом, так и на малом количестве исходных данных байесовские сети позволяют оценить вероятность наступления некоторого события при наблюдаемой последовательности явлений. Для прогнозирования заболеваемости используется простая форма скрытых марковских моделей, основной идеей которых является сопоставление каждой случайной величины $Y_t$ (например, количество выявленных, госпитализированных случаев с COVID-19) с ненаблюдаемой случайной величиной $S_t$ (например, общее количество инфицированных индивидуумов), определяющей условное распределение $Y_t$~\cite{Strat_1999}. Таким образом, величина $Y_t$ зависит только от значения скрытой переменной $S_t$ в момент времени $t$, а последовательность $S_t$ обладает марковским свойством, то есть величина $S_t$ зависит только от $S_{t-1}$ (рис.~\ref{fig:ANN_Markov-chain_scheme}а).

Искусственные нейронные сети (ИНС) представляют собой направленный взвешенный граф, вершины которого моделируют функционирование биологических нейронов (рис.~\ref{fig:ANN_Markov-chain_scheme}б). Обучение ИНС заключается
в вычислении коэффициентов связей между вершинами, определяющих силу входящих сигналов, и выполняется на основе эмпирических данных: статистики заболеваемости и при наличии, значений факторов, ее предопределяющих. Для корректного обучения ИНС необходим большой объем исторических данных. В
исследовании~\cite{Wieczorek_2020} рассматриваются возможности применения нейронных сетей для прогнозирования распространения COVID-19. Результаты работы сети разработанной архитектуры для некоторых регионов достигали 87\%.

\subsection{Модели на основе фильтрации}\label{sec_filtr_models}
Любые временные ряды заболеваемости можно рассматривать как случайный процесс, состоящий из сигнала, отражающего реальную эпидемическую обстановку, и высокочастотного шума. Фильтрация шума позволяет уточнить прогноз и может выполняться как в ходе предварительной обработки исходных данных, так и в составе самого алгоритма прогнозирования.

Одним из подходов является вейвлет-декомпозиция, в которой временной ряд представляется с помощью вейвлет-функций~\cite{Shmueli_2006}. Однако, такой подход используется совместно с другими моделями.

Одной из таких моделей является экспоненциальное сглаживание, представляющее собой частный случай взвешенного скользящего среднего, а именно значение заболеваемости $y_t$ в момент времени $t$ описывается взвешенной суммой последних наблюдений: $l_t = b y_t + (1 - b)y_{t-1}$. Здесь $b\in (0,1)$ -- коэффициент сглаживания, который обеспечивает уменьшение веса по мере старения данных, которое может рассматриваться как отражение естественного процесса обучения. Такой метод построения модели не подходит для рядов, в поведении которых присутствуют отчетливый тренд или сезонность. Для этих целей используются обобщенные модели~\cite{Gardner_1985}, например, сезонная модель Хольта-Винтерса~\cite{HoltWinters_1987}. Результаты прогнозирования ряда ежедневных ПЦР-тестов в Новосибирской области в рамках данной модели представлены на рис.~\ref{fig:forecasting}.

Любые эпидемические процессы можно описать следующей системой разностных уравнений:
\begin{eqnarray*}
\begin{array}{ll}
    {\bf x}_t = {\bf A}{\bf x}_{t-1} + {\bf w}_t,\\
    {\bf y}_t = {\bf H}{\bf x}_t + {\bf D}{\bf f}_t + {\bf v}_t,
\end{array}
\end{eqnarray*}
где ${\bf x}_t$~-- вектор переменных состояний системы в момент времени $t$, ${\bf y}_t$~-- вектор наблюдений, ${\bf f}_t$~-- вектор значений внешних факторов, ${\bf w}_t$ и ${\bf v}_t$~-- белый шум. Матрицы параметров ${\bf A}$, ${\bf H}$, ${\bf D}$ определяют модель эпидемического процесса и выбираются исходя из решаемой задачи -- краткосрочного или долгосрочного прогнозирования.

Такая форма записи позволяет предложить обобщенные модели распространения заболевания, в частности, модели на основе калмановской фильтрации~\cite{Hamilton_1994}.

\subsection{Выводы}
Результаты, полученные на основе временных рядов, использованы нами в дифференциальных и агентных моделях. Прогнозирование ежедневных ПЦР-тестов и индекса самоизоляции позволяют строить сценарии развития COVID-19 в регионе в зависимости от введения ограничительных мер, а именно количество ожидаемых выявленных, умерших, госпитализированных, критических случаев и индекса репродукции вируса (численные расчеты приведены в~Разделе~\ref{sec_num_results}).

%% file: text/2equation_based_model.tex
\section{Модели, основанные на дифференциальных уравнениях}\label{sec_equation_models}
В данном разделе приведен обзор математических моделей, основанных на обыкновенных дифференциальных уравнениях (ОДУ)\nomenclature{ОДУ}{обыкновенные дифференциальные уравнения}, приведенных в Разделе~\ref{sec_ODE_models}, на уравнениях в частных производных (УРЧП)\nomenclature{УРЧП}{уравнения в частных производных}, описанных в Разделе~\ref{sec_conv-diff}, и на стохастических дифференциальных уравнениях (СДУ)\nomenclature{СДУ}{стохастические дифференциальные уравнения}, приведенных в Разделах~\ref{sec_SDE} и~\ref{sec_MFG_model}, а также описание ключевой характеристики распространения заболевания -- индекс репродукции вируса (п.~\ref{sec_reproductionNo}). В качестве примера будет представлена SEIR-HCD модель, описанная в п.~\ref{sec_R0-SEIR-HCD}.

\subsection{Введение}
\begin{sidewaysfigure}
    \centering
    \includegraphics[width=\textwidth]{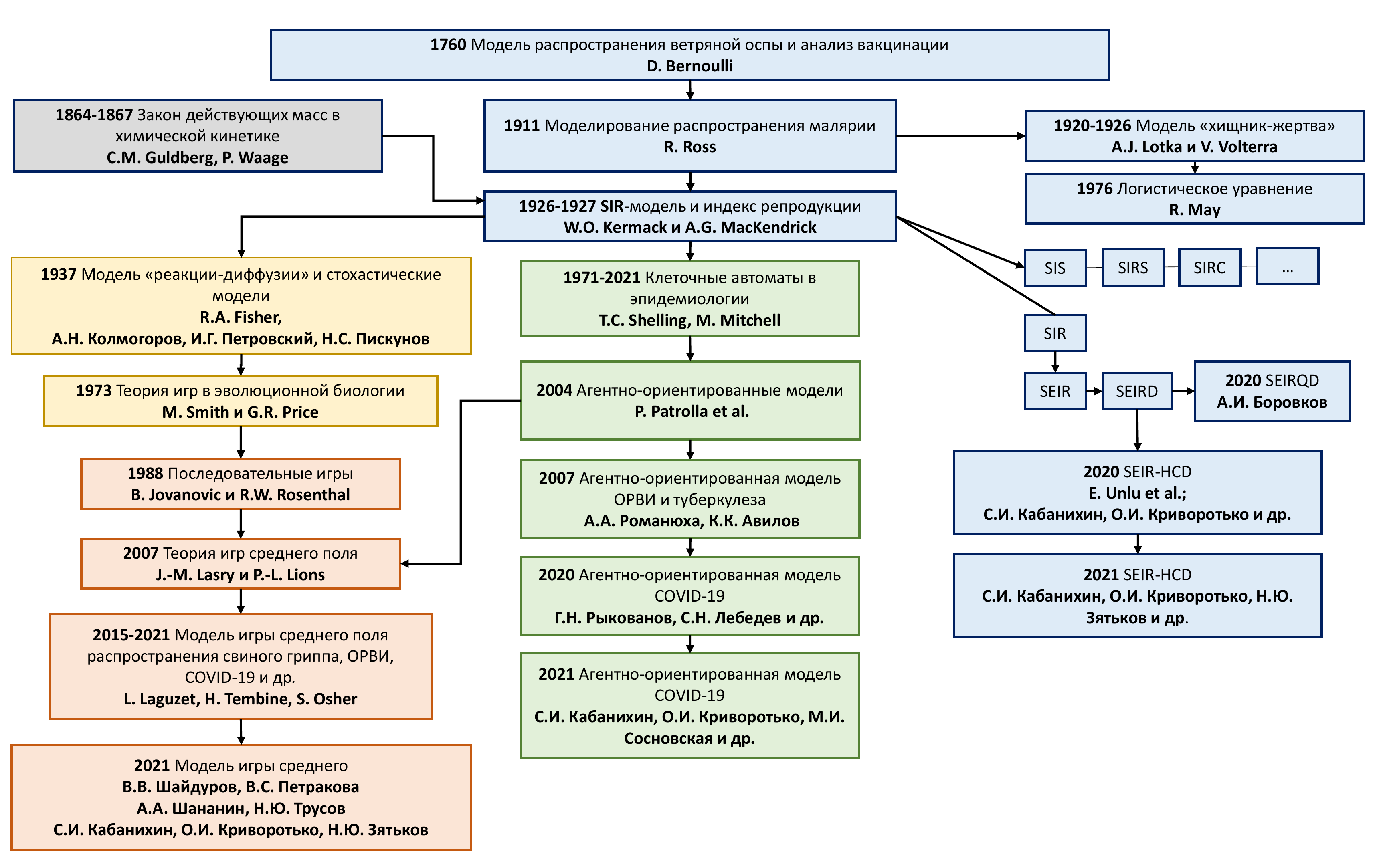}
    \vspace*{-5mm}
    \caption{Диаграмма развития математических моделей эпидемиологии с 1760 года на основе камерных, имитационных моделей и их комбинации. В синих рамках приведены результаты по применению ОДУ к описанию эпидемий, в желтых рамках -- УРЧП, в оранжевых -- модели ИСП и в зеленых рамках -- АОМ.}
    \label{fig:hist_review}
\end{sidewaysfigure}

\begin{figure}[!ht]
    \centering
    \includegraphics[width=0.8\textwidth]{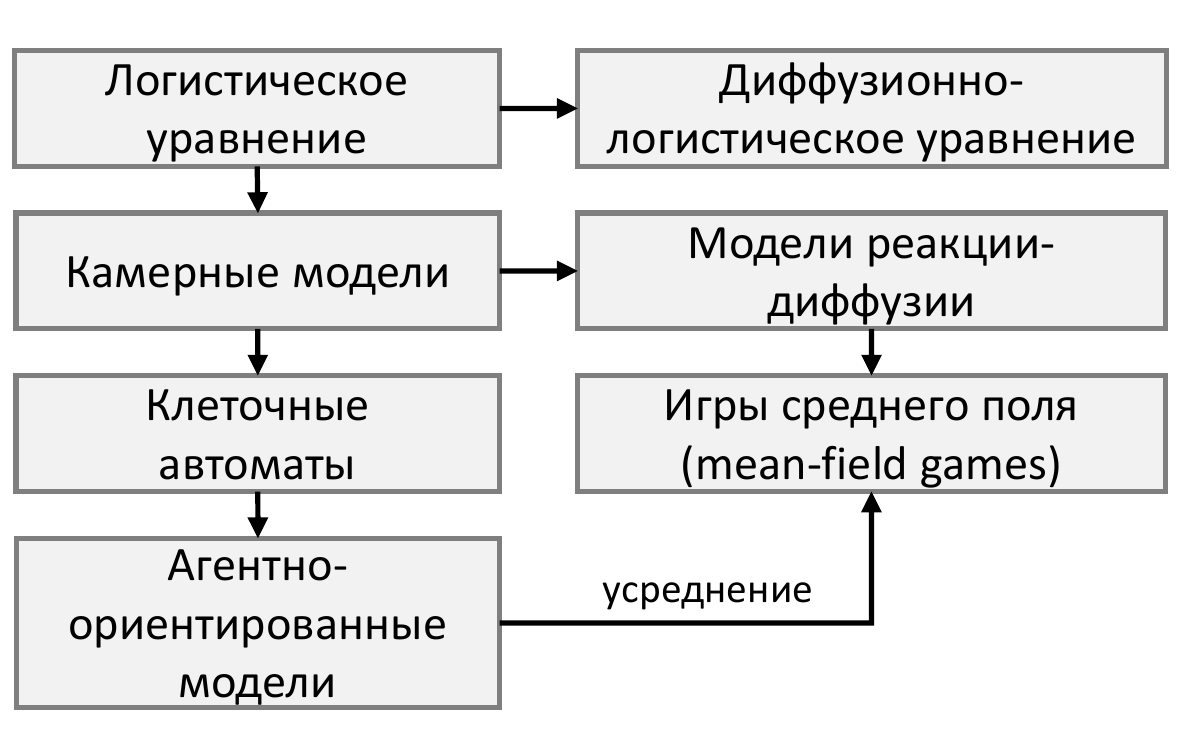}
    \caption{Взаимосвязь математических моделей распространения COVID-19 на основе камерного подхода и имитационного моделирования.}
    \label{fig:COVID_review}
\end{figure}

Дифференциальные модели основаны на законе сохранения масс и особенностях передачи инфекции. Диаграмма~\ref{fig:hist_review} иллюстрирует развитие моделей эпидемиологии с 1760 по настоящее время.
Для описания распространения новой коронавирусной инфекции, вызванной вирусом штамма SARS-CoV-2 были использованы все основные достижения, отраженные в диаграмме~\ref{fig:hist_review}. На рис.~\ref{fig:COVID_review} приведена классификация существующих математических моделей динамики COVID-19 и их взаимосвязи. Одна из простейших моделей вспышки COVID-19 описывается логистическим уравнением~(\ref{eq:logistic_eq}). Добавление состояний агентов (бессимптомные, госпитализированные, критические, умершие случаи и т.д.) к логистическому уравнению приводит к камерным SIR-моделям~(\ref{model_KermackMackendrick}) и~(\ref{eq:SEIR-HCD}). Учет пространственной неоднородности приводит к диффузионно-логистическому уравнению~(\ref{model_PDE}). Дальнейшие преобразования камерных моделей можно условно разделить на два направления: учет пространственной структуры в непрерывной (модель реакции-диффузии) и дискретной (модель конечных автоматов) постановках. Добавление управления системами состояний формирует новый класс моделей ИСП~(\ref{eq_3})-(\ref{eq_4}), а учет индивидуальных характеристик агентов -- к АОМ. Отметим, что усреднение в АОМ приводит к моделям среднего поля, а именно нелинейным цепям Маркова, в которых вероятности перехода зависят от распределения состояний агентов (подробнее см.~Раздел~\ref{sec_MFG_model}).

\subsection{SIR-модели}\label{sec_ODE_models}
\begin{sidewaysfigure}
    \begin{minipage}[h]{0.26\linewidth}
    \center{\includegraphics[width=0.95\textwidth]{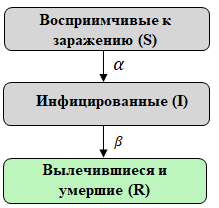}\\ а) SIR\\[5pt] \includegraphics[width=\textwidth]{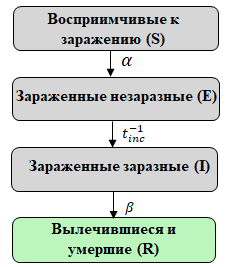}\\ б) SEIR} 
    \end{minipage}
    \hfill
    \begin{minipage}[h]{0.6\linewidth}
    \center{\includegraphics[width=\textwidth]{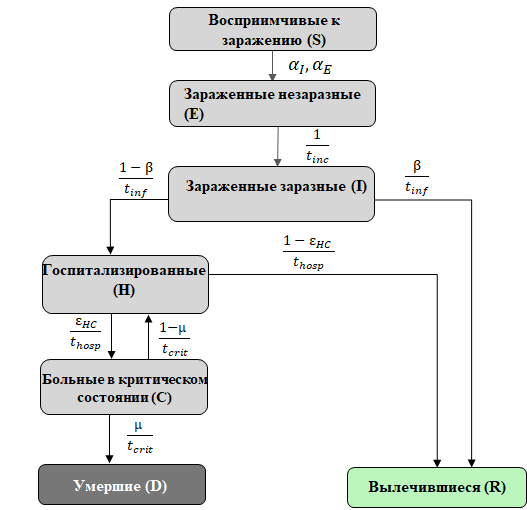}\\ в) SEIR-HCD} 
    \end{minipage}
    \caption{Схемы камерных моделей (а) SIR, (б) SEIR и (в) SEIR-HCD.}
    \label{fig:SEIR-HCD_scheme}
\end{sidewaysfigure}

Математическое моделирование в эпидемиологии началось с работы D.~Bernoulli в 1760 году, в которой была продемонстрирована эффективность вакцинации населения против ветряной оспы~\cite{Bernoulli_1760}. Впоследствии появилась серия математических моделей, основанные на законе баланса масс (см.~обзорные статьи~\cite{Bacaer_2011, Brauer_2017} и приведенную в них литературу). Работы R.~Ross в 1911~\cite{Ross_1911}, A.J.~Lotka в 1920~\cite{Lotka_1920} и V.~Volterra в 1926~\cite{Volterra_1926} (модель <<хищник-жертва>>), привели к созданию камерной \nom{SIR}{модель, основанная на законе сохранения масс, в которой вся популяция разделена на 3 группы: $S$~-- восприимчивые (Susceptible), $I$~-- инфицированные (Infected) и $R$~-- вылеченные и умершие (Removed)}-модели W.O.~Kermack и A.G.~McKendrick \cite{McKendrick_1926, KermackMcKendrick_1927}
\begin{eqnarray}\label{model_KermackMackendrick}
\left\{\begin{array}{ll}
    \dfrac{dS}{dt} = -\alpha \dfrac{SI}{N}, & t> 0,\\[5pt]
    \dfrac{dI}{dt} = \alpha \dfrac{SI}{N} - \beta I,\\[5pt]
    \dfrac{dR}{dt} = \beta I,
\end{array}\right.
\end{eqnarray}
в которой популяция из $N$ особей разделена на три группы (камеры): $S$~-- восприимчивые, $I$~-- инфицированные ($I << N$) и $R$~-- вылеченные и умершие, связанные между собой вероятностными переходами $\alpha, \beta\in (0,1)$ (схема модели приведена на Рис.~\ref{fig:SEIR-HCD_scheme}а). Одним из важных результатов работы~\cite{KermackMcKendrick_1927} было введение индекса репродукции (заразности)
\[\mathcal{R}_0 = \dfrac{\alpha}{\beta},\]
которая является важнейшей характеристикой заболевания и параметром распространения эпидемии (см. подробнее раздел~\ref{sec_reproductionNo}).

Отметим, что частным случаем модели~(\ref{model_KermackMackendrick}) при условии отсутствия иммунитета после перенесенного заболевания в рамках исследуемых временных промежутков (например, сезонный грипп) является SI-модель, решение которой сводится к логистическому уравнению:
\begin{eqnarray}\label{eq:logistic_eq}
    \dfrac{dI}{dt} = (\alpha-\beta)I(t) + \dfrac{\alpha}{N}I^2(t).
\end{eqnarray}

Важным свойством модели~(\ref{model_KermackMackendrick}) является выполнение закона действующих масс (закон сохранения), в рамках которого моделируемая популяция постоянна в течение всего времени $N = S(t)+I(t)+R(t)$.

В работах Е.Н.~Пелиновского и его коллег~\cite{Pelinovsky_2020, Wang_2020, Koltcova_2020} для моделирования распространения COVID-19 применяется обобщенное логистическое уравнение, описывающее рост численности заболевших. Предположение о единственности пика вспышки $I(t)$ эпидемии ограничивает применение логистической модели для описания длительного периода пандемии и учета ограничительных мер.

Для учета инкубационного периода течения COVID-19 используется модификация модели Кермака-Маккендрика SEIR-типа (схема SEIR-модели приведена на Рис.~\ref{fig:SEIR-HCD_scheme}б), которых на сегодняшний день разработано более 100 моделей~(см. например, работы~\cite{Cheng_2020, Tamm_2020, Unlu_2020, KOI_KSI_2020, Borovkov_2020, Yang_2021, Kolpakov_2021} и ссылки в них). В этих моделях популяция разделяется на группы (кроме $S$, $I$, $R$ добавляются $E$ -- бессимптомные носители, $H$ -- госпитализированные, $C$ -- критические случаи, требующие подключения аппарата искусственной вентиляции легких (ИВЛ), $D$ -- умершие в результате COVID-19, $Q$~-- помещенные на карантин и другие). Это позволяет уточнить эпидемиологическую картину в регионе за счет варьирования более детального набора коэффициентов в уравнениях. Недостатком SIR-моделей является отсутствие гибкости -- невозможность учета изменения параметров (новые мутации вируса и штамма, ограничительные меры, вакцинация). При попытке ввести в SIR-модели указанные изменения (например, сделать переменной скорость передачи инфекции $\alpha=\alpha (t)$)~\cite{Popivanov_2021}, мы сталкиваемся с неединственностью и неустойчивостью решения обратной задачи идентификации этого параметра $\alpha(t)$.

Отметим, что SIR-модели используют также для прогнозирования результатов управления развития пандемии~\cite{Silva_2021}, т.е. в правую часть уравнений добавляют кусочно-постоянную функцию управления (ограничительные меры: ношение масок, социальная дистанция, карантин). Однако и в этих случаях проблема уточнения коэффициентов моделей SIR остается открытой и требует применения методов теории обратных задач.

Также в SIR-моделях можно учитывать влияние суперраспространителей на распространение COVID-19 (инфицированные индивидуумы, у которых повышенная вирусная концентрация)~\cite{Ndairou_2020}. Однако теоретическое определение этого явления требует моделирования в масштабе отдельных людей (АОМ, см.~Раздел~\ref{sec_ABM}).

Дальнейшее развитие математических моделей можно разделить на две составляющие: введение пространственной координаты в логистические уравнения и учет дискретной пространственной неоднородности. В первом случае мы получаем новый класс математических моделей <<реакции-диффузии>> (см.~Раздел~\ref{sec_conv-diff}), а во втором -- новый подход, в котором системы дифференциальных уравнений соединяются в пространстве графовой структурой (см.~Раздел~\ref{sec_CA-ABM}).

\subsection{Модели <<реакции-диффузии>>}\label{sec_conv-diff}
В 1937 году британский ученый R.A.~Fisher~\cite{Fisher_1937} и советские математики А.Н.~Колмогоров, И.Г.~Петровский и Н.С.~Пискунов~\cite{KolmPetrPiskunov_1937} предложили и обосновали математическую модель, основанную на уравнении в частных производных параболического типа, получившую впоследствии название модели <<реакции-диффузии>>, которая была также применена при описании процессов в биологии, экологии и других приложениях:
\begin{eqnarray}\label{model_PDE}
    \dfrac{\partial u}{\partial t} = f(u) + d\dfrac{\partial^2 u}{\partial x^2}.
\end{eqnarray}
Здесь $u(x,t)$ -- вектор плотности распределения групп популяции в точке пространства $x$ и времени $t$, $d$~-- коэффициент диффузии, $f(u)$~-- функция, характеризующая характер распространения заболевания в популяции, удовлетворяющая закону сохранения масс и условиям
\[f(0)=f(1)=0,\quad f(u)>0,\;\; \text{если}\;\; 0<u<1,\]
\[f^\prime (0)>0\;\; \text{и} \;\; f^\prime (u)<f^\prime (0), \;\; \text{если}\;\; 0<u\leqslant 1.\]
Авторы работы~\cite{KolmPetrPiskunov_1937} строго доказали, что если начальное условие удовлетворяет следующим ограничениям
\[0\leqslant u(x,0)\leqslant 1,\; u(x,0)=0\;\; \forall\, x<x_1 \;\; \text{и}\;\; u(x,0)=1\;\; \forall\, x>x_2\geqslant x_1,\]
то динамика популяции по переменной $t$ описывается скоростью $v^*=2\sqrt{f^\prime(0)d}$.

Учет пространственной неоднородности позволяет более точно моделировать эпидемию от очага распространения (крупного города в стране, столицы в регионе и пр.) при известных начальных условиях. Так, в работах~\cite{Viguerie_2020, Aristov_2021, Barwolff_2021, Lau_2021_arXiv} получены оценки распространения COVID-19 в первые месяцы с начала эпидемии с учетом пассажиропотоков. Показано, что учет неоднородности влияет на характер распространения COVID-19 в крупных регионах и странах. В разделе~\ref{sec_MFG_model} будут рассмотрены стохастические дифференциальные уравнения, учитывающие пространственную неоднородность и элементы управления (задачи ИСП). Однако использование модели для описания второй и последующих волн эпидемии требует добавления уравнений в~(\ref{model_PDE}), введения множественных известных источников распространения заболевания $u(x,0)$ (обратная задача определения источника) и вычислительных ресурсов. 

\subsection{Стохастические модели}\label{sec_SDE}
Использование СДУ позволяет учитывать и анализировать случайные флуктуации эпидемиологического процесса (процессы заражения, тестирования, выздоровления и др.). В работе~\cite{Kolokoltsov_2012_arXiv} показано, что нелинейные кинетические уравнения с мерой, описывающие динамический закон предела больших чисел для системы с большим числом $N$ агентов, разрешимы и что их решения представляют собой $1/N$-равновесия по Нэшу для аппроксимирующих систем из $N$ агентов. Указанное заключение получено в предположении, что в основе динамики репрезентативных агентов лежит управляемый нелинейный марковский процесс, связанный с интегро-дифференциальными генераторами типа Леви-Хинчина (с переменными коэффициентами), с основным упором на приложения к устойчивым и устойчивоподобным процессам. В частности, показано, что функция управления $\gamma$ задает нелинейный марковский процесс $\mu_t$, порожденный семейством операторов $A[t,\mu_t,\gamma (t,\cdot)]$, через общее кинетическое уравнение в слабой форме~\cite{Kampen_1990}:
\[\dfrac{d}{dt}(g,\mu_t) = \left( A[t,\mu_t, \gamma(t,\cdot)]g,\mu_t \right),\quad \mu_0 = \mu.\]

Более подробное описание моделей, основанных на СДУ, а также управление такими процессами приведено в Разделе~\ref{sec_MFG_model}.

Известно, что плотность потока частиц в размножающей среде при достаточно широких условиях асимптотически экспоненциальна по времени $t$ с некоторым параметром $\lambda$, т.е. с показателем $\lambda t$. В работах~\cite{Lotova_2020, Lotova_2021} показано, что если среда случайна, то параметр $\lambda$ -- случайная величина, и для оценки временной асимптотики среднего (по реализациям среды) числа частиц можно в некотором приближении осреднять экспоненту по распределению $\lambda$. В предположении гауссовости этого распределения таким образом получается асимптотическая “сверхэкспоненциальная” оценка среднего потока, выражаемая экспонентой с показателем $tE\lambda + t^2D\lambda/2$. Для численной экспериментальной проверки такой оценки разработано вычисление вероятностных моментов случайного параметра $\lambda$ на основе рандомизации фурье-приближений специальных нелинейных функционалов.

Отмечено, что результаты в статье имеют широкое приложение. В частности, согласно статистике \nom{ВОЗ}{Всемирная организация здравоохранения}, сверхэкспоненциальное поведение проявляла пандемия COVID-19, развивающаяся во всем мире. А именно, количество наблюдений (по дням) аппроксимируется с точностью до 2\% с 9 марта 2020 года по 21 марта 2020 года.

\subsection{Индекс репродукции вируса $\mathcal{R}_0$ для дифференциальных моделей}\label{sec_reproductionNo}
Общее определение индекса репродукции (basic reproduction number)~\cite{R0_wiki}: индекс репродукции вируса $\mathcal{R}_0$ определяется как среднее количество людей, которых заражает активный инфицированный, попавший в полностью неиммунизированное окружение при отсутствии специальных эпидемиологических мер, направленных на предотвращение распространения заболевания.

В разделе~\ref{sec_R0} мы покажем, что индекс репродукции $\mathcal{R}_0$ является границей устойчивости состояния равновесия SIR-системы при отсутствии инфицированных. Если $\mathcal{R}_0>1$, то на начальном этапе число заболевших будет расти экспоненциально. Если $\mathcal{R}_0 \in (0,1)$, то небольшое количество инфицированных людей, попавших в полностью восприимчивую популяцию, в среднем не смогут сохранить свою группу, и эпидемии не будет.

\subsubsection{Алгоритм вычисления индекса репродукции}\label{sec_R0}
Опишем метод вычисления индекса репродукции, предложенный van den Driessche и Watmough~\cite{Driessche_2008} для SIR-моделей. Разделим всю популяцию на две категории: $n$ групп инфицированных и $m$ групп не инфицированных. Векторы $x\in \mathbb{R}^n$ и $y\in\mathbb{R}^m$ определяют количество индивидуумов в каждой из двух категорий, например, в модели SEIR-HCD $x=(E,I,H,C)\in \mathbb{R}^4$, $y=(S,R,D)\in \mathbb{R}^3$ (см. Раздел~\ref{sec_R0-SEIR-HCD}). Тогда SIR-модель распространения инфекционного заболевания принимает вид
\begin{eqnarray}\label{eq:ODE_2compartment}
\begin{array}{ll}
    \dot{x}_i = \mathcal{F}_i(x,y) - \mathcal{V}_i(x,y),\quad &  i=1,\ldots,n,\\
    \dot{y}_j = g_j(x,y), & j=1,\ldots,m.
\end{array}    
\end{eqnarray}
Здесь $\mathcal{F}_i$ -- скорость изменения количества инфицированных и $\mathcal{V}_i$ -- скорость изменения умерших, выздоровевших и тех, кто заболевает в $i$-й группе.

Вывод индекса репродукции заключается в линеаризации системы~(\ref{eq:ODE_2compartment}) в окрестности состояния равновесия в случае отсутствия инфекции $(0,y_0)$. В работе~\cite{Driessche_2008} были введены пять условий существования этого состояния равновесия, согласованных с законом сохранения масс для модели модели~(\ref{eq:ODE_2compartment}). Вычислим собственные значения матрицы $FV^{-1}$, где
\[F_{ij} = \dfrac{\partial \mathcal{F}_i}{\partial x_j}(0,y_0),\quad V_{ij} = \dfrac{\partial \mathcal{V}_i}{\partial x_j}(0,y_0).\]
Тогда индекс репродукции $\mathcal{R}_0 = \max \lambda_i(FV^{-1})$ является неотрицательным и соответствующий ему собственный вектор $\omega$ состоит из неотрицательных компонент~\cite{Berman_1970}. Компоненты вектора $\omega$ можно интерпретировать как распределение инфицированных индивидуумов, вызывающих наибольшее количество $\mathcal{R}_0$ вторичных инфекций в поколении.

В~\cite{Driessche_2008} доказана теорема о локальной устойчивости неинфицированного состояния равновесия системы~(\ref{eq:ODE_2compartment}), а именно состояние равновесия $(0, y_0)$ локально асимптотически устойчиво, если $\mathcal{R}_0<1$, и неустойчиво, если $\mathcal{R}_0>1$. В следующем разделе приведен вывод индекса репродукции для SEIR-HCD модели распространения COVID-19.

\subsubsection{Индекс репродукции вируса для модели SEIR-HCD}\label{sec_R0-SEIR-HCD}
Используя описанный алгоритм, выведем индекс репродукции для модели SEIR-HCD распространения COVID-19~\cite{Unlu_2020, KOI_KSI_2020}
\begin{eqnarray}\label{eq:SEIR-HCD}
    \left\{\begin{array}{ll}
        \dfrac{dS}{dt} = -\dfrac{5-a(t-\tau)}{5}\left( \dfrac{\alpha_I(t) S(t)I(t)}{N} + \dfrac{\alpha_E(t) S(t)E(t)}{N}\right), \\
        \dfrac{dE}{dt} = \dfrac{5-a(t-\tau)}{5}\left( \dfrac{\alpha_I(t) S(t)I(t)}{N} + \dfrac{\alpha_E(t) S(t)E(t)}{N}\right) - \dfrac{1}{t_{inc}}E(t), \\
        \dfrac{dI}{dt} = \dfrac{1}{t_{inc}}E(t) - \dfrac{1}{t_{inf}}I(t), \\
        \dfrac{dR}{dt} = \dfrac{\beta}{t_{inf}}I(t) - \dfrac{1-\varepsilon_{HC}}{t_{hosp}}H(t), \\
        \dfrac{dH}{dt} = \dfrac{1-\beta}{t_{inf}}I(t) + \dfrac{1-\mu}{t_{crit}}C(t) - \dfrac{1}{t_{hosp}}H(t), \\
        \dfrac{dC}{dt} =  \dfrac{\varepsilon_{HC}}{t_{hosp}}H(t) - \dfrac{1}{t_{crit}}C(t), \\
        \dfrac{dD}{dt} =  \dfrac{\mu}{t_{crit}}C(t)
    \end{array}\right.
\end{eqnarray}
с начальными условиями
\begin{equation}\label{eq:SEIR-HCD_init_cond}
\begin{split}
    S(t_0)=N-E_0-I_0-R_0-H_0-C_0-D_0,\; E(t_0)=E_0,\\
    I(t_0)=I_0,\; R(t_0)=R_0,\; H(t_0)=H_0,\; C(t_0)=C_0,\; D(t_0)=D_0.
\end{split}
\end{equation}

Схема модели~(\ref{eq:SEIR-HCD}) приведена на Рис.~\ref{fig:SEIR-HCD_scheme}в, а описание и значения параметров и начальных условий для Новосибирской области приведены в таблице~\ref{tab_parameters} (начальный момент времени полагается 15.04.2020). В SEIR-HCD модели происходит перемещение бессимптомной популяции $E(t)$ после $t_{inc}$ дней в симптоматическую $I(t)$. Инфицированные индивидуумы после $t_{inf}$ дней выздоравливают с вероятностью $\beta$ и госпитализируются $H(t)$ с вероятностью $1-\beta$. Затем госпитализированные могут выздоравливать или нуждаться в подключении аппарата ИВЛ $C(t)$. В модели только критические случаи могут умереть $D(t)$ с вероятностью $\mu$.

Разделяя всю популяцию $N$ в модели~(\ref{eq:SEIR-HCD}) на инфицированных $x=(E,I,H,C)\in \mathbb{R}^4$ и неинфицированных $y=(S,R,D)\in \mathbb{R}^3$, получим следующие вектор-функции, согласно~(\ref{eq:ODE_2compartment}):
\begin{eqnarray*}
\begin{array}{c}
\mathcal{F}=\left(\begin{array}{c}
    \tilde{a}(t) \left(\frac{\alpha_I(t) S(t)I(t)}{N} + \frac{\alpha_E(t) S(t)E(t)}{N}\right)\\
    0 \\ 0 \\ 0
\end{array}\right), \quad
\mathcal{V}=\left(\begin{array}{c}
    \frac{1}{t_{inc}}E(t)\\
    -\frac{1}{t_{inc}}E(t) + \frac{1}{t_{inf}}I(t) \\
    -\frac{1-\beta}{t_{inf}}I(t) - \frac{1-\mu}{t_{crit}}C(t) + \frac{1}{t_{hosp}}H(t) \\
    -\frac{\varepsilon_{HC}}{t_{hosp}}H(t) + \frac{1}{t_{crit}}C(t)
\end{array}\right),
\end{array}
\end{eqnarray*}
где $\tilde{a}(t) = \dfrac{5-a(t-\tau)}{5}$.

\begin{center}
\begin{table}[h!]
\caption{Описание и значения параметров SEIR-HCD модели}\label{tab_parameters}
 \begin{tabular}{|p{2cm}|p{11cm}|p{2cm}| } 
 \hline
 Параметр & Описание & Значение\\ 
  \hline
$a(t)$ & Индекс самоизоляции от Яндекса & $(0, 5)$ \\ 
$\alpha_E(t)$ & Параметр заражения между бессимптомной и восприимчивой группами населения ($\alpha_E >> \alpha_I$) & $(0, 1)$ \\
$\alpha_I(t)$ & Параметр заражения между инфицированным и восприимчивым населением & $(0,1)$\\
$\beta$ & Доля инфицированных, которая переносит заболевание без осложнений & $(0,1)$\\
$\varepsilon_{HC}$ & Доля госпитализированных случаев, которым требуется подключение ИВЛ & $(0,1)$\\
$\mu$ & Доля смертельных случаев в результате COVID-19 & $(0,0.5)$\\
$\tau$ & Латентный период (характеризует запаздывание наступления заразности) & 2 дня \\
$t_{inc}$ & Длительность инкубационного периода & 2-14 дня \\
$t_{inf}$ & Длительность периода инфицирования & 2.5-14 дня \\
$t_{hosp}$ & Длительность периода госпитализации & 4-5 дня \\
$t_{crit}$ & Длительность использования аппарата ИВЛ & 10-20 дня \\
$N$ & Население в Новосибирской области (человек) & 2798170 \\
$E_0$ & Начальное количество бессимптомных носителей & (1, 5000) \\
$I_0$ & Начальное количество инфицированных случаев & (1, 5000) \\
$R_0$ & Начальное количество вылеченных случаев & (1, 100) \\
$H_0$ & Начальное количество госпитализированных & 133 \\
$C_0$ & Начальное количество критических случаев & 1 \\
$D_0$ & Начальное количество смертей & 1 \\
\hline
\end{tabular}
\label{interventions}
\end{table}
\end{center}

Состояние равновесия системы~(\ref{eq:SEIR-HCD})-(\ref{eq:SEIR-HCD_init_cond}) в случае отсутствия инфицированных индивидуумов есть $(N,0,0,0,0,0,0)$. Тогда матрицы $F$ и $V$ имеют вид:
\begin{eqnarray*}
F=\left(\begin{array}{cccc}
    \tilde{a}\alpha_E &  \tilde{a}\alpha_I & 0 & 0\\
    0 & 0 & 0 & 0\\
    0 & 0 & 0 & 0\\
    0 & 0 & 0 & 0
\end{array}\right), \quad
V=\left(\begin{array}{cccc}
    \frac{1}{t_{inc}} & 0 & 0 & 0\\
    -\frac{1}{t_{inc}} & \frac{1}{t_{inf}} & 0 & 0\\
    0 & -\frac{1-\beta}{t_{inf}} & \frac{1}{t_{hosp}} & -\frac{1-\mu}{t_{crit}}\\
    0 & 0 & -\frac{\varepsilon_{HC}}{t_{hosp}} & \frac{1}{t_{crit}}
\end{array}\right).
\end{eqnarray*}

Вычисляя максимальное собственное значение матрицы $FV^{-1}$, получим индекс репродукции для SEIR-HCD модели
\begin{eqnarray}\label{eq:R0_SEIR-HCD}
    \mathcal{R}_0(t) = \tilde{a}(t) \left( \alpha_E(t) t_{inc} + \alpha_I(t) t_{inf} \dfrac{1+\varepsilon_{HC}(1-\mu)}{1-\varepsilon_{HC}(1-\mu)}\right),
\end{eqnarray}
первое слагаемое в котором характеризует заразность от бессимптомной части популяции, а второе слагаемое -- от симптомной с учетом госпитализированных случаев, пребывающих на изоляции.

\subsection{Выводы}
Преимущество использования дифференциальных моделей для описания распространения эпидемий (в том числе COVID-19) состоит в учете особенностей инфекции (наличие инкубационного периода, взаимосвязь инфицированных и критических случаев и т.д.) в параметрах моделей и закона сохранения масс (размера популяции), а в случае добавления переменной $x$ (модели <<реакции-диффузии>>) -- учет пространственной неоднородности. Такие модели качественно описывают вспышку эпидемии, но не обладают достаточной гибкостью. Учет изменения параметров для описания ограничительных мер, новых мутаций вируса приводит к неединственности и неустойчивости решения задачи их идентификации.

%% file: text/3CA_ABM.tex
\section{Агентно-ориентированные модели}\label{sec_CA-ABM}
В данном разделе будет приведен краткий обзор имитационных моделей, основанные на клеточных автоматах, и АОМ. Подробное построение АОМ распространения COVID-19 в конкретной регионе приведено в конце раздела~\ref{sec_ABM_COVID}.

\subsection{Введение}
Значимым преимуществом моделей, базирующихся на аппарате дифференциальных уравнений, является возможность их аналитического исследования. Тем не менее для всех таких моделей характерно допущение -- характеристики и поведение всех индивидов, отнесенных к одной подгруппе, считаются одинаковыми. Имитационные модели (клеточные автоматы, сетевые модели и АОМ) позволяют ослабить указанные ограничения.

\subsection{Клеточные автоматы}\label{sec_cellular_automata}
\begin{figure}[!ht]
    \begin{minipage}[h]{0.49\linewidth}
    \center{\includegraphics[width=\textwidth]{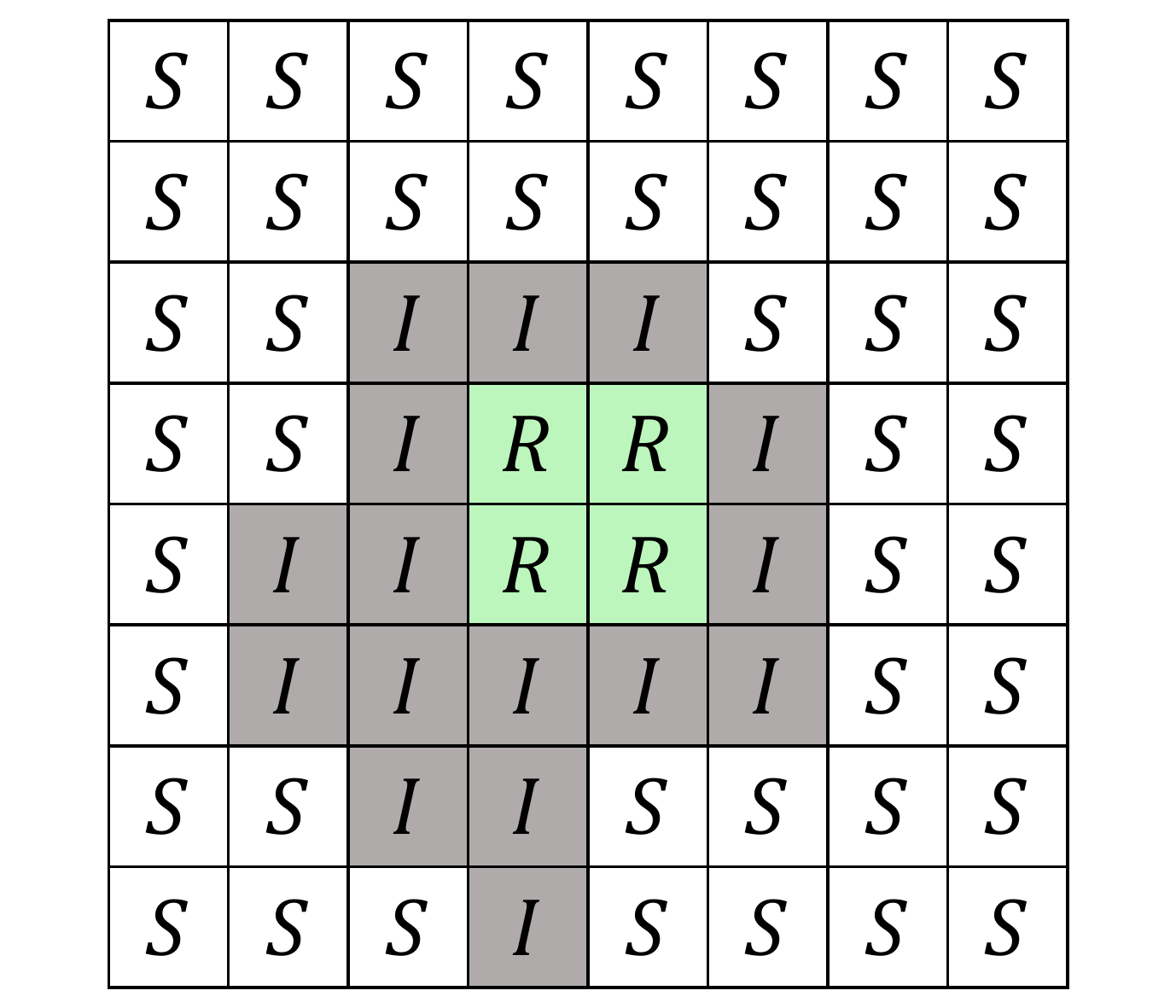}\\ а)} 
    \end{minipage}
    \hfill
    \begin{minipage}[h]{0.49\linewidth}
    \center{\includegraphics[width=\textwidth]{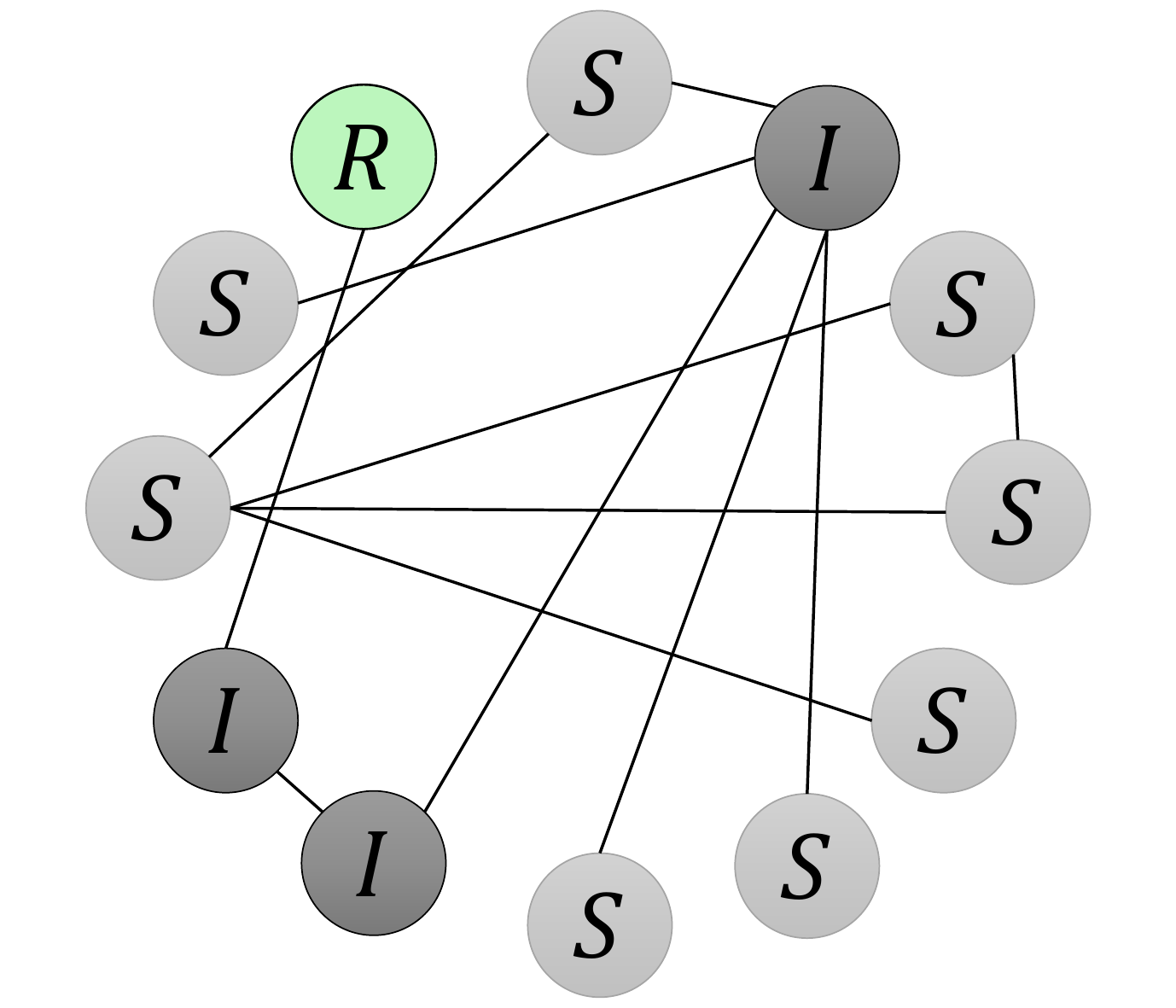}\\ б)} 
    \end{minipage}
    \caption{Представление распространения инфекции клеточным автоматом (а) и сетевой моделью (б).}
    \label{fig:CA_ABM_scheme}
\end{figure}

T.C.~Shelling в 1971~\cite{Schelling_1971} и M.~Mitchel в 1993~\cite{Mitchell_1993} предложили теорию клеточных автоматов для моделирования локальных характеристик восприимчивых популяций вместе со стохастическими параметрами, которые отражают вероятностный характер передачи болезни. Клеточные автоматы представляют собой совокупность квадратных ячеек, объединенных в прямоугольную решетку, каждая из которых принимает состояние из конечного множества. Узлы решетки моделируют индивидов, каждый из которых имеет фиксированное положение в пространстве (схема клеточного автомата представлена на Рис.~\ref{fig:CA_ABM_scheme}a). Так, анализ социальной дистанции, ношение масок при распространении COVID-19 в локальной популяции может описываться с помощью клеточных автоматов~\cite{Schimit_2021}. В работе~\cite{Dai_2021} SEIR модель описывается в терминах вероятностных клеточных автоматов и обыкновенных дифференциальных уравнений передачи COVID-19, достаточно гибких для моделирования различных сценариев социальной изоляции.

\subsection{Агентно-ориентированные модели}\label{sec_ABM}
После публикации статьи P.~Patrolla в 2004 году~\cite{Patlolla_2004} была предложена АОМ, в которой расширены возможности клеточных автоматов отслеживания распространения болезни и контактов между каждым человеком в социальной группе, расположенной в географической области. АОМ позволяют взаимодействовать между людьми и способны преодолевать ограничения различных подходов, обращаясь к естественной стохастической природе эпидемического процесса. АОМ представляют схему возможных контактов в виде динамического или статического графа, в котором вершины -- объекты с набором индивидуальных свойств, сколь угодно детализировано описывающие состояние отдельных индивидов (упрощенная схема графа в АОМ приведена на Рис.~\ref{fig:CA_ABM_scheme}б).

Группа под руководством академика Г.Н. Рыкованова~\cite{Adarchenko_2020, Romanuykha_2020} разработали и проанализировали камерную SEIR-D и АОМ для описания распространения COVID-19. Разработанная статистическая АОМ, несмотря на некоторую упрощенность модели поведения людей, позволяет расчетно анализировать такие факторы, как, например, введение карантина в отношении отдельных социальных групп или в отдельных сферах деятельности (работа, транспорт, магазины). Авторы отметили (стр.~27-28), что достоинства статистической модели влекут за собой и некоторые её недостатки. Так, чтобы достоверно моделировать те или иные факторы, в основу расчета должны закладываться адекватные исходные данные -- от численности населения и его распределения по социальным группам до загруженности различных видов транспорта или магазинов.

Группа американских ученых~\cite{Kerr_2021} разработали программный комплекс Covasim~\cite{covasim_doc}, основу которого составляет агентный подход моделирования эпидемии с учетом особенностей заболевания, фармацевтических (вакцинация) и политических (физические ограничения, ношение масок) вмешательств. Этот программный комплекс применялся для построения сценариев развития эпидемии COVID-19, изучения динамики пандемии и поддержке принятия политических решений более чем в десятке стран Африки, Азиатско-Тихоокеанского региона, Европы и Северной Америки. В статье~\cite{Aleta_2020} показано, что система реагирования, основанная на расширенном тестировании и отслеживании контактов, может играть важную роль в ослаблении интервенций социального дистанцирования при отсутствии коллективного иммунитета против SARS-CoV-2. Авторы~\cite{Lau_2020} подсчитали, что инфицированные люди не пожилого возраста (<60 лет) могут быть в 2,78 раза более заразными, чем пожилые люди, и первые, как правило, являются основной движущей силой сверхраспространения. В работах~\cite{Kucharski_2020, Hoertel_2020, Hellewell_2020} в рамках АОМ распространения COVID-19 проанализированы противоэпидемические программы в различных регионах, в результате чего получено понимание эффективных мер для разных географических и демографических условий, а также текущих штаммов SARS-CoV-2. В работе~\cite{Nielsen_2020} авторы построили АОМ, в котором источники заражения выступали суперраспространители. Они показали, что сверхраспространение резко усиливает значимость ограничений личных контактов.

В следующем разделе подробно описан процесс построения популяции на основе пакета Covasim, инфицирования вирусом штамма SARS-CoV-2 на основе графов, тестирования и вакцинации в Российской Федерации на основе статистических данных.

\subsubsection{АОМ распространения COVID-19}\label{sec_ABM_COVID}
Опишем структуру АОМ, лежащую в основе пакета Covasim~\cite{covasim_doc}:
    \begin{enumerate}
        \item \textbf{Инициация популяции.} Формируются четыре структуры контактов: домохозяйства, образовательные учреждения, рабочие и общественные места.
        \begin{enumerate}
            \item Постоянные характеристики агента:
            \begin{itemize}
                \item возраст (все агенты делятся на возрастные группы по 10 лет: 0--9 лет, 10--19, \ldots, 80+),
                \item пол,
                \item социальный статус,
                \item вероятности прогрессирования заболевания зависят от возраста агента (возникновения симптомов, тяжелых и критических случаев, смертность).
            \end{itemize}
            \item Переменные характеристики агента (пересчитываются к концу дня -- шаг по времени):
            \begin{itemize}
                \item эпидемиологический статус. Каждый агент в определенный момент времени может находиться в одной 9 стадиях заболевания: к описанным в Разделе~\ref{sec_R0-SEIR-HCD} состояниям $S$, $E$, $I$, $R$, $H$, $C$, $D$ добавлены бессимптомные больные $A$ и больные в легкой форме $M$,
                \item шанс быть протестированным.
            \end{itemize}
        \end{enumerate}
        
        Домохозяйства заполняются агентами согласно статистическим данным ООН~\cite{UN} о среднем размере семьи в регионе (2.6 человек). В зависимости от возраста агенты контактируют друг с другом в контактных сетях, представляющие собой полносвязные графы, количество вершин которых является пуассоновской случайной величиной с средним:
        \begin{itemize}
            \item для домохозяйства -- размер семьи,
            \item для общественных мест и образовательных учреждений -- 20.
            \item для работы -- 8.
        \end{itemize}
 Все агенты имеют контакты в домохозяйствах и в общественных местах, агенты в возрасте 6-21 лет также могут контактировать в образовательных учреждениях с агентами своего возраста, агенты в возрасте 22-65 лет -- на работе. Пример графов в разных структурах контактов в АОМ приведен на Рис.~\ref{fig:ABM_graph_example}.
 \begin{figure}[!ht]
    \centering
    \includegraphics[width=\textwidth]{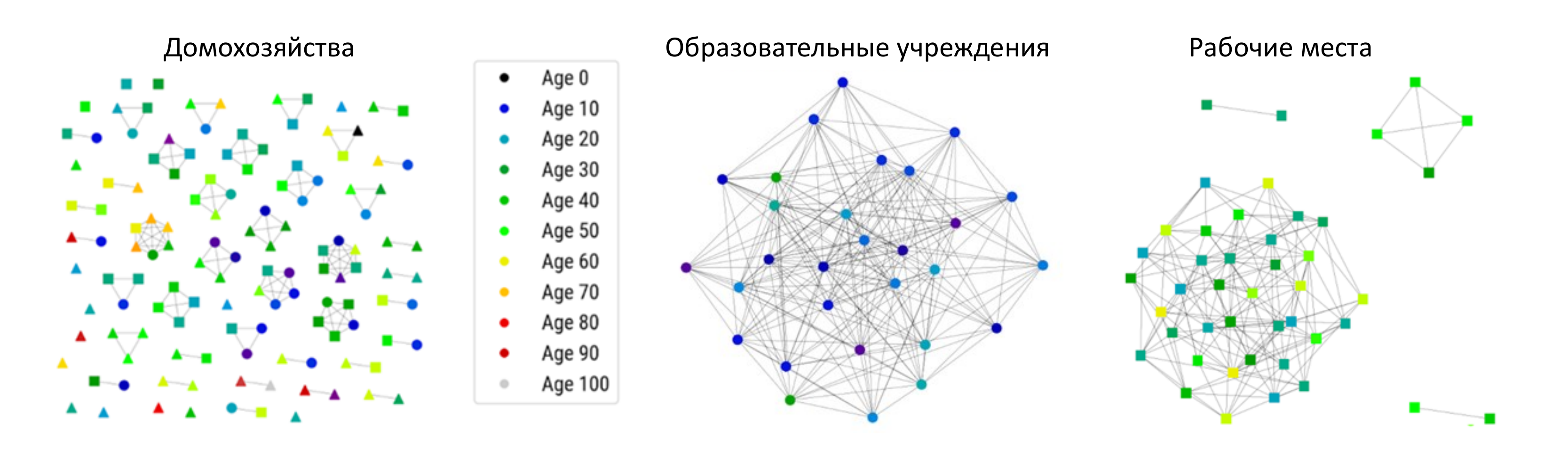}
    \caption{Пример схем связей для подвыборки 127 человек из 10 000. Все люди присутствуют в домохозяйствах (слева), в том числе некоторые не имеют связи с домом. Как правило, эти люди, включая учителей, присутствуют в школьной сети (кружки); другое подмножество присутствует в сетях рабочих мест (квадраты); некоторые люди не входят ни в школьную, ни в рабочую сеть (треугольники). Цветом отмечена возрастная категория агента.}
    \label{fig:ABM_graph_example}
\end{figure}
        
    \item \textbf{Заражение.}
        В рамках модели предполагается, что вирус передается между агентами, соединенными ребром графа. Заражение при близком контакте описывается кусочно-постоянным параметром $\alpha(t)$.
        В зависимости от структуры контакта, параметр $\alpha$ умножается на соответствующую константу $w_{\alpha}$ (для домохозяйств $w_{\alpha}=3$, для образовательных учреждений и работы $w_{\alpha}=0.6$, для общественных мест $w_{\alpha}=0.3$). Таким образом, вероятность передачи вируса для каждой контактной сети различная. Симптомные и бессимптомные агенты передают вирус одинаково.
        
        \item \textbf{Прогрессирование заболевания.} Переход из одной стадии заболевания $S,E,A,I,M,H,C,R,D$ в другую в момент времени $t$ контролируется параметрами, зависящими от возраста, т.е. чем старше агент, тем он более уязвим (например, вероятность проявлять симптомы после заражения $p_{sym} = 0.5 + 0.05i\in [0.5, 0.9]$, где $i$~-- номер возрастной группы). Взаимосвязь эпидемиологических состояний обозначена на~Рис.~\ref{fig:ABM_epid_status}. Продолжительность каждой стадии заболевания представляет собой случайную логнормальную величину с различными средними и параметрами дисперсии (см.~Таблицу~\ref{duration}).
\begin{sidewaysfigure}
    \centering
    \includegraphics[width=\textwidth]{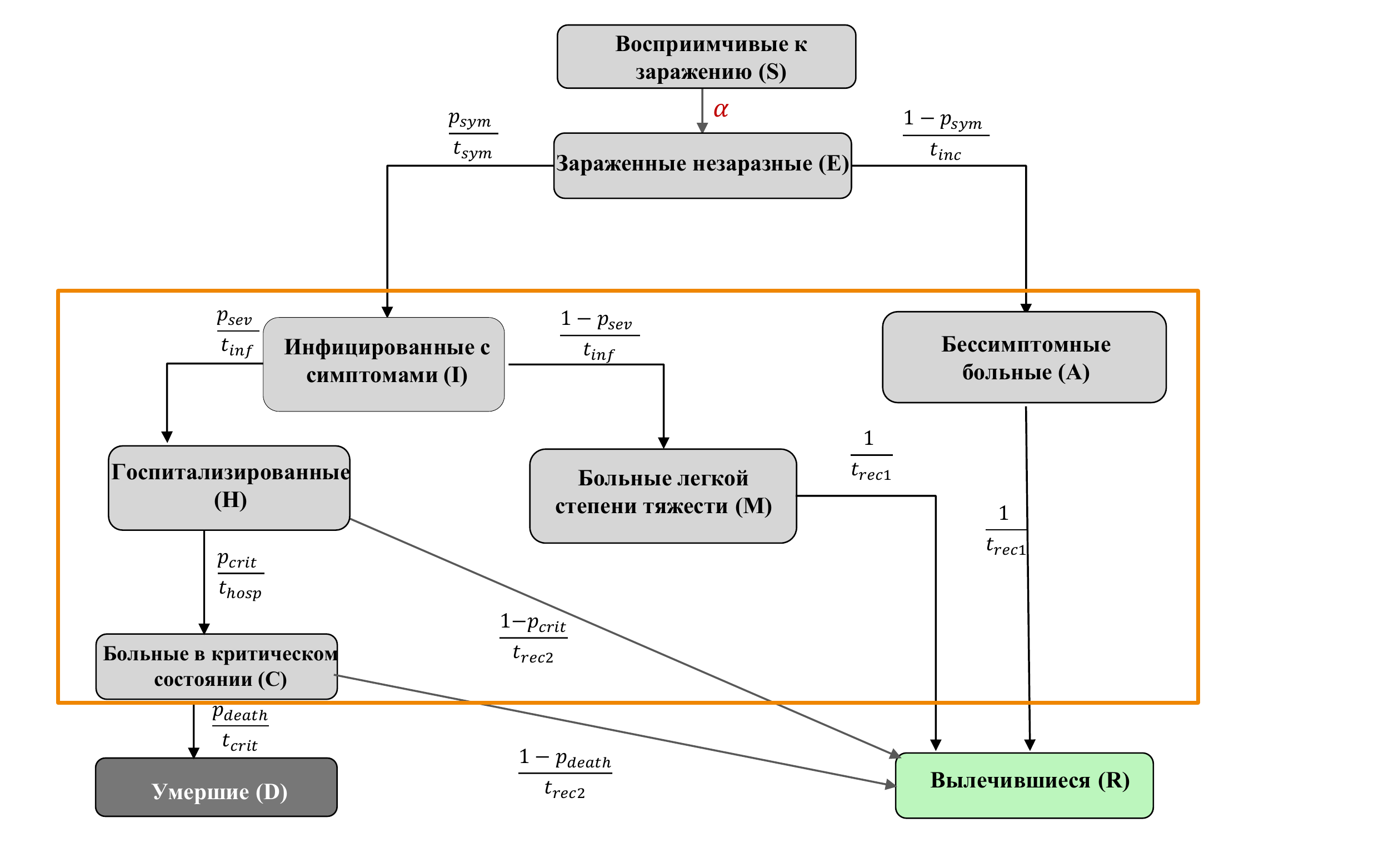}
    \caption{Диаграмма состояний агентов в АОМ. Оранжевой рамкой обозначены те состояния, находясь в которых агент имеет возможность получить положительный тест на COVID-19.}
    \label{fig:ABM_epid_status}
\end{sidewaysfigure}

\begin{table}[!ht]
\centering
\begin{tabular}{|p{2cm}|p{9cm}|p{3.7cm}| } 
 \hline
 Параметр & Описание & Распределение \\ 
  \hline
  $t_{inc}$ & Количество дней с момента контакта до того, как агент станет заразным.& $\mathrm{LogN} (4.6,4.8)$ \quad \cite{Lauer_2020}\\
  \hline
  $t_{sym}$ & Количество дней с момента, когда агент стал заразен, до проявления симптомов. & $\mathrm{LogN} (1, 0.9)$  \qquad\;\; \cite{Lauer_2020} \\
   \hline
  $t_{rec1}$ & Продолжительность болезни для бессимптомных и легких случаев. & $\mathrm{LogN} (8, 2)$  \qquad\;\;\;\; \cite{Wofel_2020} \\
  \hline
 $t_{rec2}$ & Продолжительность болезни для тяжелых и критических случаев. & $\mathrm{LogN}(14, 2.4)$  \qquad\;\; \cite{Verity_2020}\\
 \hline
 $t_{inf}$ & Количество дней, за которое агент переходит из легкого состояния в тяжелое. & $\mathrm{LogN} (6.6, 4.9)$  \qquad \cite{Lauer_2020} \\
 \hline
 $t_{hosp}$ & Количество дней, за которое агент переходит из тяжелого состояния в критическое. & $\mathrm{LogN}(3, 7.4)$  \qquad\;\; \cite{Wang_2020_clinic}\\
 \hline
 $t_{crit}$ & Длительность пребывания агента в критическом состоянии. & $\mathrm{LogN}(6.2, 1.7)$ \qquad\;\; \cite{Verity_2020} \\
 \hline
 \end{tabular}
 \caption{Параметры продолжительность болезни в днях в АОМ.}
\label{duration}
\end{table}

Модель основана на нескольких предположениях:
  \begin{enumerate}
 \item Изначально иммунитета нет ни у одного агента.
 \item Часть агентов находятся в инкубационном периоде $E(0)$.
 \item Умереть могут только пациенты, находящиеся на ИВЛ (критическое состояние $C(t)$).
\end{enumerate}

\item \textbf{Тестирование агентов} проводится согласно ежедневным статистическим данным о количестве проведенных тестов в регионе. Шанс быть протестированным на COVID-19 $p(t,i)$ зависит от эпидемиологического статуса агента и определяется в ходе решения обратной задачи (см. Раздел~\ref{sec_inv_ABM}). Положительный результат могут получить агенты, находящиеся в симптомном, бессимптомном, в легкой форме, госпитализированном, критическом состояниях (на рис.~\ref{fig:ABM_epid_status} эти состояния обведены в оранжевую рамку). В модели предполагается, что вероятность тестирования агентов с симптомами выше, чем у бессимптомных больных.

\item \textbf{Введение сдерживающих эпидемию мер}. В модели возможно введение карантинных мер как для всех контактных слоев, так и для каждого в отдельности. Это может быть сделано двумя способами: либо изменением значения параметра $\alpha(t)$ (в случае введения обязательной меры ношения масок или социального дистанцирования), либо удалением ребер в графах (в случае введения самоизоляции и дистанционной работы).
\end{enumerate}

Загружаются все необходимые параметры и статистические данные, создается искусственная популяция с учетом распределения по возрастам в регионе. Далее агенты соединяются в контактные сети. Затем начинается цикл по времени: на каждом шаге (временной интервал равен одному дню) обновляется эпидемиологический статус агента с учетом его структуры контактов и введенных ограничительных мер (самоизоляция, закрытие общественных мест, ношение масок и т.д.).

\subsection{Индекс репродукции вируса $\mathcal{R}_0$ для АОМ}
В АОМ индекс репродукции вируса $\mathcal{R}_0$ определяется следующим образом:
\[\mathcal{R}_0(t) = \dfrac{d\cdot I_{new}(t)}{I_{active}(t)}.\]
Здесь $I_{new}(t)$~-- количество новых случаев инфицирования в день $t$, $I_{active}(t)$~-- количество людей с активным инфекционным заболеванием в день $t$, $d$~-- среднее время инфицирования.

\subsection{Выводы}
АОМ позволяют преодолевать ограничения дифференциальных моделей, используя стохастическую природу эпидемического процесса, возможности введения ограничительных мер в модель, особенности моделируемого региона и учета индивидуальности индивидуума (агента). На основе результатов моделирования в рамках АОМ получено понимание эффективности мер для разных географических и демографических условий для каждого выявленного штамма SARS-CoV-2. Для повышения достоверности результатов необходимо использовать адекватные статистические данные (эпидемиологические, демографические, транспортные), уточнять границы чувствительных параметров моделей, иметь большие вычислительные мощности для реализации моделирования и прогнозирования.

%% file: text/4MFG_model.tex
\section{Модели <<игры среднего поля>>}\label{sec_MFG_model}
В данном разделе приведено описание комбинации SIR-модели (с учетом пространственной неоднородности) и агентной модели. Распределение агентов в пространственной SIR-модели описывается уравнением Колмогорова-Фоккера-Планка~(\ref{eq_8}) (Раздел~\ref{sec_KFP}), а управление агентов описывается уравнением Гамильтона-Якоби-Беллмана~(\ref{eq_23}) (Раздел~\ref{sec_GJB}). Совместное рассмотрение КФП и ГЯБ образуют систему уравнений ИСП (Раздел~\ref{sec_MFG}).

\subsection{Введение}
J.M. Smith и G.R. Price в 1973 году~\cite{SmithPrice_1973} разработали концепцию эволюционной устойчивой стратегии, которая является центральной концепцией теории игр. Затем B.~Jovanovic и R.W.~Rosenthal в 1988 году~\cite{JovanovicRosenthal_1988} доказали существование равновесия в анонимных последовательных играх. Только в 2006-2007 годах J.-M.~Lasry и P.-L.~Lions~\cite{Las} предложили и обосновали концепцию ИСП, которая используется в физике, экономике, метеорологии, социальных и биологических процессах. Основные задачи, которые рассматриваются в теории ИСП:\\
-- исследование существования и единственности решения ИСП, при этом ИСП формализуется в виде системы двух уравнений в частных производных первого порядка и задачи управления для некоторой динамической системы, используемая для выводя уравнения ГЯБ, соответствующего модели поведения агентов (подробнее см.~Раздел~\ref{sec_MFG});\\ 
-- построение приближенного равновесия в игре конечного числа агентов;\\
-- исследование сходимости приближенных равновесий в играх конечного числа лиц к решению ИСП при стремлении числа игроков к бесконечности;\\
-- анализ конкретных эпидемиологических задач при помощи методологии ИСП.

В работе~\cite{Kolokoltsov_2013b} исследуется сходимость решений для игр с нелинейными процессами устойчивого типа с переменными коэффициентами. Концепция ИСП находится на пересечении теории среднего поля и нелинейного марковского управления~\cite{Andersson_2011, Buckdahn_2009}

С 2015 года опубликовано множество работ, использующих ИСП для описания эпидемических процессов распространения гриппа, \nom{ОРВИ}{острая респираторная вирусная инфекция}, СOVID-19~\cite{Laguzet_2015, Lee_2020, Tembine_2020, Osher_2021}.

\subsection{Уравнение Колмогорова-Фоккера-Планка}\label{sec_KFP}
Рассмотрим однородные марковские случайные процессы в $\mathbb {R} ^{n}$, для которых оператор переходных вероятностей задаётся переходной плотностью $p(t,x,y)$. Зададим оператор $\mathbf{Q}$, действующий на функцию $f(x)$ в $\mathbb {R} ^{n}$:
\[(\mathbf{Q}f)(x) = \int\limits_{\mathbb {R}^n} q(x,y)f(y)\, dy.\]
Если существует предел (обобщенная функция)
\[q(x,y)=\lim\limits_{h\to 0}\dfrac{p(h,x,y)-\delta(x-y)}{h},\]
то уравнение Колмогорова принимает вид:
\[\dfrac{\partial p(t,x,y)}{\partial t} = \int\limits_{\mathbb {R}^n} p(t,x,z)q(z,y)\, dz.\]

В случае, когда $\mathbf{Q}$~-- дифференциальный оператор второго порядка с непрерывными коэффициентами (это означает, что $q(x,y)$ есть линейная комбинация первых и вторых производных $\delta (x-y)$ с непрерывными коэффициентами), а матрица коэффициентов $\sigma^{ij}$ перед вторыми производными является симметричной и положительно определенной в каждой точке, то уравнение Колмогорова будет совпадать с уравнением Фоккера-Планка~\cite{Bogoluybov_1939}:
\begin{eqnarray}\label{eq_KFP_common}
    \dfrac{\partial p(t,x,y)}{\partial t} = \dfrac{1}{2}\sum\limits_{i,j}^M\dfrac{\partial ^2}{\partial y^i \partial y^j} (\sigma^{ij}(y)p(t,x,y)) - \sum\limits_j^M \dfrac{\partial}{\partial y^j} (b^j(y)p(t,x,y)).
\end{eqnarray}
Вектор $b^{j}$ в физической литературе называется вектором сноса, а матрица $\sigma^{ij}$~-- тензором диффузии.

Уравнение КФП~(\ref{eq_KFP_common}) используется для расчёта плотности вероятности в СДУ:
\begin{eqnarray}\label{eq_SDE_common}
    d{\mathbf{X}}_{t}={\boldsymbol {b}}({\mathbf{X}}_{t},t)\,dt+{\boldsymbol {\sigma }}({\mathbf{X}}_{t},t)\,d{\mathbf{B}}_{t}.
\end{eqnarray}
Здесь $\mathbf{X}_{t}\in \mathbb{R}^{M}$~-- функция состояния системы, а $\mathbf{B}_{t}\in \mathbb{R} ^{M}$~-- стандартное $M$-мерное броуновское движение. В нашем случае вектор $\mathbf{X}_{t}$ описывает состояния агентов в каждой из камер и является решением системы~(\ref{eq_SDE_common}). Более подробно см. систему~(\ref{eq_3}). Если начальное распределение задано как $\mathbf{X}_{0}\sim p({0, \mathbf{x}})$, то плотность вероятности $p(t, {\mathbf{x}})$ состояния системы $\mathbf{X}_{t}$ является решением уравнения КФП~(\ref{eq_KFP_common}).

\subsection{Уравнение Гамильтона-Якоби-Беллмана}\label{sec_GJB}
Уравнение ГЯБ~-- дифференциальное уравнение в частных производных, играющее центральную роль в теории оптимального управления. Рассмотрим задачу оптимального управления на промежутке времени $[0,T]$:
\[\psi=\min _{u}\left\{\int_{0}^{T}C[x(t),u(t)]\,dt+G[x(T)]\right\}.\]
Здесь $C$ и $G$~-- липшиц-непрерывные функции стоимости, определяющие соответственно интегральную и терминальную часть функционала, $x(t)$~-- вектор, определяющий состояние системы в каждый момент времени с заданным начальным значением $x(0)$, $u(t)$~-- вектор управления.

Эволюция системы под действием управления $u(t)$ описывается следующим образом:
\[\dot {x}(t)=F[x(t),u(t)],\]
и уравнения ГЯБ принимают следующий вид:
\begin{eqnarray}\label{eq_GJB_common}
    \dfrac{\partial{\psi}(x,t)}{\partial t}+\min _{u}\left\{\nabla \psi(x,t)\cdot F(x,u)+C(x,u)\right\}=0,
\end{eqnarray}
с начальным значением в конечный момент времени $T$
\begin{eqnarray}\label{eq_GJB_init_cond}
    \psi(x,T)=G(x).
\end{eqnarray}
Неизвестная в этом уравнении беллмановская функция значения $\psi(x,t)$, отвечающая максимальному значению, которое можно получить, ведя систему из состояния $(x,t)$ оптимальным образом до момента времени $T$. Тогда оптимальное решение в начальный момент времени~-- значение $\psi = \psi(x(0), 0)$.

Уравнение~(\ref{eq_GJB_common}) называется уравнением Беллмана~\cite{Bellman_1957}, также известное как уравнение динамического программирования. Также Беллманом было введено понятие \textit{принципа оптимальности}: оптимальная стратегия имеет свойство, что какими бы ни были начальное состояние и начальное решение, последующие решения должны составлять оптимальный курс действий по отношению к состоянию, полученному в результате первого решения. Иными словами, оптимальная стратегия зависит только от текущего состояния и цели, и не зависит от предыстории.

При рассмотрении задачи с непрерывным временем полученные уравнения могут рассматриваться как продолжение более ранних работ в области теоретической механики, связанных с уравнением Гамильтона-Якоби.

\subsection{Модель игры среднего поля}\label{sec_MFG}
Предположение, что агенты в популяции рациональны (то есть обладают способностью находиться в состоянии $X(t)$ и иметь возможность его изменить), позволяет рассматривать задачу управления с большим количеством участников. Динамика отдельного индивидуума удовлетворяет дифференциальному уравнению Ито
\begin{equation*}
    dX^N_i(t) = b (t,X^N_i(t),\theta^N(t, X^N_i(t)),u_i(t, X^N_i(t)))dt + \sigma (t,X^N_i(t),\theta^N(t, X^N_i(t))) dW^N_i(t).
\end{equation*}
Здесь $i \in {1,..,N},\; W^N_i $ -- независимые стандартные винеровские процессы, $u_i(t,X^N_i(t))$ -- стратегия $i-$го агента и $\theta^N(t,X^N_i(t))$~-- эмпирическая мера распределения агентов в системе в момент времени $t$~\cite{Fis}. В предположении, что функции $b$ и $\sigma$ непрерывны во времени и одинаковы для всех агентов, в работе \cite{Fis} показано, что когда количество агентов в системе чрезвычайно велико $N \rightarrow \infty$, мы можем заменить массу отдельных индивидов репрезентативным агентом, состояние которого определяется следующим уравнением
\begin{equation}\label{eq_3}
    dX(t) = b (t,X(t),p(t),u(t, X(t)))dt + \sigma (t,X(t),p(t, X(t))) dW(t).
\end{equation}
Здесь $X(t):[0,T]\rightarrow \Omega$; $p(t,X(t)):\theta^N(t, X^N_i(t))\overset{N \rightarrow \infty}{\rightarrow} p(t, X(t))$~-- это распределение агента по пространству состояний $\Omega$ в момент времени $t$ и $u(t, X(t))$ -- стратегия репрезентативного агента, обеспечивающая равновесие по Нэшу системы взаимодействующих агентов (ни один агент не может увеличить выигрыш, изменив свою стратегию, если другие агенты своих стратегий не меняют) и минимизирующая функционал
\begin{equation}\label{eq_4}
    J(p, u) = \mathbf{E}\left[ \int_0^T C\left(t,X(t),p(t, X(t)),u(t, X(t))\right)dt + G \left(X(T),p(T, X(T))\right)\right].
\end{equation}

Данный подход к контролю над популяцией с большим количеством взаимодействующих агентов получил название ИСП~\cite{Las, Ben, Fis}. Модель ИСП предполагает, что каждый агент выбирает свою рациональную стратегию $u(t,x)$ с учетом своего положения и распределения $p(t,x): [0,T]\times \Omega \rightarrow \mathrm{R}$ других агентов. В работе~\cite{Ben} было показано, что при постоянном $\sigma$ в уравнении~\eqref{eq_3}, характеризующем стохастический характер равновесия процесса взаимодействия агентов, распределение агентов $p(t,x)$ подчиняется уравнению КФП  
\begin{equation}\label{eq_5}
    \dfrac{\partial p}{\partial t} - \dfrac{1}{2}\sigma^2\Delta p + \nabla (p u) = 0 \;\; \text{ в } [0,T]\times \Omega
\end{equation}
с начальными условиями 
\begin{equation}\label{eq_6}    
    p(0,x) = p_0(x) \;\;\text{ на } \Omega
\end{equation}
и граничными условиями типа Неймана 
\begin{equation}\label{eq_7}
    \dfrac{\partial p}{\partial x} = 0 \;\; \forall t \text{ и } x \in \Gamma_\Omega.  
\end{equation}

\begin{sidewaysfigure}
    \centering
    \includegraphics[width=\textwidth]{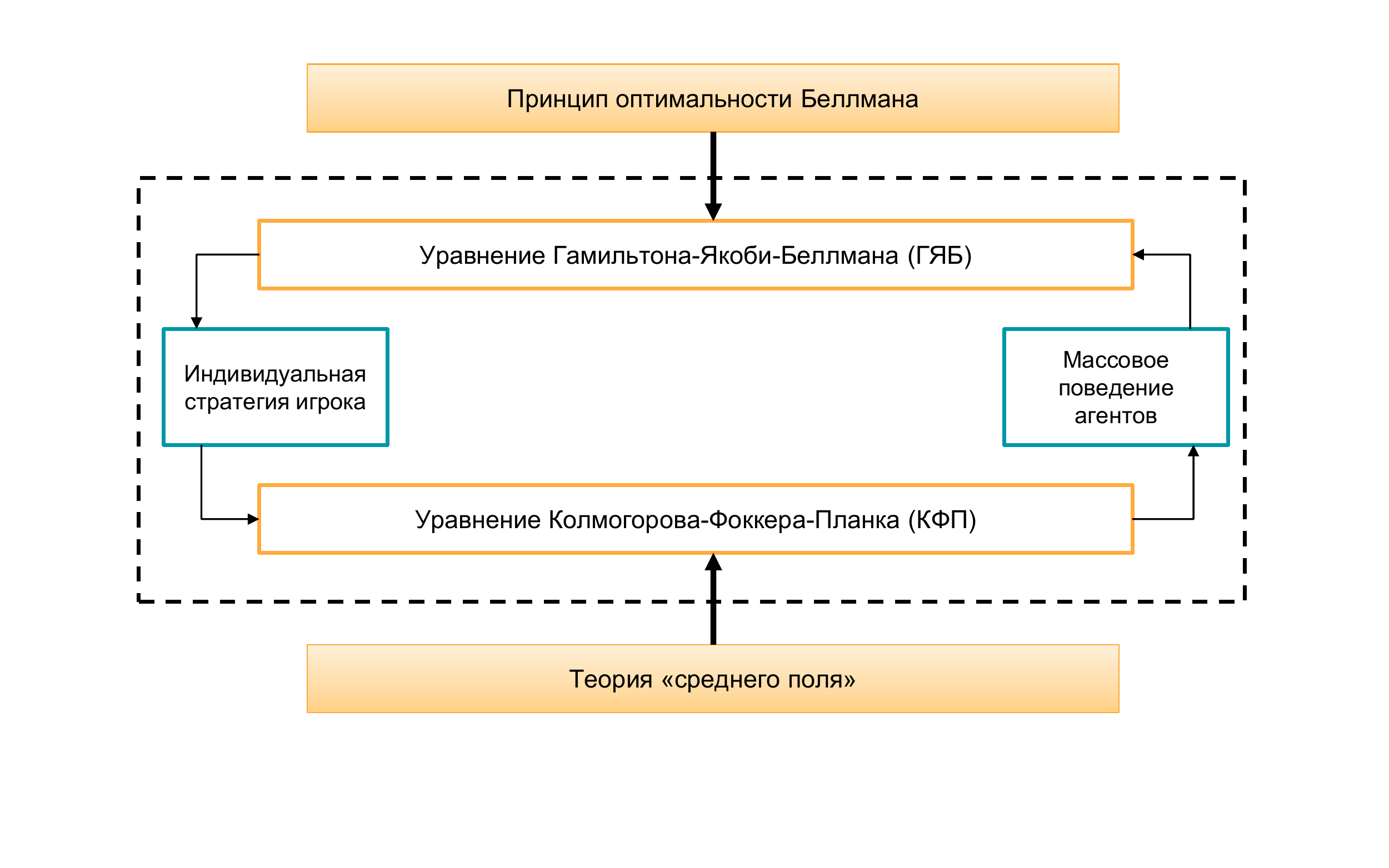}
    \vspace*{-10mm}
    \caption{Классическая схема ИСП, основанная на системе уравнений КФП и ГЯБ.}
    \label{fig:MFG_principle}
\end{sidewaysfigure}

Классическая модель ИСП в схематичном виде изображена на Рис.~\ref{fig:MFG_principle}. Исследованы модели с динамикой каждого игрока, задаваемой марковской цепью с непрерывным временем и конечным числом состояний~\cite{Gomes_2013}, а также марковским процессом общего вида~\cite{Kolokoltsov_2011, Kolokoltsov_2013}. В этом случае уравнение~(\ref{eq_5}) и ГЯБ заменяются на специально построенные задачи оптимизации. Альтернативный подход к ИСП называется вероятностным и связан с исследованием задачи управления для динамической системы, описываемой нелинейным марковским процессом. В этом случае динамика и интегральная часть выигрыша зависят от распределения игроков в текущий момент времени. Распределение игроков определяется по решению задачи оптимизации~\cite{Carmona_2013}.

Недостатком вероятностного подхода является зависимость определения решения от выбора сопутствующего вероятностного пространства. Этот недостаток пытается преодолеть минимаксный подход, в рамках которого задача поиска решения ИСП сводится к решению игры бесконечного числа лиц, при этом динамика определяется распределением оптимальных траекторий.

ИСП достаточно гибки, чтобы улавливать межклассовое взаимодействие при распространении эпидемии, при котором несколько органов власти осведомлены о рисках лиц, принимающих локальные решения, а отдельные лица осведомлены о рисках агентов (государство, региональное правительство), принимающих глобальные решения~\cite{Tembine_2020, Osher_2021}. В следующих разделах приведены уравнения КФП и ГЯБ для описания распределения и управления SIR-модели распространения эпидемии.

\subsubsection{Уравнение КФП для SIR-модели}
Для достаточно больших ($N\to\infty$) популяций SIR-модель можно интерпретировать как приближение среднего поля вероятностной модели клеточного автомата~\cite{Berec_2002, Schimit_2009}. Запишем аналог начально-краевой задачи для уравнения КФП~(\ref{eq_5})-(\ref{eq_7}) в случае SIR-модели~(\ref{model_KermackMackendrick}).

Введем плотность $p_i(t,x):[0,T]\times[0,1]\rightarrow \mathrm{\textbf{R}}$ распределения агентов в группах $S$, $I$ и $R$, где $i \in\{S,I,R\}$. Переменная $x\in\Omega = [0,1]$ характеризует физическое дистанцирование следующим образом: $x=0$ означает, что агент склонен соблюдать физическую дистанцию, а $x=1$ означает противоположное. Определим функции $u_i(t,x):[0,T]\times[0,1]\rightarrow \mathrm{\textbf{R}}$, $i \in\{S,I,R\}$, которые описывают скорость перемещения в пространственной области репрезентативного агента в каждой группе населения. Принимая во внимание стохастический характер взаимодействия агентов~(\ref{eq_3}), уравнения КФП~\eqref{eq_5}-\eqref{eq_7} для функций $p_i(t,x)$ запишутся в виде системы уравнений в частных производных:
\begin{equation}\label{eq_8}
 \hspace{-4mm}\left\{ \begin{aligned}
   & \dfrac{\partial {p}_{S}}{\partial t}+\nabla \left( {{p}_{S}}{{u}_{S}} \right)+\alpha {{p}_{S}}{{p}_{I}}-{\sigma _{S}^{2}\Delta {{p}_{S}}}/{2}\;=0, \\ 
  & \dfrac{\partial {p}_{I}}{\partial t}+\nabla \left( {{p}_{I}}{{u}_{I}} \right)-\alpha {{p}_{S}}{{p}_{I}}+\beta {{p}_{I}}-{\sigma _{I}^{2}\Delta {{p}_{I}}}/{2}\;=0, \\ 
 & \dfrac{\partial {p}_{R}}{\partial t}+\nabla \left( {{p}_{R}}{{u}_{R}} \right)-\beta {{p}_{I}}-{\sigma _{R}^{2}\Delta {{p}_{R}}}/{2}\;=0.
\end{aligned} \right.
\end{equation}
Здесь $\sigma_i \geqslant 0$, $i \in \{S,I,R,C\}$. Начальные~(\ref{eq_6}) и граничные~(\ref{eq_7}) условия остаются без изменений
\begin{equation}\label{eq_9}
    p_i (0,x)=p_{i0}(x)\quad \; \forall x\in[0,1]. 
\end{equation}
\begin{equation}\label{eq_10}
    \dfrac{\partial p_i}{\partial x} = 0\quad \;\forall t \in [0,T] \;\text{ и } \;x = 0,1.  
\end{equation}

Начальные условия $p_{i0}$ совпадают с плотностью нормального распределения со средним, демонстрирующим склонность соблюдать физическую дистанцию, и дисперсией, описывающей строгость соблюдения ограничений. Например, восприимчивая популяция $m_S$ не склонна соблюдать ограничения, поэтому среднее в распределении сместится к 1, а инфицированная $m_I$, наоборот, стремится соблюдать ограничения и карантин.

Отметим, что система~(\ref{eq_8}) удовлетворяет закону сохранения масс, а именно
\[\dfrac{\partial}{\partial t}\int\limits_0^1 \left( p_S(t,x)+p_I(t,x)+p_R(t,x)\right)\, dx = 0.\]

\subsubsection{Уравнение ГЯБ для SIR-модели}
Определим целевой функционал~(\ref{eq_4}) следующим образом~\cite{Lee_2020}:
\begin{eqnarray}\label{eq_HJB_SIRmodel}
    J(p, u) = \int\limits_0^T \int\limits_0^1 \left(\sum\limits_{i\in\{S,I,R\}} \dfrac{r_i}{2}p_i|u_i|^2 + \dfrac{c}{2} (p_S + p_I + p_R)^2\right) dxdt  + \dfrac{1}{2}\int\limits_0^1 p_I^2(T,x) dx.
\end{eqnarray}
Здесь $r_i$ и $c$ неотрицательные константы. Первое слагаемое, характеризующее кинетическую энергию, описывает цену перемещения агентов со скоростью $u_i$ за весь промежуток времени $[0,T]$. Чем выше значение $r_i$, тем более затратно агентам перемещаться между группами (например, $r_S=r_R=1$, $r_I=10$ означает, что агентам из инфицированной группы труднее передвигаться). Второе слагаемое в~(\ref{eq_HJB_SIRmodel}) контролирует скопление всего населения в одном месте. Это может повысить риск вспышек заболевания и их более быстрого и широкого распространения.

Для вывода оптимальной стратегии воспользуемся методом множителя Лагранжа~\cite{Ben}. Введем произвольные гладкие функции $\psi_i(t, x)\in C ^ {\infty} \left(\left[0, T \right] \times [0,1] \right)$, $i\in\{S,I,R\}$, и запишем минимаксную задачу для функционала Лагранжа:
\begin{eqnarray}\label{eq:minmax_pr}
    \inf\limits_{p, u} \sup\limits_{\left(\psi\right)_{i\in\{S,I,R\}}} \mathcal{L}(p, u, \psi_i),
\end{eqnarray}
где функционал Лагранжа составлен из целевого функционала~(\ref{eq_HJB_SIRmodel}) и уравнений КФП~(\ref{eq_8}), домноженных на функции $\psi_i(t, x)$ и проинтегрированных по $x$ и $t$:
\begin{eqnarray*}
    \begin{array}{cc}
         \mathcal{L}(p, u, \psi_i) = & J(p,u) - \int\limits_0^T\int\limits_0^1 \left[ \sum\limits_{i\in\{S,I,R\}} \psi_i\left( \dfrac{\partial p_i}{\partial t}+\nabla (p_iu_i)-\dfrac{\sigma_i^2}{2}\Delta p_i\right) +\right.\\
            & \left.+ \alpha p_Ip_S(\psi_I - \psi_S) + \beta p_I (\psi_R - \psi_I)\right]\, dxdt.
    \end{array}
\end{eqnarray*}
Пусть
\[\partial \psi_i/ \partial x = 0\; \quad\forall t \in [0,T] \;\text{ и } \;x = 0,1\; {\forall i \in \{S,I,R\}}\]
и 
\[\alpha_i(t,0) = \alpha_i(t,1) = 0\; \;\forall t \in [0,T] \; {\forall i \in \{S,I,R\}}.\]
Применяя интегрирование по частям и пользуясь условиями оптимальности Каруша-Куна-Таккера, получаем систему уравнения ГЯБ для оптимального управления системой агентов~(\ref{eq_8}):
\begin{equation}\label{eq_23}
 \hspace{-5mm}\left\{ \begin{aligned}
   & \dfrac{\partial \psi_S}{\partial t}+\dfrac{\sigma_S^2}{2}\Delta\psi_S+u_S \cdot \nabla\psi_S + \alpha p_I (\psi_I-\psi_S) =-\dfrac{r_S}{2}|u_S|^2+c(p_S+p_I+p_R), \\
  & \begin{aligned}
  \dfrac{\partial \psi_I}{\partial t}+\dfrac{\sigma_I^2}{2}\Delta\psi_I + u_I \cdot \nabla\psi_I  +& \alpha p_S (\psi_I-\psi_S) + \beta (\psi_R-\psi_I) =\\
  &= -\dfrac{r_I}{2}|u_I|^2+c(p_S+p_I+p_R) - \delta(T-t)p_I,
  \end{aligned} \\ 
  & \dfrac{\partial \psi_R}{\partial t}+\dfrac{\sigma_R^2}{2}\Delta\psi_R + u_R \cdot \nabla\psi_R =-\dfrac{r_R}{2}|u_R|^2+c(p_S+p_I+p_R), \\ 
\end{aligned} \right.
\end{equation}
где $\delta(T-t)$~-- дельта-функция Дирака. Начальные условия для уравнений ГЯБ~(\ref{eq_23})
\begin{equation}\label{eq_24}  
    \psi_i(T,x)=0\,\,\,\,\,\forall \text{ }x\in \left[ 0,1 \right],\quad i\in\{S,I,R\}.
\end{equation}

\subsection{Выводы}
Модели ИСП описания распространения эпидемий включают в себя классические SIR-модели с учетом произвольной пространственной неоднородности (для областей с любой геометрией), а также учитывают рациональность агентов и внешнее управление (а именно, правительство может наложить ограничения на взаимодействие для разных классов населения в зависимости от их статуса заражения). С другой стороны, модели ИСП являются предельным случаем АОМ и оптимизируют вычислительные затраты.

Как и для рассмотренных ранее моделей, коэффициенты уравнений, описывающие эпидемиологические характеристики моделируемого заболевания и особенности популяции, неизвестны. Необходимо формулировать обратную задачу для модели ИСП с целью уточнения чувствительных параметров и увеличения качества прогнозирования.

%% file: text/5algorithms.tex
\section{Алгоритмы численного решения}\label{sec_algorithms}
Математические модели в эпидемиологии характеризуются своими коэффициентами и начальными условиями, которые индивидуальны для каждой моделируемой популяции. В следующих разделах будут приведены алгоритмы, входящие в комплекс программ COVID-19 моделирования распространения коронавирусной инфекции в Новосибирской области. В Разделе~\ref{sec_DirectPr} приведены алгоритмы численного решения прямых задач, в которых при заданных эпидемиологических параметрах и начальных условиях требуется определить распределение различных категорий населения (восприимчивые, инфицированные, госпитализированные, вылеченные и т.п.) в течение всего времени моделирования. В Разделе~\ref{sec_InversePr} описываются алгоритмы численного решения обратных задач, в которых требуется определить неизвестные параметры моделей по дополнительной информации о количестве выявленных инфицированных, протестированных, госпитализированных и умерших в фиксированные моменты времени.

\subsection{Прямые задачи}\label{sec_DirectPr}
Прямая задача для модели эпидемиологии состоит в определении количества или плотности восприимчивой, протестированной, инфицированной, госпитализированной и т.п. групп населения по заданным эпидемиологическим коэффициентам и начальным условиям. Численные методы решения прямых задач для математических моделей, основанных на дифференциальных уравнениях (Раздел~\ref{sec_equation_models}), характеризуются конечно-разностной структурой (методы Эйлера, конечных разностей, элементов, объемов). Полученное численное решение чувствительно к разбиению области моделирования (по времени, по пространству и времени). Обзор численных методов решения задач игр среднего поля приведен в работах~\cite{Achdou_2010, Achdou_2013, Shaydurov_2020}.

В комплексе программ COVID-19 для численного решения прямых задач использованы:
\begin{itemize}
    \item для моделей SIRC\textsubscript{r} и SEIR-HCD: метод Рунге-Кутты 4-го порядка.
    \item для агентных моделей: методы графов, Монте-Карло и программные комплексы Covasim~\cite{Kerr_2021}.
    \item для моделей ИСП: метод конечных разностей дробных шагов, конечных объемов.
\end{itemize}

\subsection{Обратные задачи}\label{sec_InversePr}
Параметры моделей в эпидемиологии (индекс репродукции вируса $\mathcal{R}_0(t)$, вероятность госпитализации, тестирования, выздоровления, количество бессимптомных носителей и т.п.) заданы приближенно и нуждаются в уточнении в каждом конкретном регионе для увеличения точности моделирования и прогнозирования распространения эпидемии. С этой целью мы решаем обратную задачу, которая состоит в определении вектора неизвестных параметров $\mathbf{q}$ математической модели эпидемиологии по дополнительной информации $f_k^i$ о количестве выявленных ($i=1$), госпитализированных ($i=2$), умерших ($i=3$) в фиксированные моменты времени $t_k$, $k=1,\ldots, K$ (подробнее о постановке обратной задач для моделей SEIR-HCD и АОМ см.~Разделы~\ref{sec_inv_SEIR} и~\ref{sec_inv_ABM} соответственно). Обратная задача может быть сформулирована в виде задачи минимизации целевого функционала~\cite{Kaltenbacher_2008, Kabanikhin_2009}:
\begin{eqnarray}
    J(\mathbf{q}) = \left< A(\mathbf{q}) - \mathbf{f}, A(\mathbf{q}) - \mathbf{f}\right>.
\end{eqnarray}
Здесь $A$~-- нелинейный оператор обратной задачи, $\mathbf{f} = \left\{ f_k^i \right\}_{\substack{i\in \mathcal{I}, \\k=\overline{1,K}}}$, $\mathcal{I}$~-- множество измеряемых состояний системы.

Обратная задача является некорректной, а именно ее решение может быть неединственным и/или неустойчивым~\cite{Tikhonov_1983, Tikhonov_1995, Kabanikhin_2009}. Для разработки алгоритма регуляризации решения обратной задачи проводится анализ чувствительности параметров, который позволяет упорядочить  параметры $q_i$ по степени чувствительности по отношению к вариации данных обратной задачи~\cite{Hongyu_2011, Banks_2015, Andrianakis_2015, KOI_2020}.

В комплексе программ \textit{SBRAS-COVID-19} для численного решения обратных задач использованы:
\begin{itemize}
    \item Поиск глобального минимума функционала: генетический алгоритм, метод дифференциальной эволюции, метод имитации отжига, метод роя частиц, метод древовидных оценок Парзена, тензорная оптимизация, стохастический градиентный спуск, глубокие нейронные сети, обучение с подкреплением.
    \item Для уточнения минимума: методы минимальных ошибок, наискорейшего спуска, Левенберга-Марквардта, Бройдена-Флетчера-Гофбардто-Шанно, Нелдера-Мида.
\end{itemize}

Для исследования прямых и обратных задач программного комплекса \textit{SBRAS-COVID-19} использованы теоретические результаты:
\begin{enumerate}
    \item SIR-модели:
    \begin{itemize}
        \item Теория нелинейных операторных уравнений Вольтерра в банаховых и гильбертовых пространствах -- локальная корректность, корректность в окрестности точного решения, единственность и условная устойчивость, сходимость дискретных аналогов к точному решению [ссылка на статью в ЖВМиМФ].
        \item Устойчивость обратных задач -- А.Н.~Тихонов~\cite{Tikhonov_1943}, М.М.~Лаврентьев~\cite{Lavrentiev_1953}, С.К. Годунов~\cite{Godunov_1992}, С.И. Кабанихин\cite{Kabanikhin_2013}.
        \item Методы идентифицируемости -- структурная (дифференциальной алгебры, передаточной функции, разложения в ряд Тейлора) и практическая (Монте-Карло, корреляционной матрицы)~\cite{Hongyu_2011}.
        \item Методы анализа чувствительности: ортогональный, сингулярного разложения, собственных значений~\cite{Hongyu_2011, Banks_2015, KOI_2020}.
        \item Теория регуляризации -- методы М.М.~Лаврентьева~\cite{Lavrentiev_1962}, А.Н.~Тихонова~\cite{Tikhonov_1963} и В.К.~Иванова~\cite{Ivanov_1963}, итерационная регуляризация~\cite{KSI_2015, KSI_Zhang_2020}.
        \item Методы учета априорной информации -- И.И.~Еремин, В.В.~Васин, А.Г.~Ягола.
        \item Исследование сходимости природоподобных алгоримтов -- теорема стохастической сходимости~\cite{Giglyavskiy_1991}.
        \item Теория больших данных -- тензорное разложение~\cite{Tyrtyshnikov_1993, Oseledets_2011, Zheltkov_2013, Zheltkova_2018}.
\end{itemize}
    \item Агентные модели:
    \begin{itemize}
        \item Методы анализа чувствительности -- баесовский подоход, регрессионный анализ~\cite{Andrianakis_2015, KOI_JIIP_2021}.
        \item Методы высокопроизводительных вычислений -- распределенные вычисления, MPI.
    \end{itemize}
\end{enumerate}

\subsubsection{Алгоритм решения обратной задачи для SEIR-HCD}\label{sec_inv_SEIR}
Предположим, что известна дополнительная информация о количестве симптомных выявленных случаях $(1-b_k)h_k$, критических $C_k$ и умерших $g_k$ в SEIR-HCD модели в фиксированные моменты времени $t_k$, $k=1,\ldots, K$:
\begin{eqnarray}\label{eq:SEIR-HCD_data}
    E(t_k; \mathbf{q}) = (1-b_k) h_k,\quad C(t_k; \mathbf{q}) = C_k,\quad D(t_k; \mathbf{q}) = g_k,\qquad k=1,\ldots, K.
\end{eqnarray}
Здесь $\mathbf{q}=(\alpha_E(t), \alpha_I(t), \varepsilon_{HC}, \mu, E_0, I_0)$~-- вектор неизвестных параметров модели, $h_k$~-- количество выявленных случаев в Новосибирской области (см. Рис.~\ref{fig:Nsk_stat_data} в Приложении~\ref{sec_app_data}), $b_k$~-- процент бессимптомных случаев по результатам ПЦР.

Обратная задача для SEIR-HCD модели~(\ref{eq:SEIR-HCD})-(\ref{eq:SEIR-HCD_init_cond}),~(\ref{eq:SEIR-HCD_data}) состоит в определении вектора параметров $\mathbf{q}$ по дополнительной информации~(\ref{eq:SEIR-HCD_data}). Обратная задача для SEIR-HCD модели была сведена к задаче минимизации целевого функционала 
\begin{eqnarray}\label{eq:J_SEIRHCD}
\begin{array}{ll}
    J(\mathbf{q}) = &\sum\limits_{k=1}^{K} \dfrac{1}{(1-b_k)^2h_k^2}\left( \dfrac{1}{t_{inc}} E(t_{k-1};\mathbf{q}) - (1-b_k)h_k \right)^2 + \dfrac{(C(t_k;\mathbf{q}) - C_k)^2}{C_k^2} +
    \\ [4mm]
    &+ \dfrac{(D(t_k;\mathbf{q}) - g_k)^2}{g_k^2}.
\end{array}
\end{eqnarray}

Здесь первое слагаемое $t_{inc}^{-1}E(t_{k-1};\mathbf{q})$ описывает количество бессимптомных носителей вируса COVID-19, которые в день $t_k$ попадают в состояние выявленных симптомных инфицированных (см.~Рис.~\ref{fig:SEIR-HCD_scheme}в). В функционал $J(\mathbf{q})$ также включена информация о критических $C(t_k;\mathbf{q})$ и умерших $D(t_k;\mathbf{q})$ в результате СOVID-19 случаев, информация о которых доступна в открытых источниках (см.~Приложение~\ref{sec_app_data}).

Анализ идентифицируемости модели~(\ref{eq:SEIR-HCD})-(\ref{eq:SEIR-HCD_init_cond}),~(\ref{eq:SEIR-HCD_data}) показал, что параметр $\beta$, описывающий долю инфицированных индивидуумов, которые переносят заболевание без осложнений, является наименее идентифицируемым~\cite{Vavilov_ident}, поэтому в качестве дополнительной информации использованы данные медицинского центра <<Инвитро>> (см. Приложение~\ref{sec_app_data}), которые в полтора раза повысили устойчивость решения обратной задачи.

Для численного решения задачи минимизации $J(\mathbf{q})$~(\ref{eq:J_SEIRHCD}) использовалась следующая последовательность шагов: 
\begin{enumerate}
\itemsep0em 
    \item Подготовка и обработка реальных данных для вычисления обратной задачи:
    \begin{enumerate}
    \itemsep0em
        \item[1.1.] Данные по новым выявленным случаям заражения COVID-19 $h_k$, критическим (требующих подключение аппарата ИВЛ) $C_k$ и умершим $g_k$.
        \item[1.2.] Данные по проценту бессимптомных выявленных случаев $b_k$ от общего числа выявленных случаев заражения COVID-19, а также его прогноз на период моделирования, построенного с помощью нейронных сетей (см. Раздел~\ref{sec_ML_models}).
        \item[1.3.] Индекс самоизоляции от Яндекса $a(t)$, а также его прогноз на период моделирования, построенного с помощью нейронных сетей (см. Раздел~\ref{sec_ML_models}).
        \item[1.4.] Данные по проценту индивидуумов с антителами к COVID-19 от медицинского центра <<Инвитро>> $\beta (t)$, а также его прогноз на период моделирования, построенного с помощью нейронных сетей (см. Раздел~\ref{sec_ML_models}). 
    \end{enumerate}
    \item Определение границ параметров для искомого вектора $\mathbf{q}$ (см.~Таблицу~\ref{tab_parameters}, колонку 3).
    \item Уточнение вектора неизвестных параметров $\mathbf{q}$ путём решения обратной задачи алгоритмом усвоения данных. Обратная задача решается для каждого 30-дневного отрезка реальных данных (тренировочные данные). Далее, по восстановленным параметрам $\mathbf{q}$ для текущего 30-дневного отрезка осуществляется прогноз эпидемиологических данных на следующие 7 дней (валидационные данные). Прогноз осуществляется путём решения прямой задачи~(\ref{sec_R0-SEIR-HCD}) по схеме, указанной на рис.~\ref{fig:data-driven} для постоянного набора восстановленных на последнем временном промежутке параметров $\mathbf{q}$. Далее новые тренировочный и валидационный периоды смещаются на 7 дней и для этих периодов решается новая обратная задача. И так далее до тех пор пока не закончатся реальные данные.
    \begin{figure}[h!]
    \centering
    \includegraphics[width=0.7\textwidth]{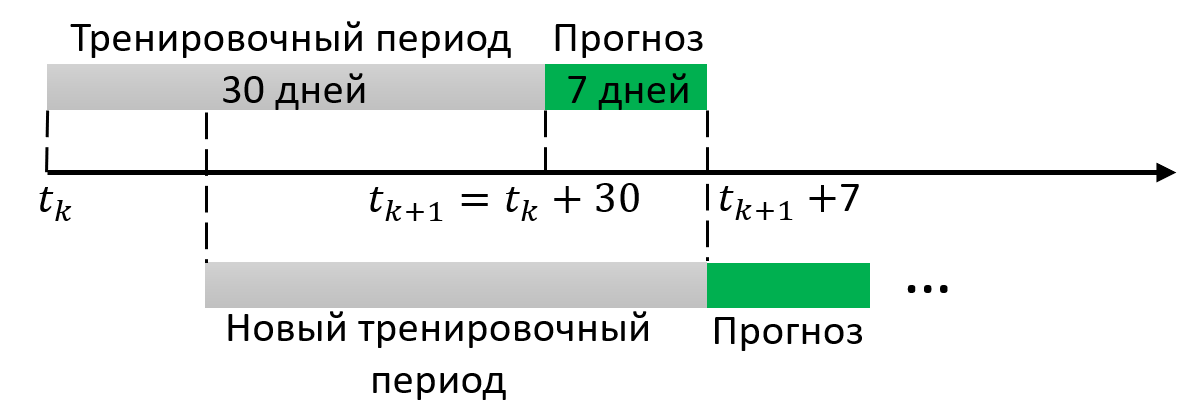}
    \caption{Алгоритм усвоения данных при решении обратной задачи для SEIR-HCD модели~(\ref{eq:SEIR-HCD})-(\ref{eq:SEIR-HCD_init_cond}),~(\ref{eq:SEIR-HCD_data}). Тренировочный период (30 дней) -- уточнение параметров $\mathbf{q}$. Прогноз (7 дней) рассчитывается при найденных $\mathbf{q}$. Новый период сдвигается на 7 дней и снова решается обратная задача.}
    \label{fig:data-driven}
\end{figure}
    \begin{enumerate}
    \itemsep0em 
        \item[3.1.] Для решения задачи минимизации целевого функционала~(\ref{eq:J_SEIRHCD}) использовался пакет глобальной оптимизации \href{https://optuna.org/}{OPTUNA}, в реализации которого лежат природоподобные алгоритмы, метод древовидных оценок Парзена и тензорная оптимизация. 
        \item[3.2.] Полученный на предыдущем шаге глобальный оптимум уточнялся с помощью локальных методов градиентного типа~\cite{KSI_2015}.
    \end{enumerate}
\end{enumerate}

\subsubsection{Алгоритм решения обратной задачи для АОМ}\label{sec_inv_ABM}
Предположим, что для АОМ, описанной в разделе~\ref{sec_ABM_COVID}, известна дополнительная информация о количестве ежедневно выявленных случаев $h_k$, проведенных ПЦР-тестов в регионе $T(t_k)$, критических $C_k$ и умерших $g_k$ случаев в фиксированные моменты времени $t_k, k=1,\ldots, K$. Обратная задача для модели АОМ состоит в определении вектора параметров $\mathbf{q}=(\alpha, \alpha_d(t), \alpha_c(t), p(t), E_0)$ по дополнительной информации $h_k, T(t_k), C_k, g_k$, $k=1,\ldots, K$. Здесь $\alpha$~-- параметр контагиозности вируса, $\alpha_d(t)$~-- дни изменения параметра $\alpha$, $\alpha_c(t)$~-- значения, на которые изменяется параметр $\alpha$ в дни $\alpha_d(t_k)$, $p(t)$~-- шанс быть протестированным (зависит от возрастной группы), $E_0$~-- начальное количество бессимптомных инфицированных.

Обратная задача сводится к задаче минимизации целевого функционала:
\begin{eqnarray}\label{eq:J_ABM}
    J(\mathbf{q}) = \sum\limits_{k=1}^{K} \dfrac{|Y(t_k;\mathbf{q}) - h_k|}{h_k} + \dfrac{|C(t_k;\mathbf{q}) - C_k|}{C_k} + \dfrac{|D(t_k;\mathbf{q}) - g_k|}{g_k}.
\end{eqnarray}
Здесь $Y(t; \mathbf{q})$~-- количество моделируемых выявленных случаев COVID-19 в результате ПЦР тестирования.

Была исследована идентифицируемость агентной модели на основе байесовского подхода~\cite{Andrianakis_2015} с тремя неизвестными параметрами: параметр контагиозности $\alpha$, начальное количество инфицированных $E_0$ и параметр тестирования $p$ по статистическим данным о количестве выявленных $h_k$, смертей $g_k$ в результате COVID-19 и пациентов, находящихся в отделении интенсивной терапии $C_k$. В результате удалось уменьшить границы поиска параметра $\alpha$ более чем в 2 раза, в то время как границы параметров $p$ и $E_0$ остались неизменными.

Алгоритм численного решения задачи минимизации функционала $J(\mathbf{q})$~(\ref{eq:J_ABM}):
\begin{enumerate}
 \item Подготовка и обработка реальных данных для вычисления обратной задачи (файлы с данными доступны на сайте \url{http://covid19-modeling.ru/} в разделе "Данные"):
     \begin{enumerate}
     \item[1.1.] Заполнение пропусков в данных о ежедневно выявленных случаев, проведенных ПЦР-тестов, критических и умерших методами обратной и прямой экстраполяции.
     \item[1.2.] Сглаживание данных с помощью Гауссовского фильтра \cite{GF} для уменьшения флуктуации данных перед использованием их в решении обратной задачи.
     \end{enumerate}
 \item Уточнение границ изменения неизвестных параметров $\mathbf{q}$ методами анализа чувствительности.
 \item Поэтапное восстановление вектора параметров $\mathbf{q}$ с шагом 30 дней, в ходе которого значения кусочно-постоянного параметра $\alpha$ $(\alpha_d(t), \alpha_c(t))$ находились последовательно один за другим. Таким образом, значения параметров, восстановленные на предыдущем шаге, использовались в последующем запуске алгоритма минимизации функционала, который представлял собой комбинацию методов глобального (библиотека \href{https://optuna.org/}{OPTUNA}) и локального (градиентные) метода оптимизации.
 \item Построение сценариев распространения количества выявленных случаев COVID-19:
 \begin{enumerate}
     \item[4.1.] Экстраполяция количества ожидаемых ПЦР-тестов в регионе $T(t)$ с помощью регрессионных моделей SARIMA (Раздел~\ref{sec_regression}) в комбинации с методами машинного обучения (Раздел~\ref{sec_ML_models}) на 45 дней.
     \item[4.2.] Для уточненных параметров и ограничительных мер (введение ограничительных мер описано в Разделе~\ref{sec_ABM_COVID}) решаем прямую задачу.
     \end{enumerate}
    \item Построение доверительных интервалов:
    \begin{enumerate}
     \item[5.1.] Решаем прямую задачу с идентифицированными параметрами 10000 раз с помощью метода Монте-Карло (вычислительное время решения прямой задачи для АОМ на кластере составляет 10 сек).
     \item[5.2.] Для каждого дня считаем квантили уровня 0.1, 0.5, 0.9 (строим функцию распределения случайной величины выявленных случаев и выбираем ее значения в точках 0.1, 0.5 и 0.9).
     \item[5.3.] Получаем 3 массива точек с шагом один день, по которым строим медианное значение (квантиль уровня 0.5, сплошная линия на графиках в Разделе~\ref{sec_num_results}).
     \end{enumerate}
\end{enumerate}

%% file: text/6conclusion.tex
\section{Заключение}\label{sec_conclusion}
В данном разделе будет проведен анализ численных расчетов для АОМ и SEIR-HCD модели в Новосибирской области (Раздел~\ref{sec_num_results}), влияние социальной дистанции на распространение эпидемии в Новосибирской области на основе ИСП (Раздел~\ref{sec_num_res_MFG}), выводы (Раздел~\ref{sec_num_conclusion}) и направления дальнейшей работы (Раздел~\ref{sec_future_work}).

\subsection{Анализ расчета сценариев распространения COVID-19}\label{sec_num_results}
Численные расчеты данного раздела приведены для Новосибирской области. Данные, использованные при анализе и построении моделей, описаны в Приложении \ref{sec_app_data}.

В Табл.~\ref{tab:reconstr_param} приводятся восстановленные параметры $\alpha$ (АОМ) и $\alpha_I, \alpha_E, \varepsilon_{HC}, \mu$ (SEIR-HCD), характеризующие распространение COVID-19 в Новосибирской области. Дата изменения $\alpha_d(t)$ параметров (результат решения обратной задачи для АОМ, см.~Раздел~\ref{sec_inv_ABM}) коррелируют с ограничительными мерами, принятыми в регионе (с интервалом 5-14 дней). Например, возобновление удаленного режима с 1 ноября 2020 года в учебных заведениях повлекло уменьшение значений параметра контагиозности $\alpha$ в структурах с 14 ноября 2020 года.

В последней колонке Табл.~\ref{tab:reconstr_param} приведены 95\% доверительные интервалы (95\% CI) полученных значений, вычисленные по 30 испытаниям оптимизационного алгоритма OPTUNA. В АОМ параметр контагиозности вируса $\alpha (t)$ имеет различные значения для каждой структуры контактов (домохозяйства, образовательные учреждения, работа и общественные места). В SEIR-HCD модели восстановленные значения параметров являются усредненными в регионе, так как разделения на структуры контактов не подразумевалось. 

\begin{longtable}{|p{1.9cm}|p{1.5cm}|p{1.8cm} p{2.2cm} p{1.9cm}| p{3.2cm}|}
 \hline
 {\textbf{Дата изменения}} & {\textbf{Символ}} & \multicolumn{3}{c|}{\textbf{Восстановленное значение}} & \textbf{Доверительный интервал} \\ 
  \cline{3-5}
   & & \textbf{Домохоз-ва} & \textbf{Образование и работа} & \textbf{Обществ. места} & \\ 
  \hline
\multirow{3}{*}{23-05-2020} & $\alpha$ & 0.02524 & 0.00505 & 0.00252 & \\
\cline{3-6}
    & $\alpha_I$ & \multicolumn{3}{c|}{0.06383} & (0.02879, 0.09887) \\
    (начало & $\alpha_E$ & \multicolumn{3}{c|}{0.09348} & (0.03673, 0.15023) \\
    периода) & $\varepsilon_{HC}$ & \multicolumn{3}{c|}{0.00824} & (0.00452, 0.01196) \\
    & $\mu$ & \multicolumn{3}{c|}{0.22189} & (0.06458, 0.3792)\\
\hline
\multirow{4}{*}{17-06-2020} & $\alpha$ & 0.0223 & 0.00446 & 0.00223 &\\
\cline{3-6}
    & $\alpha_I$ & \multicolumn{3}{c|}{0.06701} & (0.0409, 0.09313) \\
    & $\alpha_E$ & \multicolumn{3}{c|}{0.08834} & (0.02224, 0.15445) \\
    & $\varepsilon_{HC}$ & \multicolumn{3}{c|}{0.01607} & (0.01461, 0.01753) \\
    & $\mu$ & \multicolumn{3}{c|}{0.24682} & (0.09869, 0.39495)\\
\hline
\multirow{4}{*}{12-07-2020} & $\alpha$ & 0.02788 & 0.00558 & 0.00279 & \\
\cline{3-6}
    & $\alpha_I$ & \multicolumn{3}{c|}{0.0643} & (0.04158, 0.08702) \\
    & $\alpha_E$ & \multicolumn{3}{c|}{0.07852} & (0.02035, 0.13669) \\
    & $\varepsilon_{HC}$ & \multicolumn{3}{c|}{0.00896} & (0.00812, 0.0098) \\
    & $\mu$ & \multicolumn{3}{c|}{0.2681} & (0.10798, 0.42823)\\
\hline
\multirow{4}{*}{06-08-2020} & $\alpha$ & 0.02181 & 0.00436 & 0.00218 & \\
\cline{3-6}
    & $\alpha_I$ & \multicolumn{3}{c|}{0.04452} & (0.03538, 0.05367) \\
    & $\alpha_E$ & \multicolumn{3}{c|}{0.03547} & (0.00641, 0.06453) \\
    & $\varepsilon_{HC}$ & \multicolumn{3}{c|}{0.00208} & (0.00133, 0.00283) \\
    & $\mu$ & \multicolumn{3}{c|}{0.21982} & (0.07017, 0.36946)\\
\hline
\multirow{4}{*}{31-08-2020} & $\alpha$ & 0.02466 & 0.00493 & 0.00247 & \\
\cline{3-6}
    & $\alpha_I$ & \multicolumn{3}{c|}{0.06572} & (0.04372, 0.08773) \\
    & $\alpha_E$ & \multicolumn{3}{c|}{0.08158} & (0.02038, 0.14278) \\
    & $\varepsilon_{HC}$ & \multicolumn{3}{c|}{0.00115} & (0.00103, 0.00128) \\
    & $\mu$ & \multicolumn{3}{c|}{0.24302} & (0.08012, 0.40592)\\
\hline
\multirow{4}{*}{25-09-2020} & $\alpha$ & 0.02948 & 0.0059 & 0.00295 & \\
\cline{3-6}
    & $\alpha_I$ & \multicolumn{3}{c|}{0.07915} & (0.03855, 0.11974) \\
    & $\alpha_E$ & \multicolumn{3}{c|}{0.13327} & (0.05215, 0.2144) \\
    & $\varepsilon_{HC}$ & \multicolumn{3}{c|}{0.00854} & (0.00784, 0.00925) \\
    & $\mu$ & \multicolumn{3}{c|}{0.22748} & (0.07004, 0.38493)\\
\hline
\multirow{4}{*}{20-10-2020} & $\alpha$ & 0.02948 & 0.0059 & 0.00295 & \\
\cline{3-6}
    & $\alpha_I$ & \multicolumn{3}{c|}{0.09853} & (0.05844, 0.13862) \\
    & $\alpha_E$ & \multicolumn{3}{c|}{0.13321} & (0.05673, 0.20969) \\
    & $\varepsilon_{HC}$ & \multicolumn{3}{c|}{0.00979} & (0.00929, 0.01029) \\
    & $\mu$ & \multicolumn{3}{c|}{0.25076} & (0.07987, 0.42165)\\
\hline
\multirow{4}{*}{14-11-2020} & $\alpha$ & 0.02754 & 0.00551 & 0.00275 & \\
\cline{3-6}
    & $\alpha_I$ & \multicolumn{3}{c|}{0.07429} & (0.04941, 0.09917) \\
    & $\alpha_E$ & \multicolumn{3}{c|}{0.09587} & (0.0368, 0.15495) \\
    & $\varepsilon_{HC}$ & \multicolumn{3}{c|}{0.01365} & (0.01239, 0.01491) \\
    & $\mu$ & \multicolumn{3}{c|}{0.22794} & (0.08333, 0.37254)\\
\hline
\multirow{4}{*}{09-12-2020} & $\alpha$ & 0.02447 & 0.00489 & 0.00245 & \\
\cline{3-6}
    & $\alpha_I$ & \multicolumn{3}{c|}{0.066} & (0.03925, 0.09274) \\
    & $\alpha_E$ & \multicolumn{3}{c|}{0.09382} & (0.02578, 0.16187) \\
    & $\varepsilon_{HC}$ & \multicolumn{3}{c|}{0.01057} & (0.00983, 0.01131) \\
    & $\mu$ & \multicolumn{3}{c|}{0.2593} & (0.09546, 0.42313)\\
\hline
\multirow{4}{*}{03-01-2021} & $\alpha$ & 0.02447 & 0.00489 & 0.00245 & \\
\cline{3-6}
    & $\alpha_I$ & \multicolumn{3}{c|}{0.11286} & (0.09334, 0.13239) \\
    & $\alpha_E$ & \multicolumn{3}{c|}{0.05064} & (0.00117, 0.1001) \\
    & $\varepsilon_{HC}$ & \multicolumn{3}{c|}{0.00995} & (0.00948, 0.01043) \\
    & $\mu$ & \multicolumn{3}{c|}{0.20234} & (0.06234, 0.34234)\\
\hline
\multirow{4}{*}{28-01-2021} & $\alpha$ & 0.02332 & 0.00466 & 0.00233 & \\
\cline{3-6}
    & $\alpha_I$ & \multicolumn{3}{c|}{0.06664} & (0.04204, 0.09123) \\
    & $\alpha_E$ & \multicolumn{3}{c|}{0.07783} & (0.01485, 0.1408) \\
& $\varepsilon_{HC}$ & \multicolumn{3}{c|}{0.01158} & (0.01093, 0.01223) \\
    & $\mu$ & \multicolumn{3}{c|}{0.2535} & (0.11234, 0.39465)\\
\hline
\multirow{4}{*}{22-02-2021} & $\alpha$ & 0.02332 & 0.00466 & 0.00233 & \\
\cline{3-6}
    & $\alpha_I$ & \multicolumn{3}{c|}{0.05395} & (0.02886, 0.07904) \\
    & $\alpha_E$ & \multicolumn{3}{c|}{0.1069} & (0.03337, 0.18044) \\
    & $\varepsilon_{HC}$ & \multicolumn{3}{c|}{0.00858} & (0.00796, 0.0092) \\
    & $\mu$ & \multicolumn{3}{c|}{0.24984} & (0.1092, 0.39048)\\
\hline
\multirow{4}{*}{19-03-2021} & $\alpha$ & 0.0398 & 0.00796 & 0.00398 & \\
\cline{3-6}
    & $\alpha_I$ & \multicolumn{3}{c|}{0.05269} & (0.03706, 0.06832) \\
    & $\alpha_E$ & \multicolumn{3}{c|}{0.07891} & (0.03338, 0.12445) \\
    & $\varepsilon_{HC}$ & \multicolumn{3}{c|}{0.0084} & (0.00773, 0.00906) \\
    & $\mu$ & \multicolumn{3}{c|}{0.25176} & (0.11776, 0.38576)\\
\hline
\multirow{4}{*}{13-04-2021} & $\alpha$ & 0.02788 & 0.00558 & 0.00279 & \\
\cline{3-6}
    & $\alpha_I$ & \multicolumn{3}{c|}{0.04975} & (0.02219, 0.07731) \\
    & $\alpha_E$ & \multicolumn{3}{c|}{0.10684} & (0.03534, 0.17834) \\
    & $\varepsilon_{HC}$ & \multicolumn{3}{c|}{0.00381} & (0.00222, 0.00539) \\
    & $\mu$ & \multicolumn{3}{c|}{0.27165} & (0.11115, 0.43214)\\
\hline
\multirow{4}{*}{08-05-2021} & $\alpha$ & 0.02175 & 0.00435 & 0.00217 & \\
\cline{3-6}
    & $\alpha_I$ & \multicolumn{3}{c|}{0.05668} & (0.03625, 0.0771) \\
    & $\alpha_E$ & \multicolumn{3}{c|}{0.09091} & (0.03979, 0.14203) \\
    & $\varepsilon_{HC}$ & \multicolumn{3}{c|}{0.00271} & (0.00195, 0.00346) \\
    & $\mu$ & \multicolumn{3}{c|}{0.28275} & (0.14527, 0.42023)\\
\hline
\multirow{4}{*}{02-06-2021} & $\alpha$ & 0.0406 & 0.00812 & 0.00406 & \\
\cline{3-6}
    & $\alpha_I$ & \multicolumn{3}{c|}{0.08184} & (0.04872, 0.11495) \\
    & $\alpha_E$ & \multicolumn{3}{c|}{0.09535} & (0.02064, 0.17007) \\
    & $\varepsilon_{HC}$ & \multicolumn{3}{c|}{0.00527} & (0.00333, 0.0072) \\
    & $\mu$ & \multicolumn{3}{c|}{0.20956} & (0.07153, 0.34758)\\
\hline
\multirow{4}{*}{27-06-2021} & $\alpha$ & 0.03063 & 0.00613 & 0.00306 & \\
\cline{3-6}
    & $\alpha_I$ & \multicolumn{3}{c|}{0.09507} & (0.06195, 0.12818) \\
    & $\alpha_E$ & \multicolumn{3}{c|}{0.08282} & (0.02295, 0.14268) \\
    & $\varepsilon_{HC}$ & \multicolumn{3}{c|}{0.02107} & (0.02011, 0.02203) \\
    & $\mu$ & \multicolumn{3}{c|}{0.22876} & (0.07595, 0.38158)\\
\hline
\multirow{4}{*}{22-07-2021} & $\alpha$ & 0.02843 & 0.00569 & 0.00284 & \\
\cline{3-6}
    & $\alpha_I$ & \multicolumn{3}{c|}{0.06641} & (0.04038, 0.09245) \\
    & $\alpha_E$ & \multicolumn{3}{c|}{0.08157} & (0.02461, 0.13852) \\
    & $\varepsilon_{HC}$ & \multicolumn{3}{c|}{0.00774} & (0.00641, 0.00907) \\
    & $\mu$ & \multicolumn{3}{c|}{0.25786} & (0.0942, 0.42151)\\
\hline
\multirow{4}{*}{16-08-2021} & $\alpha$ & 0.02449 & 0.0049 & 0.00245 & \\
\cline{3-6}
    & $\alpha_I$ & \multicolumn{3}{c|}{0.05153} & (0.02638, 0.07667) \\
    & $\alpha_E$ & \multicolumn{3}{c|}{0.08968} & (0.0224, 0.15697) \\
    & $\varepsilon_{HC}$ & \multicolumn{3}{c|}{0.00366} & (0.00232, 0.005) \\
    & $\mu$ & \multicolumn{3}{c|}{0.21537} & (0.09108, 0.33965)\\
\hline
\multirow{4}{*}{10-09-2021} & $\alpha$ & 0.03273 & 0.00655 & 0.00327 & \\
\cline{3-6}
    & $\alpha_I$ & \multicolumn{3}{c|}{0.06993} & (0.04723, 0.09263) \\
    & $\alpha_E$ & \multicolumn{3}{c|}{0.0706} & (0.01428, 0.12692) \\
    & $\varepsilon_{HC}$ & \multicolumn{3}{c|}{0.01063} & (0.00999, 0.01128) \\
    & $\mu$ & \multicolumn{3}{c|}{0.24317} & (0.07392, 0.41242)\\
\hline
\multirow{4}{*}{08-10-2021} & $\alpha$ & 0.04294 & 0.00859 & 0.00429 & \\
\cline{3-6}
    & $\alpha_I$ & \multicolumn{3}{c|}{0.08198} & (0.05147, 0.11248) \\
    & $\alpha_E$ & \multicolumn{3}{c|}{0.13857} & (0.0659, 0.21124) \\
    & $\varepsilon_{HC}$ & \multicolumn{3}{c|}{0.01178} & (0.01126, 0.01231) \\
    & $\mu$ & \multicolumn{3}{c|}{0.24395} & (0.09619, 0.3917)\\
\hline
\caption{Восстановленные параметры распространения COVID-19 в Новосибирской области с 13.03.2020 по 30.09.2021.\\ Здесь $\alpha$~-- параметр контагиозности, $\alpha_E$~-- доля заражения между бессимптомной и восприимчивой группами населения, $\alpha_I$~-- доля заражения между инфицированным и восприимчивым населением, $\varepsilon_{HC}$~-- доля госпитализированных случаев, которым требуется подключение ИВЛ, $\mu$~-- доля смертельных случаев в результате COVID-19.}
\label{tab:reconstr_param}
\end{longtable}

\begin{figure}[H]
    \centering
    \includegraphics[width=\textwidth]{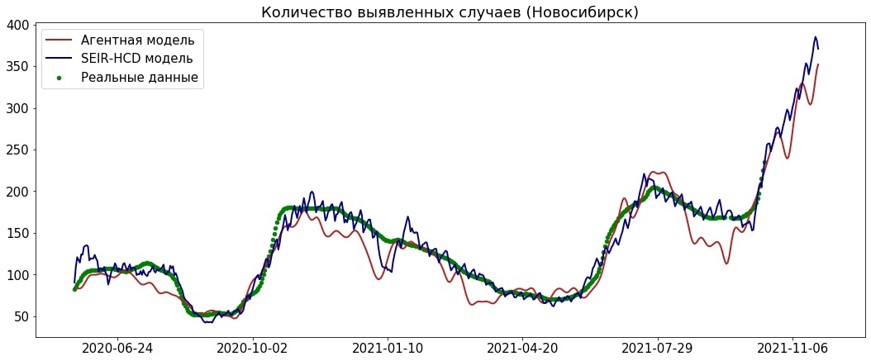}
    \caption{Моделирование распространения ежедневно выявленных случаев COVID-19 в Новосибирской области с 12.03.2020 по 30.11.2021 (расчеты проведены 16.10.2021).\\ \textcolor{blue}{Синяя линия} -- SEIR-HCD модель, \textcolor{red}{красная линия} -- АОМ, \textcolor{OliveGreen}{зеленая линия} -- реальные данные с 12.03.2020 по 16.10.2021.}
    \label{result1}
\end{figure}

На рисунке \ref{result1} представлены численные результаты моделирования распространения ежедневно выявленных случаев. Зелеными точками отмечены реальные данные с 12.03.2020 по 16.10.2021, которые участвовали в решении обратной задачи. Красной линией представлены результаты для АОМ, синей -- для камерной SEIR-HCD модели. При моделировании был сделан прогноз на 40 дней (с 17.10.2021 по 26.11.2021) с учетом сохранения карантинных мер. Результат прогнозирования был сравнен с реальными данными 08.11.2021: количество выявленных случаев согласно реальным данным -- 400 человек, результатам АОМ -- 362, камерной SEIR-HCD модели -- 387.

\begin{figure}[H]
    \centering
    \includegraphics[width=\textwidth]{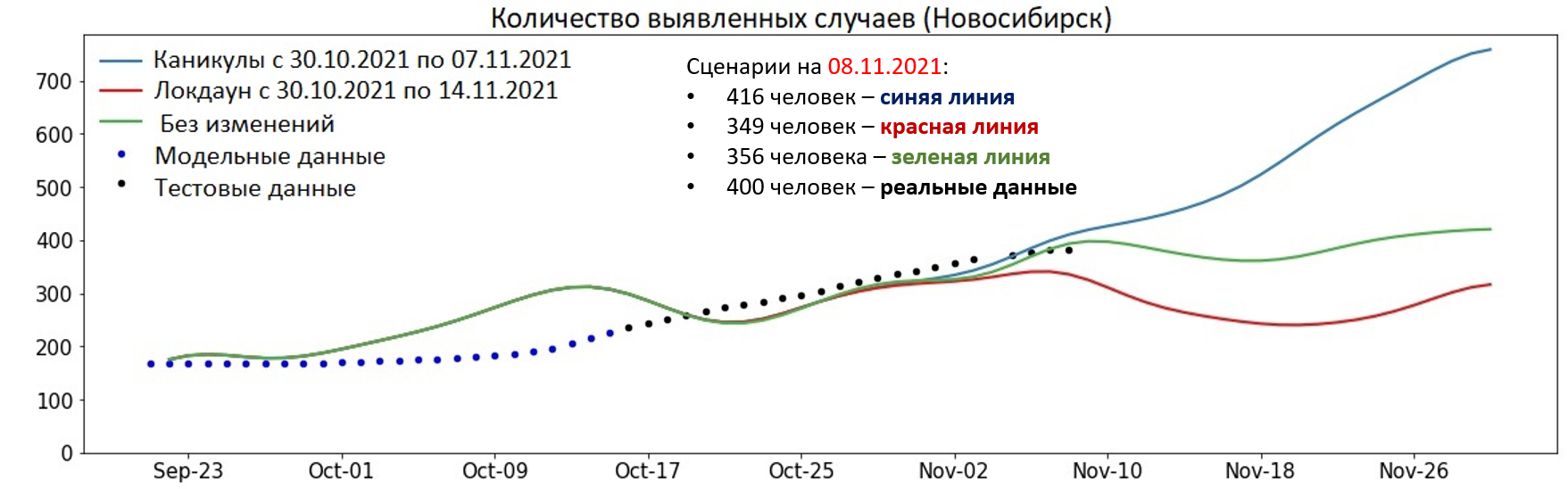}
    \caption{Сценарии распространения ежедневно выявленных случаев COVID-19 в Новосибирской области с 17.10.2021 по 30.11.2021.\\
    \textcolor{blue}{Синяя линия} -- нерабочие дни с 30.10 по 07.11 с учетом оттока 40\% населения в эндемичные районы, \textcolor{red}{красная линия} -- локдаун с 30.10.2021 по 14.11.2021, \textcolor{OliveGreen}{зеленая линия} -- базовый сценарий на 16.10.2021.\\
    \textcolor{blue}{\textbf{Синие точки}} -- реальные данные выявленных случаев COVID-19 по 16.10.2021, используемые в моделировании, \textbf{черные точки} -- реальные данные выявленных случаев COVID-19 с 17.10.2021 по 08.11.2021, используемые для валидации.}
    \label{result2}
\end{figure}

Для анализа эффективности ограничительных мер были изучены следующие сценарии распространения ежедневно выявленных случаев COVID-19 (рис.~\ref{result2}):
\begin{itemize}
    \item Нерабочие дни с 30.10 по 07.11, в течение которых уменьшается на 40\% людей на работе и учебе, потом увеличивается заболеваемость из-за привезенных случаев в 2 раза –- \textcolor{blue}{синяя линия};
    
    \item Локдаун с 30.10.2021 по 14.11.2021, в течение которого полностью закрыты образовательные учреждения, 50\% общественных мест и 50\% рабочих переведены на удаленный режим работы –- \textcolor{red}{красная линия};
    
    \item Ничего не предпринимать (базовый сценарий) -- \textcolor{OliveGreen}{зеленая линия}.
\end{itemize}

Результаты численных расчетов для сценариев развития представлены на рисунке~\ref{result2}. Так, введенная мера о нерабочих днях (синяя линия) может привести к самому неблагоприятному сценарию развития ситуации, при которой к 8 ноября ожидается 416 выявленных случаев. Базовый сценарий развития (зеленая линия) предполагает 356 выявленных случаев к 8 ноября, в рамках которого количество людей в общественных местах уменьшится из-за введения QR-кодов. Наиболее оптимистичный сценарий стоит ждать в случае более серьезного локдауна (красная линия), при котором к 8 ноября ожидается 349 выявленных случаев.

\begin{figure}[H]
    \centering
    \includegraphics[width=\textwidth]{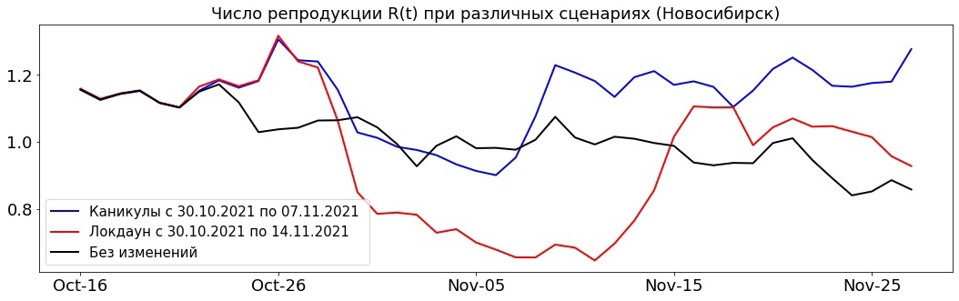}
    \caption{Сценарии изменения индекса репродукции вируса $\mathcal{R}_0(t)$ для Новосибирской области с 17.10.2021 по 30.11.2021. \\
    \textcolor{blue}{Синяя линия} -- объявление нерабочих дней с 30.10 по 07.11 и отток 40\% населения из региона, \textcolor{red}{красная линия} -- закрытие учебных заведений и 50\% общественных и рабочих мест с 30.10.2021 по 14.11.2021, \textbf{черная линия} -- базовый сценарий на 16.10.2021.}
    \label{result3}
\end{figure}

Для каждого из сценариев были проведены расчеты индекса репродукции вируса $\mathcal{R}_0$ для АОМ, описанного в разделе \ref{sec_reproductionNo}. Результаты численных расчетов представлены на рис.~\ref{result3}. В случае сценария <<объявления нерабочей недели>> индекс репродукции ожидается наибольшим по значению, что впоследствии может привести к повышенному числу выявленных и умерших случаев, а также нагрузку на систему здравоохранения (\textcolor{blue}{синяя линия}). Более серьезные ограничения до 14.11 (\textcolor{red}{красная линия}) значительно уменьшат значение индекса репродукции, однако через 2 недели он снова может увеличиться и переступить порог в 1. Базовый сценарий развития (\textbf{черная линия}) к 30 ноября предполагает стабильную ситуацию в регионе.

\subsection{Анализ влияния социальной дистанции на распространение эпидемии}\label{sec_num_res_MFG}
На основе SIR-модели была введена модель \nom{SIRC\textsubscript{r}}{камерная модель, основанная на системе из 4 обыкновенных дифференциальных уравнений, где $C_r$~-- группа населения, имеющая перекрестный иммунитет (Cross-immune)}~\cite{Casagrandi_2006}, в которой $C_r(t)$~-- группа населения, имеющая перекрестный иммунитет (т.е. иммунную память к коронавирусу):
\begin{eqnarray}\label{model_SIRC}
\left\{\begin{array}{ll}
    \dfrac{dS}{dt} = -\alpha S(t)I(t) + \gamma C_r(t), & t> 0,\\[5pt]
    \dfrac{dI}{dt} = \alpha S(t)I(t) + \varepsilon \alpha C_r(t)I(t) - \beta I(t),\\[5pt]
    \dfrac{dR}{dt} = (1-\varepsilon) \alpha C_r(t)I(t) + \beta I(t) - \delta R(t), \\[5pt]
    \dfrac{dC_r}{dt} = \delta R(t) - \alpha C_r(t)I(t) - \gamma C_r(t).
\end{array}\right.
\end{eqnarray}
Значения параметров для Новосибирской области, определенные с помощью решения обратной задачи для модели~(\ref{model_SIRC}) по дополнительной информации о количестве ежедневно выявленных в результате ПЦР-теста $h_k$ и вылеченных от COVID-19 $R_k$, $k=1,\ldots, 100$, с 1 мая 2020 года по 8 августа 2020 года приведены в Табл.~\ref{tab:param_SIRC}.
\begin{table}[!ht]
\centering
\begin{tabular}{|p{9.2cm}|p{1.8cm}|p{3.5cm}| } 
 \hline
 \textbf{Описание параметров} & \textbf{Символ} & \textbf{Среднее значение} \\ 
  \hline
  Скорость передачи инфекции & $\alpha$ & 0.2821\\
  \hline
  Скорость выздоровления инфицированного населения & $\beta$ & 0.253\\
   \hline
  Вероятность приобретения перекрестного иммунитета & $\delta$ & 0.0889\\
  \hline
  Скорость, с которой перекрестно-иммунная популяция снова становится восприимчивой & $\gamma$ & 0.0376\\
 \hline
 Средняя вероятность повторного заражения человека с перекрестным иммунитетом & $\varepsilon$ & 0.0928\\
 \hline
 \end{tabular}
 \caption{Эпидемиологические параметры для SIRC\textsubscript{r} модели~(\ref{model_SIRC}) для Новосибирской области с 01.05.2020 по 08.08.2020, полученные с помощью решения обратной задачи.}
\label{tab:param_SIRC}
\end{table}

Численные расчеты данного раздела основываются на работе~\cite{Petrakova_2021}.
Для сведения модели~(\ref{model_SIRC}) к модели ИСП (см.~Раздел~\ref{sec_MFG}) вместо численности населения в каждой группе $S(t), I(t), R(t), C_r(t)$ введем плотность распределения людей внутри этих групп $p_S(t,x)$, $p_I(t,x)$, $p_R(t,x)$, $p_{C_r}(t,x)$. Переменная $x$ изменяется в пределах [0,1] и означает соблюдение карантинных мер: $x=1$ означает, что человек полностью придерживается карантинных ограничений, включая вакцинирование, а $x=0$ означает, что он не придерживается никаких ограничений.

Введем функции $u_S(t,x)$, $u_I(t,x)$, $u_R(t,x)$, $u_{C_r}(t,x)$, обозначающие изменение состояния человека в шкале соблюдения карантинных мер, вызываемое окружающей обстановкой, воздействием СМИ, вакцинацией и организационными мероприятиями. Тогда уравнения КФП~(\ref{eq_5})-(\ref{eq_7}) примут вид:
\begin{equation}\label{eq_KFP_SIRC}
 \hspace{-4mm}\left\{ \begin{aligned}
   & \dfrac{\partial {p}_{S}}{\partial t}+\nabla \left( {{p}_{S}}{{u}_{S}} \right)+\alpha {{p}_{S}}{{p}_{I}} - \gamma p_{C_r}-{\sigma _{S}^{2}\Delta {{p}_{S}}}/{2}\;=0, \\ 
  & \dfrac{\partial {p}_{I}}{\partial t}+\nabla \left( {{p}_{I}}{{u}_{I}} \right)-\alpha {{p}_{S}}{{p}_{I}} - \varepsilon\alpha p_{C_r}p_I+\beta {{p}_{I}}-{\sigma _{I}^{2}\Delta {{p}_{I}}}/{2}\;=0, \\ 
 & \dfrac{\partial {p}_{R}}{\partial t}+\nabla \left( {{p}_{R}}{{u}_{R}} \right)-(1-\varepsilon)\alpha p_{C_r}p_I-\beta {{p}_{I}}+\delta p_R-{\sigma _{R}^{2}\Delta {{p}_{R}}}/{2}\;=0,\\
 & \dfrac{\partial {p}_{C_r}}{\partial t}+\nabla \left( {{p}_{C_r}}{{u}_{C_r}} \right)-\delta p_R + \alpha p_{C_r}p_I+\gamma {{p}_{C_r}}-{\sigma _{C_r}^{2}\Delta {{p}_{C_r}}}/{2}\;=0
\end{aligned} \right.
\end{equation}
с начальными
\begin{equation}\label{eq_KFP_SIRC_initial}
    p_i(0,x) = \frac{A_i}{B_i} \left( exp\left(-\frac{(x-x_i^c)^2}{2 (\sigma^c_i)^2}\right)\bigg/{\sigma^c_i\sqrt{2\pi}} + a_ix^2 + b_i(1-x)^2 \right),\;\; x\in [0,1]
\end{equation}
и краевыми условиями
\begin{equation}\label{eq_KFP_SIRC_boundary}
    \dfrac{\partial p_i}{\partial x}(t,0)= \dfrac{\partial p_i}{\partial x}(t,1)=0,\qquad t\in [0,T],\;\; i\in \{S,I,R,C_r\}.
\end{equation}
Здесь $A_i$~-- доля текущей фракции по отношению к общей численности населения в начальный момент времени; $B_i$~-- коэффициент нормировки, равный интегралу по $[0,1]$ от выражения в скобках; $a_i = exp\left(-\frac{(1-x_i^c)^2}{2 (\sigma^c_i)^2}\right) (1-x^c_i)/(2(\sigma^c_i)^3\sqrt{2\pi})$ and $b_i = exp\left(-\frac{(x_i^c)^2}{2 (\sigma^c_i)^2}\right) (x^c_i)/(2(\sigma^c_i)^3\sqrt{2\pi})$ для обеспечения выполнения граничных условий \eqref{eq_7} для $p_{0i}$. В численных расчетах мы использовали следующие значения $A_i$, $x^c_i$ и $\sigma^c_i$
\begin{equation*}
\begin{aligned}
   & A_S = 0.999757;   & x_S^c = 0.8 ;\;\;\sigma_S^c = 0.1; \\
   & A_I = 0.000204;   & x_I^c = 0.2;\;\;\sigma_I^c =0.1 ; \\
   & A_R = 0.000039; & x_R^c = 0.7;\;\;\sigma_R^c =0.2 ;\\
   & A_{C_r} = 0.000000; &\;\;\;\;\; x_{C_r}^c = 0.3 ;\;\;\sigma_{C_r}^c = 0.2 .\\
\end{aligned}
\end{equation*}

Функционал стоимости организационных мероприятий и социально-экономических потерь
\begin{equation}\label{eq_J_SIRC}
    J(p_{sirc}, u_{sirc}) = \int\limits_0^T \int\limits_0^1 \left( \tilde{C}(t,x,u_{sirc}(t,x))+G(t,x,p_{sirc}(t,x)) \right)\, dxdt.
\end{equation}
Здесь $\tilde{C}(t,x,u_{sirc}(t,x))$~-- стоимость проводимых организационных мероприятий: вакцинация, карантинные мероприятия, введение QR-кодов и другие, $G(t,x,p_{sirc}(t,x))$~-- стоимость социальных и экономических потерь для группы населения в позиции $x$ на момент времени $t$, $p_{sirc}=(p_S, p_I, p_R, p_{C_r})$, $u_{sirc} = (u_S, u_I, u_R, u_{C_r})$.

Минимум функционала $J(p_{sirc}, u_{sirc})$~(\ref{eq_J_SIRC}) при выполнении уравнений КФП определяется методом множителей Лагранжа $\psi_S(t,x)$, $\psi_I(t,x)$, $\psi_R(t,x)$, $\psi_{C_r}(t,x)$, которые удовлетворяют системе ГЯБ типа~(\ref{eq_GJB_common}):
\begin{equation}\label{eq_GJB_SIRC}
 \hspace{-5mm}\left\{ \begin{aligned}
   & \dfrac{\partial \psi_S}{\partial t}+\dfrac{\sigma_S^2}{2}\Delta\psi_S+u_S \cdot \nabla\psi_S = \tilde{C}_S(t,x,p_{sirc}, \psi_{sirc}), \\
  & \begin{aligned}
  \dfrac{\partial \psi_I}{\partial t}+\dfrac{\sigma_I^2}{2}\Delta\psi_I + u_I \cdot \nabla\psi_I = \tilde{C}_I(t,x,p_{sirc}, \psi_{sirc}),
  \end{aligned} \\ 
  & \dfrac{\partial \psi_R}{\partial t}+\dfrac{\sigma_R^2}{2}\Delta\psi_R + u_R \cdot \nabla\psi_R =\tilde{C}_R(t,x,p_{sirc}, \psi_{sirc}), \\ 
  & \dfrac{\partial \psi_{C_r}}{\partial t}+\dfrac{\sigma_{C_r}^2}{2}\Delta\psi_{C_r} + u_{C_r} \cdot \nabla\psi_{C_r} =\tilde{C}_{C_r}(t,x,p_{sirc}, \psi_{sirc})
\end{aligned} \right.
\end{equation}
с начальными условиями на конце периода времени оптимизации $T$
\[\psi_i (T,x) = 0, \qquad x\in [0,1],\;\; i\in \{S,I,R,C_r\}\]
и краевыми условиями
\[\dfrac{\partial \psi_i}{\partial x}(t,0)= \dfrac{\partial \psi_i}{\partial x}(t,1)=0,\qquad t\in [0,T],\;\; i\in \{S,I,R,C_r\}.\]

Дифференциальные части задач КФП и ГЯБ сопряжены между собой и их коэффициенты связаны алгебраическими уравнениями
\begin{equation}\label{eq_connection}
    \dfrac{\partial \tilde{C}}{\partial u_i}(t,x,u_{sirc}) = -\dfrac{\partial \psi_i}{\partial x}(t,x),\quad i\in \{S,I,R,C_r\}
\end{equation}
из которых определяются параметры $u_S, u_I, u_R, u_{C_r}$, отражающие текущее изменение позиций групп населения по отношению к соблюдению карантинных мер.

\begin{figure}[H]
    \centering
    \includegraphics[width=0.9\textwidth]{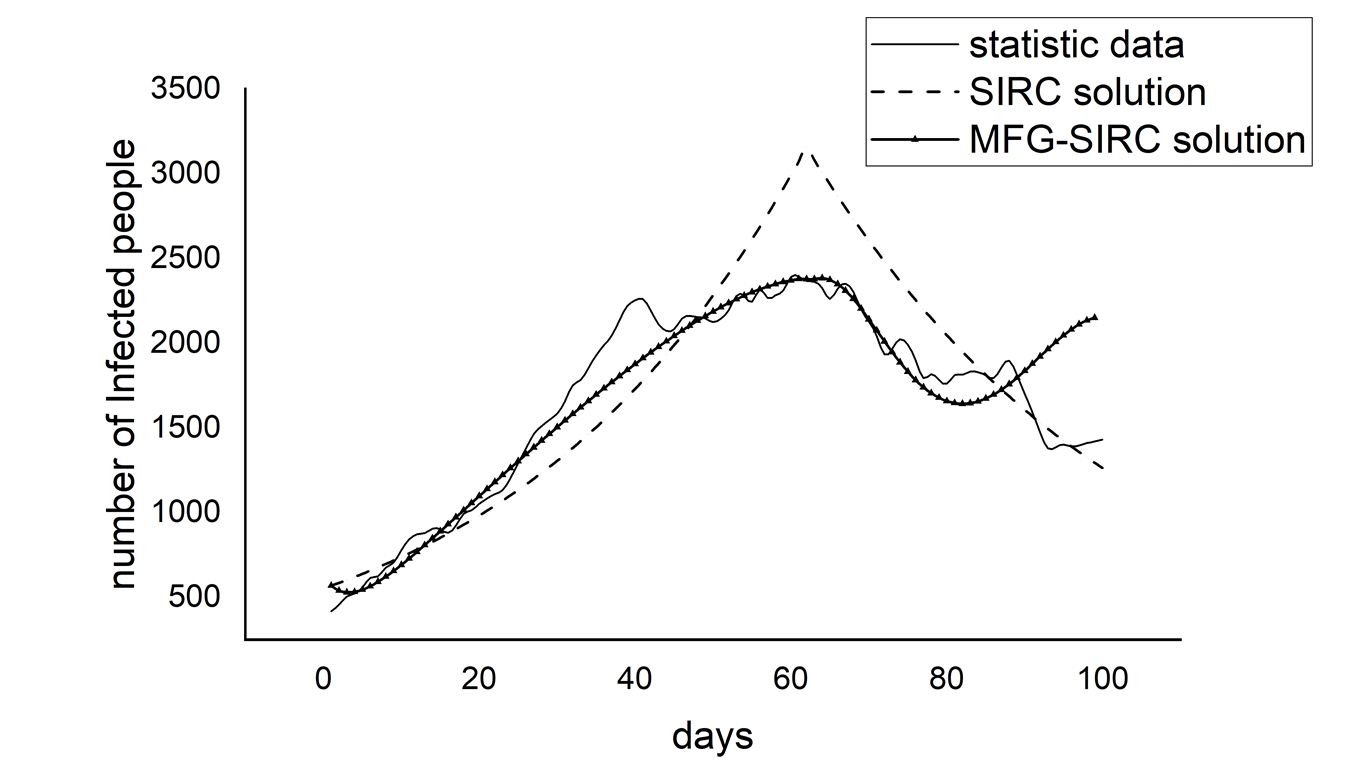}
    \caption{Сравнение результатов моделирования числа выявленных случаев COVID-19 в Новосибирской области со статистическими данными с 01.05.2020 по 08.08.2020.\\
    Сплошная линия -- статистические данные, пунктирная линия -- SIRC\textsubscript{r} модель, линия с треугольниками -- ИСП модель.}
    \label{fig:SIRC}
\end{figure}

Решение задач для уравнений КФП~(\ref{eq_KFP_SIRC})-(\ref{eq_KFP_SIRC_boundary}) и ГЯБ~(\ref{eq_GJB_SIRC}) совместно с уравнением связи~\ref{eq_connection} дает решение задачи оптимального поведения агентов при эпидемии с указанием строгости мер или возможности ослабления карантинных ограничений. На рис.~\ref{fig:SIRC} приведено сравнение результатов моделирования числа инфицированного населения для моделей SIRC\textsubscript{r}~(\ref{model_SIRC}) и ИСП~(уравнения КФП и ГЯБ) со статистическими данными в течение 100 дней с 1 мая по 8 августа 2020 года в Новосибирской области.

\subsection{Выводы}\label{sec_num_conclusion}
Сочетание моделей SIR и АОМ позволяет строить более разнообразные сценарии распространения пандемии COVID-19. А именно, используя результаты решения обратной задачи SEIR-HCD модели как дополнительную информацию для АОМ, мы уточняем сценарии с учетом ограничительных мер с помощью АОМ. Рассчитанные сценарии используются как дополнительная информации для дальнейшего уточнения параметров SEIR-HCD модели. Также можно перевести на материалистический язык разные с содержательной точки зрения сценарии и сравнить их.

При усложнении SIR-модели, вводя возрастные разграничения в популяции и пространственные перемещения, мы получим первое приближение АОМ.

Для более точных рекомендаций поддержки принятия решений и более комплексного моделирования мы применяем подход ИСП и ИСПУ (управление играми среднего поля, mean-field-type control). А именно, влияние вакцинации, характеризующее социальные настроения в регионе, учитывается при построении распределений населения в эпидемиологических группах.

История заболевания (временные ряды эпидемиологических данных) составляет обучающие множества для методов машинного обучения.

\subsection{Направления дальнейшей работы}\label{sec_future_work}
Необходимо совершенствовать методы искусственного интеллекта в приложении к моделированию COVID-19 и расчетов сценариев выхода из пандемии с учетом социальных, экономических и экологических процессов, включая анализ ситуации в различных регионах и группах, обработку данных, а также создавать комплексную модель на основе SIR, АОМ и ИСП (ИСПУ).

\addcontentsline{toc}{section}{Финансовая поддержка}
\section*{Финансовая поддержка}
Работа выполнена при поддержке Математического Центра в Академгородке, соглашение с Министерством науки и высшего образования Российской Федерации \textnumero~075-15-2019-1675, Российского научного фонда (проект \textnumero~18-71-10044) и Российского фонда фундаментальных исследований (проект \textnumero~20-51-54004).

\addcontentsline{toc}{section}{Благодарности}
\section*{Благодарности}
В различные годы исследования по моделированию в эпидемиологии (туберкулеза, \nom{ВИЧ}{вирус иммунодефицита человека}, COVID-19) выполнялись по грантам:
\begin{itemize}
    \item Российского научного фонда (проект \textnumero~18-71-10044);
    \item Российского фонда фундаментальных исследований (проекты \textnumero~21-51-10003, \textnumero~20-51-54004, \textnumero~18-31-20019, \textnumero~17-51-540004);
    \item Президента РФ (\textnumero~МК-814.2019.1, \textnumero~МК-4994.2021.1.1);
    \item Мэрии города Новосибирска на предоставление грантов в форме субсидий в сфере научной и инновационной деятельности 2021;
    \item Министерства образования Республики Казахстан (\textnumero~ИРН AP09260317).
\end{itemize}

Авторы признательны всем своим коллегам, с которыми неоднократно обсуждали вопросы и проблемы, изложенные в данной работе:
\begin{enumerate}
    \item Россия
        \begin{itemize}
            \item Москва
            \begin{itemize}
                \item РАН -- А.М. Сергеев, К.Р. Нигматуллина.
                \item ИВМ РАН -- Е.Е.~Тыртышников, А.А. Романюха, Г.А. Бочаров.
                \item МФТИ -- А.А. Шананин, А.В. Гасников, Н.В. Трусов.
                \item Сколтех -- И.В. Оселедец.
                \item ИСП РАН -- А.И. Аветисян.
                \item ВЦ им. А.А. Дородницына РАН -- Ю.Г. Евтушенко.
            \end{itemize}
            \item Санкт-Петербург
            \begin{itemize}
                \item ИТМО -- А.И. Боровков, В.Н. Леоненко.
            \end{itemize}
            \item Екатеринбург
            \begin{itemize}
                \item РФЯЦ-ВНИИТФ им. академ. Е.И. Забабахина -- Г.Н. Рыкованов, С.Н.~Лебедев, О.В. Зацепин.
            \end{itemize}
            \item Красноярск
            \begin{itemize}
                \item ИВМ СО РАН -- В.В. Шайдуров, В.С. Петракова.
            \end{itemize}
            \item Новосибирск
            \begin{itemize}
                \item НСО -- А.В. Васильев.
                \item СО РАН -- В.Н. Пармон, М.И. Воевода, С.Р. Сверчков.
                \item ИВМиМГ СО РАН -- Г.А. Михайлов, Г.З. Лотова, Н.Ю. Зятьков.
                \item ИМ СО РАН -- М.А. Шишленин, Е.П. Вдовин.
                \item ФИЦ ИВТ СО РАН -- Ф.А. Колпаков, И.Н. Киселев.
                \item ФИЦ ИЦиГ СО РАН -- А.Н. Колчанов.
                \item МЦА НГУ -- М.И. Сосновская, А.В. Неверов.
            \end{itemize}
        \end{itemize}
      \item Китай
      \begin{itemize}
        \item Шанхай, Fudan University -- Jin Cheng.
    \end{itemize}
    \item США
      \begin{itemize}
        \item Сиэттл, Institute for Disease Modeling -- Cliff Kerr.
    \end{itemize}
    \item Великобритания
      \begin{itemize}
        \item Лидс, University of Leeds -- Daniel Lesnic.
    \end{itemize}
    \item Казахстан
      \begin{itemize}
        \item КазНПУ им. Абая -- М.А. Бектемесов.
    \end{itemize}
    \item Болгария
      \begin{itemize}
        \item София, University of Sofia -- Н. Попиванов.
    \end{itemize}
\end{enumerate}

%% file: text/7references.tex
%
%
%
%

%% file: text/Appendix.tex
\newpage

\appendix
\section*{Приложения}
\addcontentsline{toc}{section}{Приложения}
\renewcommand{\thesubsection}{\Alph{subsection}}
\renewcommand\thefigure{\thesubsection.\arabic{figure}}
\setcounter{figure}{0}
\renewcommand\thetable{\thesubsection.\arabic{table}}
\setcounter{table}{0}

\subsection{Обработка эпидемиологических данных}\label{sec_app_data}
Для анализа данных, описанного в разделах \ref{sec_regression}-\ref{sec_ML_models} и дальнейшего построения SEIR-HCD (Раздел~\ref{sec_R0-SEIR-HCD}) модели и АОМ (Раздел~\ref{sec_ABM_COVID}), были использованы следующие показатели для Новосибирской области:
\begin{enumerate}
    \item Официальные статистические данные о выявленных случаях заражения COVID-19, проведенных ПЦР тестов, критических пациентах (подключение аппарата ИВЛ) и умерших были получены более чем из 1000 различных открытых источников с помощью разработанной нами программы автоматического сбора данных. Эта информация доступна на сайте группы COVID-19 СО РАН по ссылке:
          \url{http://covid19-modeling.ru/data/novosibirsk-region-data.csv}
    \begin{figure}[h!]
        \centering
        \includegraphics[width=0.9\textwidth]{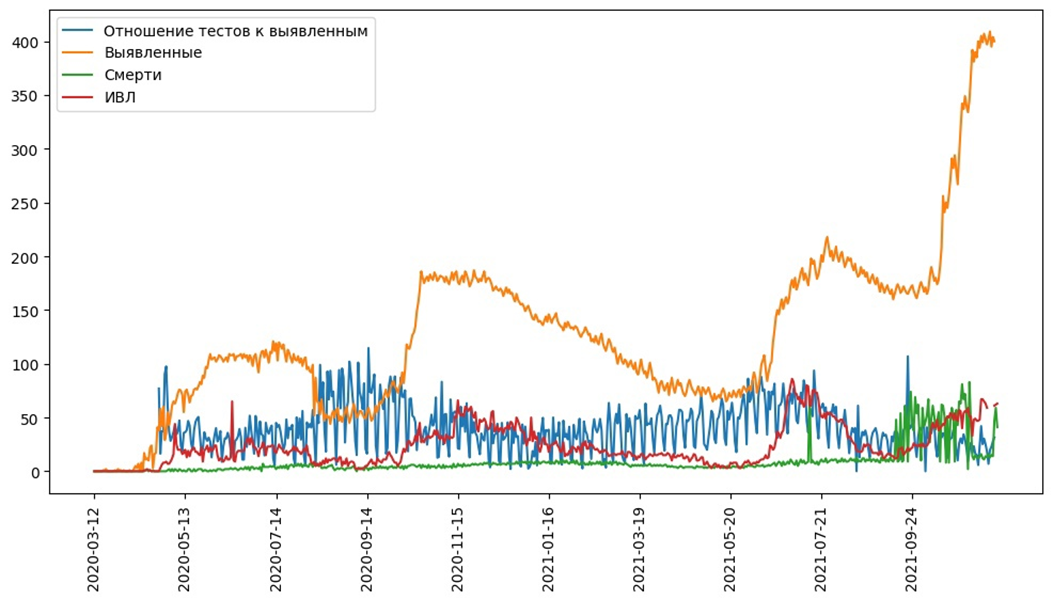}
        \caption{Статистические данные COVID-19 по Новосибирской области с 12.03.2020 по 30.11.2021 (стопкоронавирус.рф):\\
        \textcolor{blue}{Синяя линия} -- отношение ежедневных тестов ПЦР к выявленным случаям, \textcolor{orange}{оранжевая линия} -- выявленные случаи COVID-19, \textcolor{OliveGreen}{зеленая линия} -- умершие от COVID-19, \textcolor{red}{красная линия} -- критические случаи, находящиеся на ИВЛ.}
        \label{fig:Nsk_stat_data}
    \end{figure}
    
    \item Информация о распределении населения по возрастным группам доступна на сайте Федеральной службы государственной статистики~\cite{age_Novosibirsk});
\begin{table}[!ht]
\centering
\begin{tabular}{|p{5cm}|p{4cm}| } 
 \hline
 Возрастная группа, лет &\ Количество человек\\ 
  \hline
  0-9 & 357814  \\
  \hline
  10-19 & 279706 \\
   \hline
  20-29 & 316949 \\
  \hline
 30-39 & 493491\\
 \hline
 40-49 & 391877\\
 \hline
 50-59 & 343950\\
 \hline
 60-69 & 353261\\
 \hline
 70-79 & 157762\\
 \hline
 80$+$ & 103360\\
 \hline
  \end{tabular}
\caption{Распределение численности населения Новосибирской области по возрастным группам
на 1 января 2020 года.}
\label{ages}
\end{table}
    
    \item Сведения о среднем размере семьи: 2,5 человека согласно отчету Росстата  \url{https://novosibstat.gks.ru/storage/mediabank/p54_PRESS147_2020.pdf)}.
    
\end{enumerate}

Для построения SEIR-HCD модели, описанной в разделе~\ref{sec_equation_models}, были использованы следующие данные по Новосибирской области:
\begin{enumerate}
    \item Данные по новым выявленным случаям заражения COVID-19, госпитализированным, критическим (требующих подключение аппарата ИВЛ) и умершим. Информация была агрегирована с открытых источников сети Интернет с помощью разработанной программы для автоматического сбора данных и доступна для скачивания по ссылке: \url{http://covid19-modeling.ru/data/novosibirsk-region-data.csv}.
    \item Данные по проценту бессимптомных выявленных случаев от общего числа выявленных случаев заражения COVID-19, а также его прогноз на период моделирования. Данные получены из ежедневных сводок оперативного штаба Москвы, публикуемые в их официальном Телеграм-канале и доступные по ссылке: \url{https://t.me/COVID2019_official}.
    \item Индекс самоизоляции от Яндекса, а также его прогноз на период моделирования. Индекс публикуется компанией Яндекс и доступен по ссылке: \url{https://yandex.ru/company/researches/2020/podomam}.
    \item Данные по проценту индивидуумов с антителами к COVID-19 от медицинского центра «Инвитро», а также его прогноз на период моделирования. Ежедневная динамика процента индивидуумов с антителами к COVID-19 по лабораторным данным медицинского центра «Инвитро» в Новосибирске доступна по следующей ссылке: \url{https://www.invitro.ru/l/invitro_monitor/}.
\end{enumerate}

\begin{figure}[h!]
        \centering
        \includegraphics[width=0.9\textwidth]{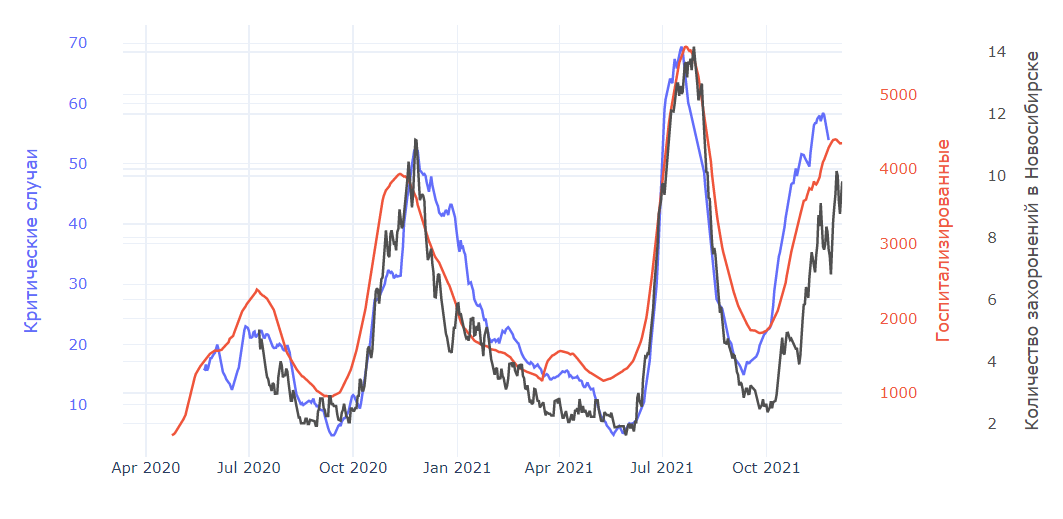}
        \caption{Количество госпитализированных (\textcolor{red}{красная линия}) и критических (\textcolor{blue}{синяя линия}) случаев COVID-19 по Новосибирской области с 24.04.2020 по 06.12.2021 и количество захоронений в городе Новосибирске (\textcolor{black}{черная линия}).}
        \label{fig:Nsk_Hosp-ICU-death_stat_data}
    \end{figure}

\begin{figure}[h!]
        \centering
        \includegraphics[width=0.9\textwidth]{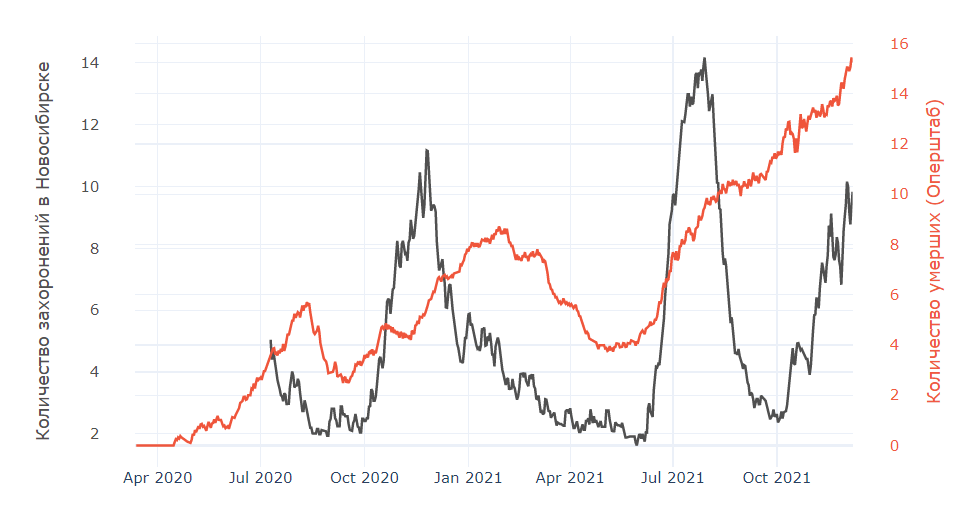}
        \caption{Количество умерших в результате COVID-19 по Новосибирской области (данные Оперштаба стопкоронавирус.рф, \textcolor{red}{красная линия}) и захоронений в городе Новосибирске (\textcolor{black}{черная линия}) в результате COVID-19 с 24.04.2020 по 06.12.2021.}
        \label{fig:Nsk_death_stat_data}
    \end{figure}

\begin{figure}[h!]
        \centering
        \includegraphics[width=0.9\textwidth]{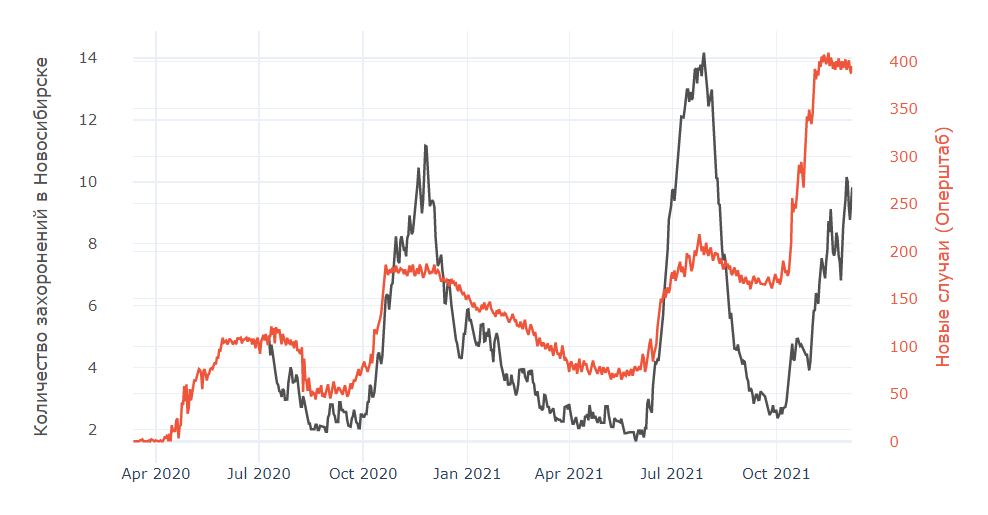}
        \caption{Количество захоронений в городе Новосибирске (\textcolor{black}{черная линия}) в результате COVID-19 и новые выявленные случаи заражения COVID-19 (данные Оперштаба стопкоронавирус.рф, \textcolor{red}{красная линия}) с 24.04.2020 по 06.12.2021.}
        \label{fig:Nsk_newcases_death_stat_data}
    \end{figure}

В силу того, что данные по смертности с официальных источников не коррелируют с остальными статистиками, в численных расчетах использовались данные по захоронениям в Новосибирске Муниципальной информационной системы <<Ритуал>> (Мэрия города Новосибирска), предоставленным директором Центра по взаимодействию с органами власти и индустриальными партнерами Новосибирского государственного университета, к.ф.-м.н. А.Н.~Люлько (см.~Рис.~\ref{fig:Nsk_Hosp-ICU-death_stat_data},~\ref{fig:Nsk_death_stat_data} и~\ref{fig:Nsk_newcases_death_stat_data}). На Рис.~\ref{fig:Nsk_Hosp-ICU-death_stat_data} изображены статистические данные с сайта стопкоронавирус.рф по количеству госпитализированных (красная линия) и критических (синяя линия) случаев COVID-19 и данные по захоронениям в городе Новосибирске по причине COVID-19 (черная линия), сглаженные 14-дневной экспоненциальной скользящей средней. Наблюдается хорошая корреляция данных. Однако данные по смертности от COVID-19 (красная линия на Рис.~\ref{fig:Nsk_death_stat_data}) в Новосибирской области и количество захоронений в городе Новосибирске (черная линия) слабо коррелируют друг с другом (см.~Табл.~\ref{tab:correlation_data}).
    
Сравнение графиков данных по захоронениям в городе Новосибирске по причине COVID-19 и новым выявленных случаям заражения COVID-19 (красная линия на Рис.~\ref{fig:Nsk_newcases_death_stat_data}) приведено на Рис.~\ref{fig:Nsk_newcases_death_stat_data}. Наблюдается частичная корреляция данных в смысле градиентного роста/падения. Однако в периоды 20.10.2020--06.12.2020 и 25.06.2021--30.07.2021 количество выявленных случаев в Новосибирской области резко перестало расти, а количество захоронений продолжало увеличиваться. Данный эффект возможен из-за небольшого снижения числа проведенных ПЦР-тестов в указанные периоды (см.~Рис.~\ref{fig:Nsk_stat_data}, синяя линия).

\begin{table}[H]
\centering
\begin{tabular}{|p{4cm}|p{2cm}|p{2cm}|p{2cm}|p{2cm}|p{2cm}| } 
 \hline
  & Госпитали-зированные & Критичес-кие & Захороне-ния в Нск & Смерти (Оперштаб) & Новые случаи\\ 
  \hline
  Госпитализированные & 1.00 & 0.89 & 0.95 & 0.54 & 0.73 \\
  \hline
  Критические & 0.89 & 1.00 & 0.88 & 0.58 & 0.79 \\
   \hline
  Захоронения в Нск & 0.95 & 0.88 & 1.00 & 0.37 & 0.58 \\
  \hline
  Смерти (Оперштаб) & 0.54 & 0.58 & 0.37 & 1.00 & 0.87 \\
 \hline
 Новые случаи & 0.73 & 0.79 & 0.58 & 0.87 & 1.00\\
 \hline
  \end{tabular}
\caption{Коэффициенты корреляция скользящих средних по статистическим данным о количестве госпитализированных, критических, выявленных и смертей от COVID-19 в Новосибирской области и количества захоронений от COVID-19 в городе Новосибирске (графики изображены на Рис.~\ref{fig:Nsk_Hosp-ICU-death_stat_data}, \ref{fig:Nsk_death_stat_data} и \ref{fig:Nsk_newcases_death_stat_data}).}
\label{tab:correlation_data}
\end{table}

Для более полного анализа и последующего построения более точных сценариев развития эпидемиологической ситуации в регионе необходим анализ следующих данных, которые не публикуются в открытых источниках:
\begin{itemize}
    \item Информация о количестве завезенных случаев COVID-19 (из других регионов РФ или из-за границы);
    \item Информация о доле больничных коек, занятых больными COVID-19, от всех доступных в учреждениях здравоохранения;
    \item Определение штамма вируса при выявлении заражения (см.~Табл.~\ref{tab_strains});
    \begin{table}[!ht]
\centering
\begin{tabular}{|p{3cm}|p{2.9cm}|p{4.4cm}|p{2.8cm}| } 
 \hline
 {\bf Название штамма} & {\bf Линия Панго} & {\bf Место обнаружения} & {\bf Дата}\\ 
  \hline
  Альфа & B.1.1.7 & Великобритания & 20.09.2020\\
  Бета & B.1.351 & ЮАР & 05.2020\\
  Гамма & P.1 & Бразилия & 11.2020\\
  Дельта & B.1.617.2 & Индия & 10.2020\\
  Омикрон & B.1.1.529 & ЮАР, Ботсвана & 09.11.2021\\
  \hline
\end{tabular}
\caption{Наиболее значимые штаммы вируса SARS-CoV-2 по версии ВОЗ на 31.12.2021.}
\label{tab_strains}
\end{table}
    \item Идентификация суперраспространителей COVID-19 при проведении тестирования.
\end{itemize}

\subsection{Комплекс программ}\label{sec_app_software}

Для построения агентно-ориентированной модели, описанной в разделе \ref{sec_ABM_COVID}, был использован программный комплекс Covasim (\url{https://github.com/InstituteforDiseaseModeling/covasim}), разработанный институтом Institute for Disease Modeling (\url{https://www.idmod.org/}). Данная библиотека написана на языке Python и создана для исследования агентных моделей COVID-19 с нетривиальными структурами. Covasim использовался для анализа эпидемиологической ситуации в более чем десяти странах, а также являлся одним из инструментов для принятия решений о введении ограничительных мер в США, Великобритании и Австралии.

Для решения обратных задач для агентно-ориентированной модели (раздел \ref{sec_InversePr}, был подготовлен программный код на языке Python для интеграции разработанного поэтапного метода восстановления неизвестных параметров с библиотекой Covasim. Исходные файлы и примеры использования доступны по ссылке на Github (\url{https://github.com/msosnovskaya/autocalibration-covasim-pub}). Также данный программный код был зарегистрирован в Роспатент (Номер свидетельства: 2021614740).

%% file: text/Authors.tex
\newpage

\section*{Об авторах}
\addcontentsline{toc}{section}{Об авторах}

\begin{tabular}{l l}
\begin{minipage}{0.22\textwidth}
\centering
\includegraphics[width=\textwidth]{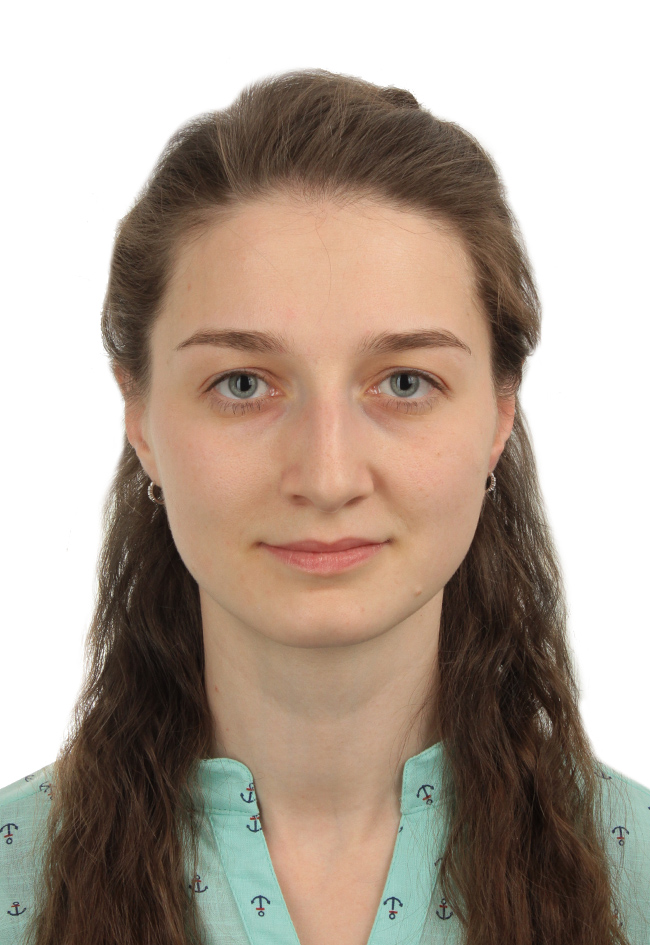}
\end{minipage}
&
\begin{minipage}{0.59\textwidth}
{\bf КРИВОРОТЬКО ОЛЬГА ИГОРЕВНА}, к.ф.-м.н. (2015), старший научный сотрудник лаборатории обратных задач естествознания Института вычислительной математики и математической геофизики СО РАН, доцент кафедры математических методов геофизики механико-математического факультета НГУ и кафедры высшей математики физического факультета НГУ.

Основные научные интересы -- численные методы решения обратных задач эпидемиологии, иммунологии, биологии, цифровой экономики, медицины, социальных процессов, идентифицируемость, оптимизация, анализ и обработка данных методами машинного обучения.
\end{minipage}
\end{tabular}

\vspace*{2mm}

Руководитель 7 научных проектов Российского научного фонда, Российского фонда фундаментальных исследований, Президентских грантов для молодых ученых с 2016 года. Лауреат премии мэрии города Новосибирска в сфере науки и инноваций в номинации ''Лучший молодой исследователь в организациях науки'' за разработку карты прогноза распространения социально-значимых заболеваний в городе Новосибирске (2020) и премии имени Г.И.~Марчука за работу ''Идентифицируемость математических моделей иммунологии и эпидемиологии'' (2021).

\vspace*{5mm}

\begin{tabular}{l l}
\begin{minipage}{0.21\textwidth}
\centering
\includegraphics[width=\textwidth]{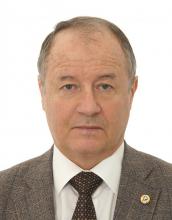}
\end{minipage}
&
\begin{minipage}{0.6\textwidth}
{\bf КАБАНИХИН СЕРГЕЙ ИГОРЕВИЧ}, профессор (1993), член-корреспондент РАН (2011), главный научный сотрудник лаборатории обратных задач естествознания ИВМиМГ СО РАН, главный научный сотрудник ИМ СО РАН, заведующий кафедрой математических методов геофизики ММФ НГУ, заведующий лабораторией методов создания, исследования и идентификации математических моделей естествознания НГУ.

Основные научные интересы -- теория и численные методы решения обратных задач естествознания.
\end{minipage}
\end{tabular}

\vspace*{2mm}

Член бюро Отделения математических наук РАН, член Президиума Сибирского отделения РАН, член бюро объединенного Ученого совета СО РАН по математике и информатике. Главный редактор журналов \textit{Journal of Inverse and Ill-Posed Problems} (WoS Q1) и \textit{Сибирский журнал вычислительной математики} (WoS).

{\bf Основные научные публикации (книги):}
\begin{enumerate}
\itemsep0em 
    \item S.I. Kabanikhin. Inverse and Ill-Posed Problems. Theory and Applications. Berlin/Boston, 2012.
    \item Кабанихин С.И. Обратные и некорректные задачи. Сибирское научное издательство, Новосибирск, 2008.
    \item S.I. Kabanikhin, A.D. Satybaev, M.A. Shishlenin. Direct methods of solving inverse hyperbolic problems.  VSP/BRILL, the Netherlands, 2004.
    \item S.I. Kabanikhin, A. Lorenzi. Identification Problems for Wave Phenomena, Utrecht, the Netherlands, VSP, 1999. 
    \item S.I. Kabanikhin, V.G. Romanov. Inverse Problems for Maxwell’s Equations, Utrecht, The Netherlands, VSP, 1994. 
    \item Кабанихин С.И. Проекционно-разностные методы определения коэффициентов гиперболических уравнений, Новосибирск, Наука, Сибирское отделение, 1988. 
\end{enumerate}

\section*{Основные совместные публикации авторов}
\addcontentsline{toc}{section}{Основные совместные публикации авторов}

\begin{enumerate}
\itemsep0em
    \item O.I. Krivorotko, S.I. Kabanikhin, M.I. Sosnovskaya, D.V. Andornaya. Sensitivity and identifiability analysis of COVID-19 pandemic models // Vavilov Journal of Genetics and Breeding. 2021. V. 25(1). P. 82-91. DOI: 10.18699/VJ21.010
    \item S. Kabanikhin, O. Krivorotko, A. Takuadina, D. Andornaya, S. Zhang. Geo-information system of tuberculosis spread based on inversion and prediction // Journal of Inverse and Ill-Posed Problems 2020 V. 29(1). P. 65-79. DOI: 10.1515/jiip-2020-0022
    \item O.I. Krivorotko, S.I. Kabanikhin, M.Yu. Zyatkov, A.Yu. Prikhodko, N.M. Prokhoshin, M.A. Shishlenin. Mathematical modeling and forecasting of COVID-19 in Moscow and Novosibirsk region // Numerical Analysis and Applications. 2020. V. 13(4). P. 332-348. DOI: 10.1134/S1995423920040047
    \item O.I. Krivorotko, D.V. Andornaya, S.I. Kabanikhin. Sensitivity Analysis and Practical Identifiability of Some Mathematical Models in Biology // Journal of Applied and Industrial Mathematics. 2020. V. 14(1). P. 115-130. DOI: 10.1134/S1990478920010123
    \item O. Krivorotko, S. Kabanikhin, Sh. Zhang, V. Kashtanova. Global and local optimization in identification of parabolic systems // Journal of Inverse and Ill-Posed Problems 2020 V. 28(6). P. 899-913. DOI: 10.1515/jiip-2020-0083
    \item S.I. Kabanikhin, O.I. Krivorotko, M.A. Bektemessov, Zh.M. Bektemessov, Sh. Zhang. Differential evolution algorithm of solving an inverse problem for the spatial Solow mathematical model // Journal of Inverse and Ill-Posed Problems 2020 V. 28(5). P. 761-774. DOI: 10.1515/jiip-2020-0108
    \item S.I. Kabanikhin, O.I. Krivorotko. Mathematical modeling of the Wuhan COVID-2019 epidemic and inverse problems // Computational Mathematics and Mathematical Physics. 2020. V. 60(11). P. 1889-1899, DOI: 10.1134/S0965542520110068
    \item S.I. Kabanikhin, O.I. Krivorotko. Optimization methods for solving inverse immunology and epidemiology problems // Computational Mathematics and Mathematical Physics. 2020. V. 60(4). P. 580-589. DOI: 10.1134/S0965542520040107
    \item H.T. Banks, S.I. Kabanikhin, O.I. Krivorotko, D.V. Yermolenko. A numerical algorithm for constructing an individual mathematical model of HIV dynamics at cellular level // Journal of Inverse and Ill-Posed Problems. 2018. V. 26(6). P. 859-873. DOI: 10.1515/jiip-2018-0019
    \item S.I. Kabanikhin, O.I. Krivorotko. An algorithm for source reconstruction in nonlinear shallow-water equations // Computational Mathematics and Mathematical Physics. 2018. V. 58(8). P. 1334-1343. DOI: 10.1134/S0965542518080109
    \item S. Kabanikhin, O. Krivorotko, V. Kashtanova. A combined numerical algorithm for reconstructing the mathematical model for tuberculosis transmission with control programs // Journal of Inverse and Ill-Posed Problems. 2018. V. 26(1). P. 121-131. DOI: 10.1515/jiip-2017-0019
    \item S.I. Kabanikhin, O.I. Krivorotko. A numerical algorithm for computing tsunami wave amplitudes // Numerical Analysis and Applications. 2016. V. 9(2). P. 118–128.
    \item S.I. Kabanikhin, O.I. Krivorotko. Identification of biological models described by systems of nonlinear differential equations // Journal of Inverse and Ill-Posed Problems. 2015. V. 23(5). P. 519-527.
    \item S. Kabanikhin, A. Hasanov, I. Marinin, O. Krivorotko, D. Khidasheli. A variational approach to reconstruction of an initial tsunami source perturbation // Applied Numerical Mathematics. 2014. V. 83. P. 22–37.
\end{enumerate}